\documentclass[%
reprint,
amsmath,amssymb,
aps,
]{revtex4-2}

\usepackage{graphicx}
\usepackage{dcolumn}
\usepackage{bm}
\usepackage{multirow}
\usepackage{longtable}
\usepackage{color}

\begin{document}

\title{Finite-nuclear-size effect in hydrogen-like ions with relativistic nuclear structure}

\author{Hui Hui Xie}
\author{Jian Li}\email{jianli@jlu.edu.cn}
\author{Li Guang Jiao}\email{lgjiao@jlu.edu.cn}
\affiliation{College of Physics, Jilin University, Changchun 130012}

\author{Yew Kam Ho}
\affiliation{Institute of Atomic and Molecular Sciences, Academia Sinica, Taipei 10617}

\date{\today}

\begin{abstract}

The finite-nuclear-size (FNS) effect has a large contribution to the atomic spectral properties especially for heavy nuclei. By adopting the microscopic nuclear charge density distributions obtained from the relativistic continuum Hartree-Bogoliubov (RCHB) theory, we systematically investigate the FNS corrections to atomic energy levels and bound-electron $g$ factors of hydrogen-like ions with nuclear charge up to $118$. The comparison of the present numerical calculations with the predictions from empirical nuclear charge models, the non-relativistic Skyrme-Hartree-Fock calculations, and the results based on experimental charge densities indicate that both the nuclear charge radius and the detailed shape of charge density distribution play important roles in determining the FNS corrections. The variation of FNS corrections to energy levels and $g$ factors with respect to the nuclear charge are investigated for the lowest several bound states of hydrogen-like ions. It is shown that they both increase by orders of magnitude with increasing the nuclear charge, while the ratio between them has a relatively weak dependence on the nuclear charge. The FNS corrections to the $s_{1/2}$ and $p_{1/2}$ bound state energies from the RCHB calculations are generally in good agreement with the analytical estimations by Shabaev [J. Phys. B, 26, 1103 (1993)] based on the homogeneously charged sphere nuclear model, with the discrepancy indicating the distinct contribution of microscopic nuclear structure to the FNS effects.

\end{abstract}


\maketitle

\section{Introduction}
\label{sec-1}

It has become increasingly important to include the finite-nuclear-size (FNS) effect, i.e., the difference due to replacement of point-like nucleus by finite-size nucleus, in sophisticated electronic structure calculations of atoms and molecules \cite{andrae2002nuclear,ANDRAE2000413}. Especially for systems with heavy nuclei, e.g., the highly charged ions, the contribution of FNS effect to system energies is comparable to or even larger than the quantum electrodynamics (QED) corrections \cite{Drake,PhysRevA.77.032501}. In addition, the FNS effects in electromagnetic interaction and weak decay of elementary particles that involve electrons have been taken into account for a long time, such as the internal conversion \cite{church1960nuclear,PhysRev.104.1382}, $\beta$-decay \cite{BEHRENS1970481,PhysRev.83.190.2,RevModPhys.90.015008}, and electron-capture decay \cite{Sarriguren_2020,RevModPhys.49.77}.

The study of FNS effects on the spectral properties of few-electron multi-charged or muonic ions is of special interest as such effects could provide critical entries for a variety of fundamental research. Since the experimental \cite{PhysRevA.87.030501,Kraft_Bermuth_2017,PhysRevLett.80.3022,PhysRevLett.85.5308,PhysRevLett.92.093002,PhysRevLett.94.223001,PhysRevLett.107.023002,PhysRevLett.122.253001,PhysRevLett.123.123001} and theoretical \cite{MOHR1998227,yerokhin2015lamb,GLAZOV2002408,johnson1985lamb,IEIDES200163,BEIER200079} determination of atomic energy levels and bound-electron $g$ factors have been considerably improved in recent years, various applications of these quantities are being available in the literature, which include stringent test of the QED theory \cite{shabaev2018stringent,PhysRevLett.126.173001,volotka2014nuclear,Indelicato_2019,Blaum_2020,Sailer2022}, determination of the fine-structure constant \cite{PhysRevLett.114.150801,PhysRevLett.109.111807,PhysRevLett.105.120801,PhysRevLett.96.253002,PhysRevLett.94.243002} and electron's mass \cite{PhysRevLett.88.011603,PhysRevA.96.030501,sturm2014high}, as well as tests for new physics beyond the standard model \cite{DEBIERRE2020135527,RevModPhys.90.025008}. With the fast development of high-precision spectroscopic measurements in few-electron atoms and highly charged ions, it is crucial from the theoretical aspect to accurately calculate the FNS corrections to system energy levels and bound-electron $g$ factors.

Hydrogen-like ions are ideal systems for theoretical study of FNS effects due to the absence of intractable electron correlations, and in the past decades they have been extensively investigated in the literature \cite{Shabaev_1993,Deck_2005,PhysRevA.72.042101,PhysRevA.99.032508,ANDRAE2000413,babak2018,BORISOGLEBSKY1979175,PRA101,VISSCHER1997207,lee2007g}. Earlier efforts were made to perform a model-independent statement of FNS effects for light muonic atoms and hydrogen-like ions using perturbation theory, and an analytic formula expressed in terms of the moments of nuclear charge distribution was proposed \cite{PhysRev.86.288,ERICKSON1965271,FRIAR1979151}. However, the nuclear moments are model-dependent and, therefore, such an analytic formula becomes inaccurate for nuclei with complex charge distributions. On the other hand, the validity of the analytic formula based on perturbation theory has not been verified in heavy muonic atoms and hydrogen-like ions. The further development of approximate formula for FNS corrections in the lowest several bound states of hydrogen-like ions were performed based on empirical nuclear charge models, such as the homogeneously charged sphere, Gaussian distribution, and the two-parameter Fermi charge distribution \cite{Shabaev_1993,GLAZOV2002408,Deck_2005,PhysRevA.72.042101}. Although these approximate formulas have been widely used in atomic spectral investigations, accurate calculation of FNS corrections based on microscopic and sophisticated nuclear charge distributions, i.e., a more realistic description of the nuclear structure, is highly anticipated. Such an effort was recently carried out by Valuev \textit{et al.} \cite{PRA101} using the Hartree-Fock method based on the Skyrme-type nuclear interaction. The obtained FNS corrections to energy levels and bound-electron $g$ factors for hydrogen-like ions of $^{40}{\rm Ca}^{19+}$, $^{116}{\rm Sn}^{49+}$, and $^{208}{\rm Pb}^{81+}$ show slight discrepancies with those estimated from empirical nuclear charge models. An important conclusion derived by the authors in their work was that the nuclear charge radius has great influence on the magnitude of the FNS effects. In view of the non-relativistic nature of the Skyrme-Hartree-Fock method as well as of the Skyrme-type nuclear interaction  which includes adjustable parameters to reproduce the experimental nuclear charge radii, it is worthwhile to perform alternative nuclear structure calculations and investigate the corresponding FNS effects in atomic spectral properties.

In recent years, the covariant (relativistic) density functional theory (CDFT) has attracted considerable attention on account of its successful description of the complex nuclear structure and reaction dynamics \cite{RING1996193,MENG2006470,Meng2015,VRETENAR2005101,NIKSIC2011519,meng2013progress}. For instance, it can well reproduce the isotopic shifts in the Pb isotopes \cite{SHARMA19939} and naturally give the origin of the pseudospin and spin symmetries in the anti-nucleon spectrum \cite{PhysRevLett.91.262501,liang2010spin,LIANG20151}, as well as provide a good description of the nuclear magnetic moments \cite{lijian1,lijian2}. Based on the CDFT with Bogoliubov transformation in the coordinate representation, the relativistic continuum Hartree-Bogoliubov (RCHB) method was developed to provide a proper treatment of the pairing correlations and mean-field potentials in the presence of continuum \cite{PhysRevLett.77.3963,Meng1998NPA}. The RCHB method has also achieved great success in a variety of aspects about exotic nuclei, such as providing a microscopic self-consistent description of halo in $^{11}$Li \cite{PhysRevLett.77.3963}, predicting the giant halo phenomena in light and medium-heavy nuclei \cite{PhysRevLett.80.460,PhysRevC.65.041302,Shuang_Quan_2002,Meng_2015}, as well as reproducing the interaction cross sections and charge-changing cross sections in sodium isotopes and other light exotic nuclei \cite{MENG19981,MENG2002209}.

In the present work, the FNS effects on the atomic energy levels and bound-electron $g$ factors of hydrogen-like ions are studied by employing the realistic nuclear charge distributions constructed in the framework of RCHB method. The paper is organized as follows. In Sec. \ref{sec-2}, we first introduce the RCHB method for solving the nuclear structure and then construct the nuclear charge densities to obtain the electrostatic interaction, followed by the definition of FNS corrections to atomic energy levels and bound-electron $g$ factors. In Sec. \ref{sec-3}, we take a systematic investigation of the FNS corrections with varying the nuclear charge and compare the present results with previous numerical calculations and analytical estimations. Finally, a brief summary and outlook is presented in Sec. \ref{sec-4}.

\section{Theoretical method}\label{sec-2}

\subsection{Relativistic continuum Hartree-Bogoliubov method}

The RCHB method with contact interaction between nucleons is employed to calculate the proton and neutron densities and from which the nuclear charge density and electrostatic potential can be properly constructed. In the RCHB method, the conventional finite-range meson-exchange interaction between nucleons is replaced by the corresponding local four-point interaction.
Starting from the Lagrangian density of the point-coupling model as\cite{PhysRevC.82.054319}
\begin{eqnarray}\label{EQ:LAG}
	{\cal L} &=& \bar\psi(i\gamma_\mu\partial^\mu-M)\psi-\frac{1}{2}\alpha_S(\bar\psi\psi)(\bar\psi\psi)\nonumber\\
	& &-\frac{1}{2}\alpha_V(\bar\psi\gamma_\mu\psi)(\bar\psi\gamma^\mu\psi)
	-\frac{1}{2}\alpha_{TV}(\bar\psi\vec{\tau}\gamma_\mu\psi)(\bar\psi\vec{\tau}\gamma^\mu\psi)\nonumber\\
	& &-\frac{1}{2}\alpha_{TS}(\bar\psi\vec{\tau}\psi)(\bar\psi\vec{\tau}\psi)
	-\frac{1}{3}\beta_S(\bar\psi\psi)^3\nonumber-\frac{1}{4}\gamma_S(\bar\psi\psi)^4\nonumber\\
	& &-\frac{1}{4}\gamma_V[(\bar\psi\gamma_\mu\psi)(\bar\psi\gamma^\mu\psi)]^2-\frac{1}{2}\delta_S\partial_\nu(\bar\psi\psi)\partial^\nu(\bar\psi\psi)\nonumber\\
	& &	-\frac{1}{2}\delta_V\partial_\nu(\bar\psi\gamma_\mu\psi)\partial^\nu(\bar\psi\gamma^\mu\psi)\nonumber\\
	& &
	-\frac{1}{2}\delta_{TV}\partial_\nu(\bar\psi\vec\tau\gamma_\mu\psi)\partial^\nu(\bar\psi\vec\tau\gamma_\mu\psi)-\frac{1}{2}\delta_{TS}\partial_\nu(\bar\psi\vec\tau\psi)\partial^\nu(\bar\psi\vec\tau\psi)\nonumber\\
	& &	-\frac{1}{4}F^{\mu\nu}F_{\mu\nu}-e\frac{1-\tau_3}{2}\bar\psi\gamma^\mu\psi A_\mu,
\end{eqnarray}
where $M$ is the nucleon mass; $A_{\mu}$ and  $F_{\mu\nu}$ are respectively the four-vector potential and field strength tensor of the
electromagnetic field.
Here $\alpha_S,~\alpha_V, ~\alpha_{TS}$ and $\alpha_{TV}$ represent the
coupling constants for four-fermion point-coulping terms, $\beta_{S},
~\gamma_S$ and $\gamma_{V}$ are those for the higher-order terms which are responsible for the effects of medium dependence, and $\delta_{S}, ~\delta_{V}, ~\delta_{TS}$ and $\delta_{TV}$ refer to those for the gradient terms which are included to simulate the finite range effects. The subscripts $S$, $V$ and $T$ stand for scalar, vector and isovector, respectively.

The energy density functional of the nuclear system is constructed under the mean-field and no-sea approximations.By minimizing the energy density functional 
, one obtains the Dirac equation for nucleons within the framework of relativistic mean-field theory as~\cite{Meng2015} 

\begin{equation}\label{eq-single-dirac}
	\left[\bm \alpha\cdot\bm p+V(\bm r)+\beta M+S(\bm r)\right]\phi_k(\bm r)=\varepsilon\phi_k(\bm r),
\end{equation}
in which $\bm\alpha$ and $\beta$ are the traditional $4\times 4$ matrices of Dirac operators, and $\phi_k(\bm r)$ is the corresponding single-particle wave function for a nucleon in the state $k$. The local scalar $S(\bm r)$ and vector $V(\bm r)$ potentials are given by
\begin{eqnarray}
	S(\mathbf{r})&=&\alpha_S \rho_S + \beta_S \rho_S^2 +\gamma_S\rho_S^3 +\delta_S\triangle \rho_S,\\
	\label{eq:vaspot}
	V^{\mu}(\mathbf{r})&=&\alpha_V j^{\mu} + \gamma_V(j_{\mu}j^{\mu})j^{\mu} +\delta_V \triangle j^{\mu}+e A^{\mu}\nonumber\\
	& &+\alpha_{TV}\tau_3 \vec{j}^{\mu}_{TV}+\delta_{TV} \tau_3 \triangle \vec{j}^{\mu}_{TV}.
	\label{eq:vavpot}
\end{eqnarray}


In order to describe open-shell nuclei, pairing correlations are crucial. A self-consistent description of pairing correction in the presence of the continuum (exactly, the contribution from the resonance states with positive energy) when treated in coordinate representation can be completed by the Bogoliubov quasi-particle transformation, which transforms the single particle basis, $a_i$, $a_i^\dagger$, into the quasiparticle basis, $\beta_i$, $\beta_i^\dagger$ ($i=1,\mathcal{M}$, for a $\mathcal M$-dimension basis) via
\begin{equation}
	\begin{pmatrix}
		a_i\\a_i^\dagger
	\end{pmatrix}=\int\mathrm d^3\bm r\begin{pmatrix}
	U_i(\bm r)&V_i(\bm r)^*\\-V_i(\bm r)&U_i(\bm r)^*
\end{pmatrix}\begin{pmatrix}
\beta_i\\\beta_i^\dagger
\end{pmatrix},
\end{equation}
where $i=(nlj)$ represents the quantum number, $U(\bm r)$ and $V(\bm r)$ are the quasiparticle wave function. 

 The relativistic Hartree-Bogoliubov model employs the unitary Bogoliubov transformation of the single-nucleon creation and annihilation operators in constructing the quasiparticle operator and provides a unified description of both the mean-field approximation and the pairing correlations \cite{VRETENAR2005101}. 
Following the standard procedure of Bogoliubov transformation, the relativistic Hartree-Bogoliubov equation 
could be derived as following: 
\begin{equation}
\label{RHB}
\left(\begin{matrix}
h_D-\lambda_\tau & \Delta \\
-\Delta^* &-h_D^*+\lambda_\tau
\end{matrix}\right)
\left(\begin{matrix}
U_k\\
V_k
\end{matrix}\right)=E_k
\left(\begin{matrix}
U_k\\
V_k
\end{matrix}\right),
\end{equation}
where $E_k$ is the quasiparticle energy, $\lambda_\tau (\tau=n,p)$ are the chemical potentials for neutrons and protons, 
and $h_D$ refers to the Dirac Hamiltonian in Eq.~(\ref{eq-single-dirac}), in which the densities can be constructed by quasiparticle wave functions,
\begin{eqnarray}\nonumber
	\rho_S(\mathbf{r})     &=&\sum_{k>0 }\bar V_k(\mathbf{r})V_k(\mathbf{r}),\\\label{eq:mesonsource}
	\rho_{V}(\mathbf{r})   &=&\sum_{k>0 } V_k^{\dagger}(\mathbf{r})V_k(\mathbf{r}),\\ \nonumber
	\rho_{T V}(\mathbf{r}) &=&\sum_{k>0 } V_k^{\dagger}(\mathbf{r})\tau_3 V_k(\mathbf{r}).
\end{eqnarray}


The pairing potential $\Delta$ in Eq. (\ref{RHB}) is given by
\begin{equation}
\label{Delta}
\Delta(\bm r_1,\bm r_2) = V^{\mathrm{pp}}(\bm r_1,\bm r_2)\kappa(\bm r_1,\bm r_2),
\end{equation}
where
\begin{equation}
\label{pair}
V^{\mathrm{pp}}(\bm r_1,\bm r_2)= V_0 \frac{1}{2}(1-P^\sigma)\delta(\bm r_1-\bm r_2)\left(1-\frac{\rho(\bm r_1)}{\rho_{\mathrm{sat}}}\right),
\end{equation}
represents the density-dependent force of zero range and $\kappa(\bm r_1,\bm r_2)$ refers to the pairing tensor \cite{Ring1980}. In Eq. (\ref{pair}), $V_0$ is the interaction strength and $\rho_\mathrm{sat}$ is the saturation density of the nuclear matter.

With spherical symmetry imposed, the quasiparticle wave function in the coordinate space can be written as 
\begin{eqnarray}
	U_k=\frac{1}{r}\left(
	\begin{array}{c}
		iG_U^{k}(r)Y_{jm}^l(\theta,\phi)\\
		F_U^{k}(r)(\pmb{\sigma\cdot \hat{r}})Y_{jm}^l(\theta,\phi)
	\end{array}\right)\chi_t(t)\nonumber\\
   V_k=\frac{1}{r}\left(
	\begin{array}{c}
		iG_V^{k}(r)Y_{jm}^l(\theta,\phi)\\
		F_V^{k}(r)(\pmb{\sigma\cdot \hat{r}})Y_{jm}^l(\theta,\phi)
	\end{array}\right)\chi_t(t).
\end{eqnarray}
The corresponding RHB equation can be expressed as the following radial integral-differential equations in coordinate space~\cite{Meng1998NPA}:
\begin{eqnarray}\nonumber &
	\frac{\mathrm dG_U}{\mathrm dr}+\frac{\kappa}{r}G_U(r)-(E+\lambda-V(r)+S(r))F_U(r)\\ \nonumber&
	+r\int r'dr'\Delta_F(r,r')F_V(r')=0,\\ \nonumber&
	\frac{\mathrm dF_U}{\mathrm dr}-\frac{\kappa}{r}F_U(r)
	+(E+\lambda-V(r)-S(r))G_U(r)\\ \nonumber&
	+r\int r'dr'\Delta_G(r,r')G_V(r')=0,\\ \nonumber&
	\frac{\mathrm dG_V}{\mathrm dr}+\frac{\kappa}{r}G_V (r)+(E-\lambda+V(r)-S(r))F_V (r)\\ \nonumber&
	+r\int r'dr'\Delta_F(r,r')F_U(r')=0,\\ \nonumber&
	\frac{\mathrm dF_V}{\mathrm dr}-\frac{\kappa}{r}F_V (r)-(E-\lambda+V(r)+S(r))G_V (r)\\ &
	+r\int r'dr'\Delta_G(r,r')G_U(r')=0.\label{rchb-eq}
\end{eqnarray}
Due to the zero-range pairing force, the above coupled integral-differential equations can be reduced to  differential ones, which can be solved in coordinate space using the shooting method with Runge-Kutta algorithms~\cite{Meng1998NPA}. 
After the solution, new densities and fields are obtained from e.g., Eq.~(\ref{eq:mesonsource}), which are iterated in the differential equations until convergence is achieved.

More implementation details of the RCHB method with point-coupling density functionals are available in Refs. \cite{PhysRevC.82.054319,Xia2018} 
. After solving the relativistic Hartree-Bogoliubov equation of Eq. (\ref{rchb-eq}), the point proton and neutron densities are obtained by summing the norm of the corresponding quasiparticle wave functions ($\tau\in\{p,n\}$),
\begin{equation}
	\label{eq:rhov}
	\rho_{\tau}(r) =\rho_{V,\tau}(r)=\sum_{k\in\tau } \frac{n_k}{4\pi r^2}\left\{\left[G_V^k(r)\right]^2+\left[F_V^k(r)\right]^2 \right\},
\end{equation}
where $n_k$ refer to the occupation number of the orbit $k$.

\subsection{Nuclear charge density}

The nuclear charge density is uniquely related to the nuclear charge form factor (namely the charge density distribution in the momentum space), which naturally include the point proton and neutron densities, the proton and neutron spin–orbit densities, and the single-proton and single-neutron charge densities \cite{Friar1975,PhysRevC.62.054303,PhysRevC.103.054310,kurasawa2019n}. 
The nuclear charge form factor $F_c$ is the ground-state expectation value of the zeroth component of the charge current $\hat J_0$, i.e., 
\begin{equation}
	F_c(\bm q)=\int\mathrm d^3\bm r e^{i\bm q\cdot\bm r}\sum_{\tau\in\{p,n\}}\left[G_{\mathrm E\tau}(q^2)\rho_\tau( r)+F_{2\tau}(q^2)W_\tau( r)\right],
\end{equation}
where $G_{\mathrm E\tau}$ and $F_{2\tau}$ denote, respectively, Sachs electric and Pauli form factors of the nucleon. The nucleon density $\rho_\tau(r)$ is obtained from Eq.~(\ref{eq:rhov}), and the spin-orbit density $W_\tau(r)$ is generally given by~\cite{PhysRevC.62.054303,kurasawa2019n}
\begin{equation}
	W_\tau(r)=\frac{\mu_\tau}{2M}\left(-\frac{\nabla^2\rho_\tau(r)}{2M}+i\nabla\cdot\langle 0|\sum_{k\in\tau}\delta(\bm r-\bm r_k)\bm\gamma_k|0\rangle \right),
\end{equation}
where $\mu_\tau$ is the anomalous magnetic moment.

The relativistic nuclear charge density is obtained by the inverse Fourier transformation of the nuclear charge form factor $F_c$ and formally written as
\begin{equation}
\label{eq-rhoc}
\rho_c(r)=\sum_{\tau\in\{p,n\}}\left[\rho_{c\tau}(r)+W_{c\tau}(r)\right],
\end{equation}
where
\begin{align}
&\rho_{c\tau}(r)=\frac1r\int_0^\infty\mathrm dx x\rho_\tau(x)\left[g_\tau(|r-x|)-g_\tau(|r+x|)\right],\\
&W_{c\tau}(r)=\frac1r\int_0^\infty\mathrm dx xW_\tau(x)\left[f_{2\tau}(|r-x|)-f_{2\tau}(|r+x|)\right].
\end{align}
The functions $g_\tau(x)$ and $f_{2\tau}(x)$ are given by
\begin{eqnarray}
g_\tau(x)=\frac1{2\pi}\int_{-\infty}^{\infty}\mathrm dq e^{iqx}G_{\mathrm E\tau}(\bm q^2),\\
f_{2\tau}(x)=\frac{1}{2\pi}\int_{-\infty}^{\infty}\mathrm dq e^{iqx}F_{2\tau}(\bm q^2),
\end{eqnarray}
In this work, we follow Ref. \cite{PhysRevC.62.054303} and adopt the following form factors
\begin{align}
\label{eq-form-factor}
&G_{\mathrm Ep}=\frac{1}{\left(1+\frac{r_p^2\bm q^2}{12}\right)^2}, \ G_{\mathrm En}=\frac{1}{\left(1+\frac{r_+^2\bm q^2}{12}\right)^2}-\frac{1}{\left(1+\frac{r_-^2\bm q^2}{12}\right)^2},\nonumber\\
&F_{2p}=\frac{G_{Ep}}{1+\bm q^2/4M_p^2},\ F_{2n}=\frac{G_{Ep}-G_{En}/\mu_n}{1+\bm q^2/4M_n^2},
\end{align}
with the proton charge radius $r_p= 0.8414$ fm \cite{RevModPhys.93.025010} and $r^2_\pm= r_{\mathrm{av}}^2\pm \frac12 \langle r_n^2\rangle $, where $r_{\mathrm{av}}^2= 0.9$ fm$^2$ is the average of the squared radii for positive and negative charge distributions and $\langle r_n^2\rangle =-0.11$ fm$^2$ \cite{atac2021measurement} is the mean squared charge radii of neutron.

\subsection{FNS corrections to atomic energy levels and bound-electron $g$ factors}

The relativistic Dirac equation for one-electron system reads (atomic units $\hbar=m_e=e=1$ are used),
\begin{equation}
\label{eq-dirac-one-electron}
\left[c\bm\alpha\cdot \bm p+\beta c^2+V(\bm r)\right]\psi(\bm r)=\varepsilon \psi(\bm r),
\end{equation}
where the eigenenergy $\varepsilon$ includes both the electronic state energy $E$ and the rest energy of electron $c^2$, i.e., $\varepsilon=E+ c^2$. When the electrostatic potential $V(\bm r)$ is in spherical symmetry, the electron wave function can be formally written as
\begin{equation}
\label{eq-dirac-s}
\psi_{n\kappa m}(\bm r)=\frac1r
\begin{pmatrix}
iP_{n\kappa}(r)Y_{jm}^l(\theta,\varphi)\\
Q_{n\kappa}(r)\left(\bm\sigma\cdot\bm{\hat r}\right)Y_{jm}^l(\theta,\varphi)
\end{pmatrix},
\end{equation}
where $\kappa$ is the Dirac quantum number, $Y_{jm}^l(\theta,\varphi)$ is the spin spherical harmonics, and $P_{n\kappa}(r)$ and $Q_{n\kappa}(r)$ are the large and small components of the radial wave function, respectively.

After separating out the angular component of wave function, the radial Dirac equation for the one-electron system is given by
\begin{align}
\label{eq-dirac}
&\begin{pmatrix}
V(r) & -c\left(\frac{\mathrm d}{\mathrm dr}-\frac{\kappa}{ r}\right)\\
c\left(\frac{\mathrm d}{\mathrm dr}+\frac \kappa r\right) &  V(r)-2c^2
\end{pmatrix}
\begin{pmatrix}
P_{n\kappa}(r)\\Q_{n\kappa}(r)
\end{pmatrix}&=
E\begin{pmatrix}
P_{n\kappa}(r)\\Q_{n\kappa}(r)
\end{pmatrix}.
\end{align}
The electrostatic potential $V(r)$ which describes the electron-nucleus interaction is constructed based on the nuclear charge density $\rho_c(r)$ shown in Eq. (\ref{eq-rhoc}) via \cite{engel2002relativistic}
\begin{equation}
\label{eq-v}
V(r)=-4\pi\left[\int_0^r\rho_c(r')\frac{r'^2}{r}\mathrm dr'+\int_r^\infty\rho_c(r')r'\mathrm dr'\right].
\end{equation}

It is noted that for the Coulomb potential of a point-like nucleus, i.e., $V(r)=-\frac Zr$, the electronic energy levels can be analytically solved in the form
\begin{equation}
\label{eq-point-energy}
E_{\rm point}[n\kappa]=c^2\left[1+\frac{\left(\alpha Z\right)^2}{\left(n-|\kappa|+\gamma\right)^2}\right]^{-1/2}-c^2,
\end{equation}
where $\gamma=\sqrt{\kappa^2 - \left(\alpha Z\right)^2}$. The FNS correction to atomic energy levels is then obtained by taking the difference between the Dirac energy for finite-size nucleus ($E_{\rm finite}[n\kappa]$) and that for point-like nucleus ($E_{\rm point}[n\kappa]$)
\begin{equation}
\Delta E_{\rm FNS}[n\kappa]=E_{\rm finite}[n\kappa]-E_{\rm point}[n\kappa].
\end{equation}

In the present work, we are also interested in the FNS effect on the bound-electron $g$ factors which connect the magnetic moment of electron $\bm \mu$ with its angular momentum $\bm J$ through
\begin{equation}
\bm \mu=-g\mu_B\bm J,
\end{equation}
where $\mu_B=\frac{1}{2}$ is the Bohr magneton. When the electron was placed into a static homogeneous magnetic field $\bm B$, the effective Hamiltonian for field-electron interaction reads
\begin{equation}
H_{\rm int}=-\bm \mu\cdot\bm B.
\end{equation}
If $\bm B$ is chosen in the \textit{z}-direction, then the first-order Zeeman splitting of the eigenenergy $\Delta E$ can be derived by using the standard perturbation theory as
\begin{equation}
\label{eq-delta-E-1}
\Delta E=g\mu_B B_z\langle J_z\rangle =g \mu_B B_zm.
\end{equation}
On the other hand, the effective Hamiltonian in the framework of relativistic quantum theory can also be expressed as \cite{grant2007relativistic}
\begin{equation}
H_{\rm int}=\frac{c}{2}\bm\alpha\cdot\left(\bm B\times\bm r\right).
\end{equation}
Employing the electron wave function defined in Eq. (\ref{eq-dirac-s}), the corresponding energy shift is given by
\begin{equation}
\label{eq-delta-E-2}
\Delta E=B_z\frac{\kappa cm}{j(j+1)}\int_0^\infty P_{n\kappa}(r)Q_{n\kappa}(r)r\mathrm dr.
\end{equation}
The bound-electron $g$ factor can then be obtained by comparing Eq. (\ref{eq-delta-E-2}) with Eq. (\ref{eq-delta-E-1}), which yields
\begin{equation}
\label{eq-g}
g[n\kappa]=\frac{2\kappa c}{j(j+1)}\int_0^\infty P_{n\kappa}(r)Q_{n\kappa}(r)r\mathrm dr.
\end{equation}
For comparison, the bound-electron $g$ factor for a point-like nucleus is analytically available as
\begin{equation}
g_{\rm point}[n\kappa]=\frac{\kappa}{j(j+1)}\left(\kappa\frac{E_{\rm point}[n\kappa]}{c^2}-\frac12\right),
\end{equation}
and, correspondingly, the FNS correction to the $g$ factor is given by
\begin{equation}
\Delta g_{\rm FNS}[n\kappa]=g_{\rm finite}[n\kappa]-g_{\rm point}[n\kappa].
\end{equation}

\section{Results and discussion}
\label{sec-3}

The nuclear ground state wave functions for even-even nuclei with $8\le Z\le 118$ are calculated based on the RCHB method. For convenience we only choose the most abundant nuclide in each isotopic chain \cite{rosman1999table}. The density functional PC-PK1 \cite{PhysRevC.82.054319}, which provides one of the best density-functional descriptions of infinite nuclear matter and finite nuclei in the ground or excited states, is employed in the RCHB calculations. In practical calculations, we use a box size $R_{\rm box}=25$ fm with the mesh size $\Delta r=0.1$ fm, and the angular momentum cutoff $J_{\rm max}=\frac{19}{2}$, which are large enough for all nuclei considered in this work. More computational details can be found in Ref. \cite{Xia2018}.

The electron Dirac equation is solved using the kinetically balanced generalized pseudospectral (KB-GPS) method developed recently by us \cite{PhysRevA.104.022801}. As a numerical method implemented in discrete variable representation, the GPS method has shown its fast convergence and high flexibility in solving the non-relativistic Schr\"{o}dinger equation \cite{SIChu,Xie_IJQC,Jiao_2021}. In our recent work \cite{PhysRevA.104.022801}, we have successfully extended the GPS method to solve the relativistic Dirac equation by incorporating the kinetically balanced condition. The derived KB-GPS method removes all spurious states in the numerical solution of the Dirac equation in discrete variable representation and, furthermore, improves the convergence of both the system eigenenergies and wave functions. The computational details of the KB-GPS method are available in Ref. \cite{PhysRevA.104.022801}. Throughout the present calculations, a total number of $N=450$ mesh points are used to solve the radial Dirac equation which ensures that the lowest several bound state energies and corresponding wave functions are converged with more than $10$ significant digits.

\begin{figure*}
\centering
\includegraphics[width=0.9\textwidth]{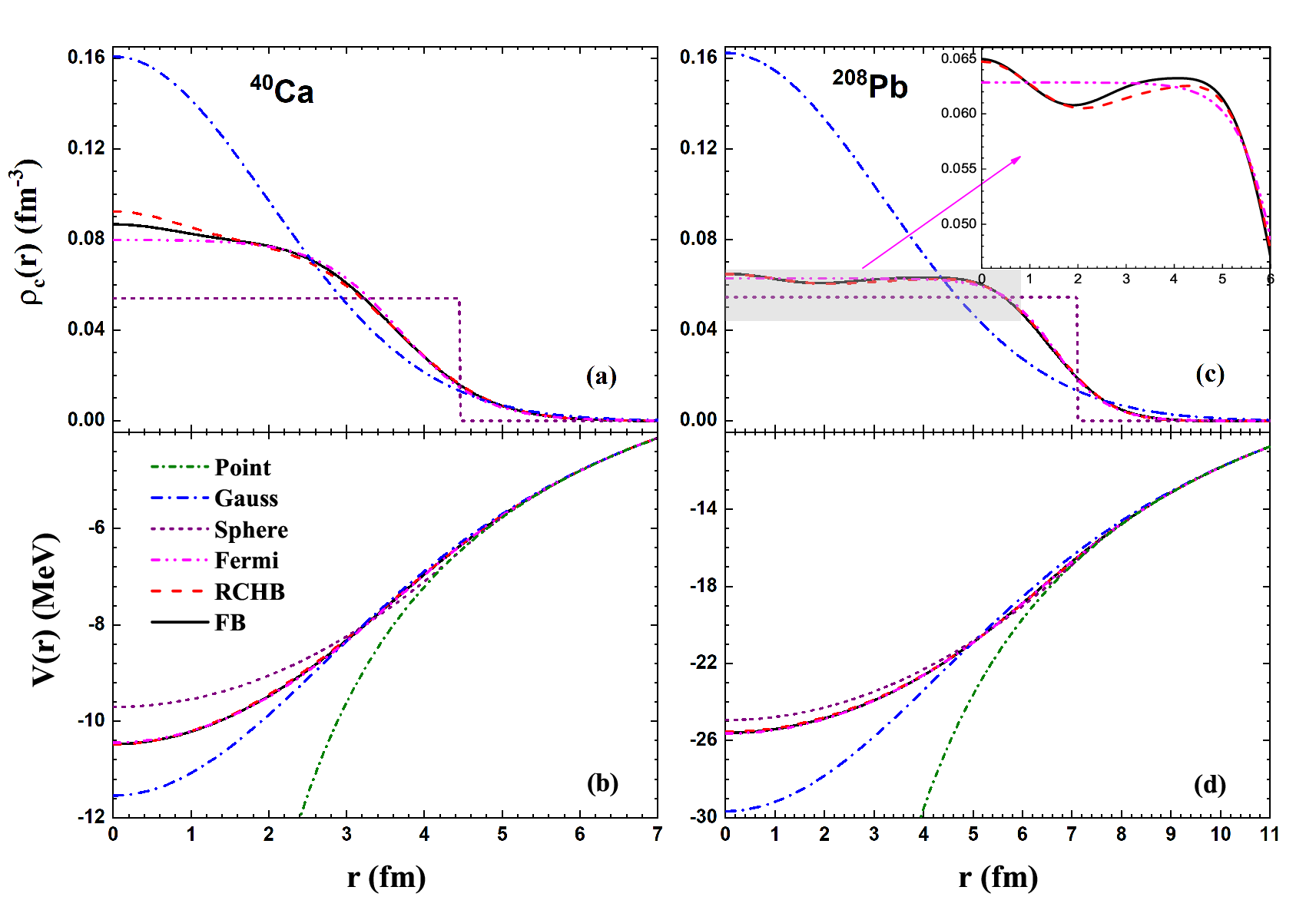}
\caption{Nuclear charge density distributions $\rho_{\rm c}(r)$ of $^{40}$Ca (a) and $^{208}$Pb (c) obtained from the RCHB calculations (``RCHB", \textit{red dash}), in comparison with several empirical charge distribution models such as homogeneously charged sphere (``Sphere", \textit{purple short dot}), Gaussian charge distribution (``Gauss", \textit{blue dash dot}), and the two-parameter Fermi (``Fermi", \textit{magenta dash dot dot}), as well as the Fourier-Bessel analysis of experimental scattering data~\cite{DEVRIES1987495} (``FB", \textit{black solid}). In panels (b) and (d), the corresponding electrostatic potentials $V(r)$ are also compared with the point-nucleus Coulomb potential (``Point", \textit{green short dash dot}). See texts for more details.}\label{fig-1}
\end{figure*}

\subsection{FNS corrections to energy levels and $g$ factors in hydrogen-like ions}
\label{sec-3-a}

In Figs. \ref{fig-1} (a) and (c), the nuclear charge density distributions for $^{40}$Ca and $^{208}$Pb nuclei obtained from the RCHB calculations are compared with three empirical nuclear charge models, i.e., the homogeneously charged sphere, the Gaussian charge distribution, and the two-parameter Fermi charge distribution, which are denoted by ``Sphere", ``Gauss", and ``Fermi", respectively. The explicit forms of the three nuclear charge models are given in the Appendix. The corresponding parameters are fitted by reproducing the experimental nuclear charge radii \cite{ANGELI201369,DEVRIES1987495}. The Fourier-Bessel analysis of the experimental scattering data \cite{DEVRIES1987495} (denoted by ``FB") are provided as the referenced nuclear charge density distributions. It can be seen that the present RCHB calculations are in better agreement with the experimental results than those obtained using the empirical models and, especially in the central region of nucleus, the RCHB model provides a correct description of the concave structure in the charge density distribution (e.g., around $2$ fm for $^{208}$Pb in Fig. \ref{fig-1} (c)). For the three empirical models, only the two-parameter Fermi distribution shows good agreement with the experimental results, while the Sphere and Gaussian charge distributions are both qualitatively and quantitatively different from the referenced charge distribution. The comparison of corresponding electrostatic potentials are displayed in Figs. \ref{fig-1} (b) and (d) for $^{40}$Ca and $^{208}$Pb nuclei, respectively. Due to the relatively large magnitude of the electrostatic potential inside the nucleus, both the present RCHB calculation and that from the two-parameter Fermi model potential are indistinguishable from the referenced FB potential based on experimental charge distribution. From the comparison of finite-nucleus potentials with the point-nucleus Coulomb potential shown in Figs. \ref{fig-1} (b) and (d), it is obvious that the FNS correction would systematically lift the electrostatic potential near and inside the nucleus and, as a result, increase the bound state energies.

\begin{table*}[!ht]
\caption{FNS corrections to bound-electron energy levels $\Delta E_{\rm FNS}$ (in units of eV) and $g$ factors $\Delta g_{\rm FNS} $ for the $1s_{1/2}$, $2s_{1/2}$, and $2p_{1/2}$ states of $^{40}{\rm Ca}^{19+}$, $^{208}{\rm Pb}^{81+}$, and $^{238}{\rm U}^{91+}$ highly charged hydrogen-like ions. Nuclear charge distributions obtained from different empirical nuclear models, FB \cite{DEVRIES1987495}, and microscopic RCHB and Skyrme-Hartree-Fock \cite{PRA101} nuclear models are employed in the calculations. The root-mean-square charge radii $r_c$ for different nuclei are given in units of fm. The relative FNS corrections in the ground state are defined as $\delta[1s_{1/2}]=\frac{\Delta E_{\rm FNS}[1s_{1/2}]}{\left|E_{\rm point}[1s_{1/2}]\right|}$ for energy level and $\frac{\Delta g_{\rm FNS}[1s_{1/2}]}{g_{\rm point}[1s_{1/2}]}$ for $g$ factor. Numbers in parentheses represent the power of ten.}
\label{tab-1}
\centering
\resizebox{\textwidth}{!}{
\begin{tabular}{lll|llll|llll}
\toprule
&&&\multicolumn{4}{c}{{$\Delta E_{\rm FNS}(\rm eV)$}}\vline&\multicolumn{4}{c}{$\Delta g_{\rm FNS}$}\\
\hline
&&\multicolumn{1}{c}{$r_c$}    \vline &\multicolumn{1}{c}{$1s_{1/2}$}    &\multicolumn{1}{c}{$\delta[1s_{1/2}]$}       &\multicolumn{1}{c}{$2s_{1/2}$}           &\multicolumn{1}{c}{$ 2p_{1/2}$}\vline&\multicolumn{1}{c}{$1s_{1/2}$}      &\multicolumn{1}{c}{$\delta[1s_{1/2}]$}      &\multicolumn{1}{c}{$2s_{1/2}$}         &\multicolumn{1}{c}{$ 2p_{1/2}$}\\
\hline
\multirow{6}{*}{$^{40}{\rm Ca}^{19+}$}
&Sphere  &3.4500  &1.4343(-2)  &2.6214(-6)  &1.8280(-3)  &7.3927(-6)&1.1139(-7)&1.7075(-7)&1.4188(-8)&5.7442(-11)\\
&Gauss   &3.4500  &1.4331(-2)  &2.6192(-6)  &1.8265(-3)  &7.3870(-6)&1.1129(-7)&1.7059(-7)&1.4175(-8)&5.7397(-11)\\
&Fermi   &3.4500  &1.4336(-2)  &2.6201(-6)  &1.8271(-3)  &7.3894(-6)&1.1134(-7)&1.7066(-7)&1.4181(-8)&5.7416(-11)\\
&FB      &3.4500  &1.4337(-2)  &2.6203(-6)  &1.8274(-3)  &7.3899(-6)&1.1135(-7)&1.7067(-7)&1.4182(-8)&5.7420(-11)\\
&RCHB    &3.4750  &1.4543(-2)  &2.6579(-6)  &1.8523(-3)  &7.4960(-6)&1.1294(-7)&1.7312(-7)&1.4385(-8)&5.8244(-11)\\
&Skyrme  &3.4776  &1.4565(-2)  &2.6619(-6)  &1.8551(-3)  &7.5066(-6)&1.1311(-7)&1.7338(-7)&1.4406(-8)&5.8504(-11)\\
\hline
\multirow{6}{*}{$^{208}{\rm Pb}^{81+}$}
&Sphere  &5.5032  &67.346	&6.6315(-4)	&11.694	&1.0016		&4.5405(-4)	&1.1306(-3)&7.8777(-5)&6.7852(-6)\\
&Gauss   &5.5032  &66.666	&6.5628(-4) &11.575	&9.9181(-1)	&4.4939(-4)	&1.1190(-3)&7.7963(-5)&6.7184(-6)\\
&Fermi   &5.5032  &67.218	&6.6172(-4)	&11.671	&9.9977(-1)	&4.5317(-4)	&1.1284(-3)&7.8624(-5)&6.7726(-6)\\
&FB      &5.5032  &67.223	&6.6176(-4)	&11.672	&9.9984(-1)	&4.5321(-4)	&1.1285(-3)&7.8630(-5)&6.7731(-6)\\
&RCHB    &5.5055  &67.278	&6.6231(-4)	&11.682	&1.0007		&4.5358(-4)	&1.1294(-3)&7.8695(-5)&6.7787(-6)\\
&Skyrme  &5.5012  &67.181	&6.6135(-4)	&11.665	&9.9982(-1)	&4.5291(-4)	&1.1277(-3)&7.8579(-5)&6.7687(-6)\\
\hline
\multirow{4}{*}{$^{238}{\rm U}^{91+}$}
&Sphere  &5.8571  &199.04   &1.5047(-3)&37.809&4.4209&1.2765(-3)&3.9703(-3)&2.4211(-4)&2.8536(-5)\\
&Gauss   &5.8571  &196.72   &1.4872(-3)&37.368&4.3707&1.2615(-3)&3.9236(-3)&2.3924(-4)&2.8210(-5)\\
&Fermi   &5.8571  &198.65   &1.5017(-3)&37.736&4.4125&1.2740(-3)&3.9626(-3)&2.4163(-4)&2.8482(-5)\\
&RCHB    &5.8462  &198.24   &1.4986(-3)&37.659&4.4033&1.2714(-3)&3.9545(-3)&2.4115(-4)&2.8422(-5)\\
\toprule
\end{tabular}
}
\end{table*}

To further investigate the FNS effects on atomic spectral properties, we present in Table \ref{tab-1} the comparison of FNS corrections to the energy levels and bound-electron $g$ factors calculated by using different charge density distributions for the $1s_{1/2}$, $2s_{1/2}$, and $2p_{1/2}$ states of $^{40}{\rm Ca}^{19+}$, $^{208}{\rm Pb}^{81+}$, and $^{238}{\rm U}^{91+}$ ions. For $^{40}{\rm Ca}^{19+}$ and $^{208}{\rm Pb}^{81+}$, the most recent calculations based on the Skyrme-Hartree-Fock nuclear charge density distributions \cite{PRA101} are also included for comparison. It is noted that the Skyrme-type interaction between nucleons in the Skyrme-Hartree-Fock method includes some adjustable parameters to reproduce the experimental nuclear charge radii.

By comparing the FNS corrections of energy levels and $g$ factors in the $1s_{1/2}$, $2s_{1/2}$, and $2p_{1/2}$ electronic states, it is readily shown that the FNS corrections in the ground state are more significant than in other bound states. With continuously exciting the bound state, the FNS corrections decrease rapidly by orders of magnitude. The comparison among different hydrogen-like ions indicates that the FNS corrections are more visible in heavier nuclei. For example, the FNS correction to the ground state energy increases from $1.4\times10^{-2}$ eV in $^{40}{\rm Ca}^{19+}$ to $67$ and $198$ eV in $^{208}{\rm Pb}^{81+}$ and $^{238}{\rm U}^{91+}$, respectively, which contribute about $2.6\times10^{-6}$, $6.6\times10^{-4}$, and $1.5\times10^{-3}$ in proportion to the total ground state energies of the three systems. Further comparisons among different electronic states of hydrogen-like ions with different nuclear charges will be presented in the following section.

From the comparison shown in Table \ref{tab-1}, it is interesting to note that the Sphere and Fermi empirical nuclear distributions give similar FNS corrections to the energy levels and bound-electron $g$ factors, and that they are both close to the referenced values obtained from the experimental (FB) nuclear charge distribution. The Gauss model gives the worst prediction which can be understood from its poor description of the nuclear charge densities as demonstrated in Fig. \ref{fig-1}. The discrepancies among these empirical models indicates that the detailed shape of the nuclear charge distribution plays a non-negligible role in the calculation of FNS corrections. The comparison between two microscopic nuclear models (RCHB and Skyrme) reveal more interesting phenomena. For $^{40}{\rm Ca}^{19+}$, both the present RCHB and the previous Skyrme calculations produce relatively larger root-mean-square (rms) nuclear charge radius and systematically larger FNS corrections compared to the FB results. However, for $^{208}{\rm Pb}^{81+}$, the RCHB and Skyrme methods give, respectively, a slightly larger and smaller value of $r_{c}$ than the experimental value. The comparison of FNS corrections to energy levels and bound-electron $g$ factors follows a similar trend. From the discussion we may simply conclude that the FNS corrections are more sensitive to the average nuclear charge radius. The empirical two-parameter Fermi distribution model, when it fairly reproduces the experimental rms radius of the nucleus, provides a reasonable estimate of the FNS corrections to both energy levels and $g$ factors. The present RCHB calculations predict relatively larger values of rms radius and, as a result, slightly overestimate the FNS corrections. Compared to the non-relativistic Skyrme-Hartree-Fock calculations, the RCHB method provides a full-relativistic, parameter-free microscopic description of the nuclear structures and shows slightly better agreement with the experimental (FB) results.

For the $^{238}{\rm U}^{91+}$ ion, there is currently no experimental scattering data for nuclear charge distribution, so we only compare our results with the three empirical models. As is shown in Table \ref{tab-1} the present RCHB calculations are in better agreement with the two-parameter Fermi distribution results than those for the other two models. The validity of the FNS corrections to atomic energy levels can be estimated through the Lamb shift, i.e., the difference between the exact binding energy and the Dirac point-nucleus energy. The Lamb shift mainly consists of three parts: the QED corrections \cite{PhysRevLett.89.143001,PhysRevLett.91.073001,PhysRevA.64.062507,MOHR197426,PhysRevA.3.1267,PhysRevA.26.2338}, the nuclear recoil corrections \cite{shabaev1985mass,Shabaev_1998,PhysRevA.52.1884,PhysRevA.57.4235}, and the nuclear structure corrections in which the leading contribution comes from the FNS effects. Other contributions to the Lamb shift include the remaining nuclear structure effects (e.g., the nuclear deformation \cite{PhysRevLett.108.063005,PhysRevA.77.032501,PhysRevA.99.012505} and nuclear polarization \cite{NEFIODOV1996227,PhysRevA.43.5853,PhysRevA.75.032521,PhysRevA.103.032811,PhysRevLett.113.023002,PhysRevA.53.4614.2}), as well as the cross term between the FNS and QED corrections \cite{PhysRevA.58.954,bondarev2010finite,PhysRevA.56.R2507,PhysRevA.69.022114,PhysRevA.83.012507}. It has been proven that such contributions can only produce a very small energy shift. The theoretical calculations of QED corrections ($265.19$ eV \cite{MOHR1998227,PhysRevLett.91.073001}), nuclear recoil ($0.46$ eV \cite{PhysRevA.57.4235}), and nuclear polarization ($-0.20$ eV \cite{NEFIODOV1996227,PhysRevA.53.4614.2}) contribute a total of $265.45$ eV for the ground state of $^{238}{\rm U}^{91+}$. The combination of these contributions with the FNS correction of $198.24$ eV from the present RCHB calculation yields a total Lamb shift of $463.69$ eV, which is in good agreement with the recent experimental measurement of $460.2\pm 4.6$ eV \cite{PhysRevLett.94.223001}.

\subsection{FNS corrections for the $s_{1/2}$ and $p_{1/2}$ states from light to heavy hydrogen-like ions}
\label{sec-3-b}

\begin{figure}[!ht]
\centering
\includegraphics[width=0.5\textwidth]{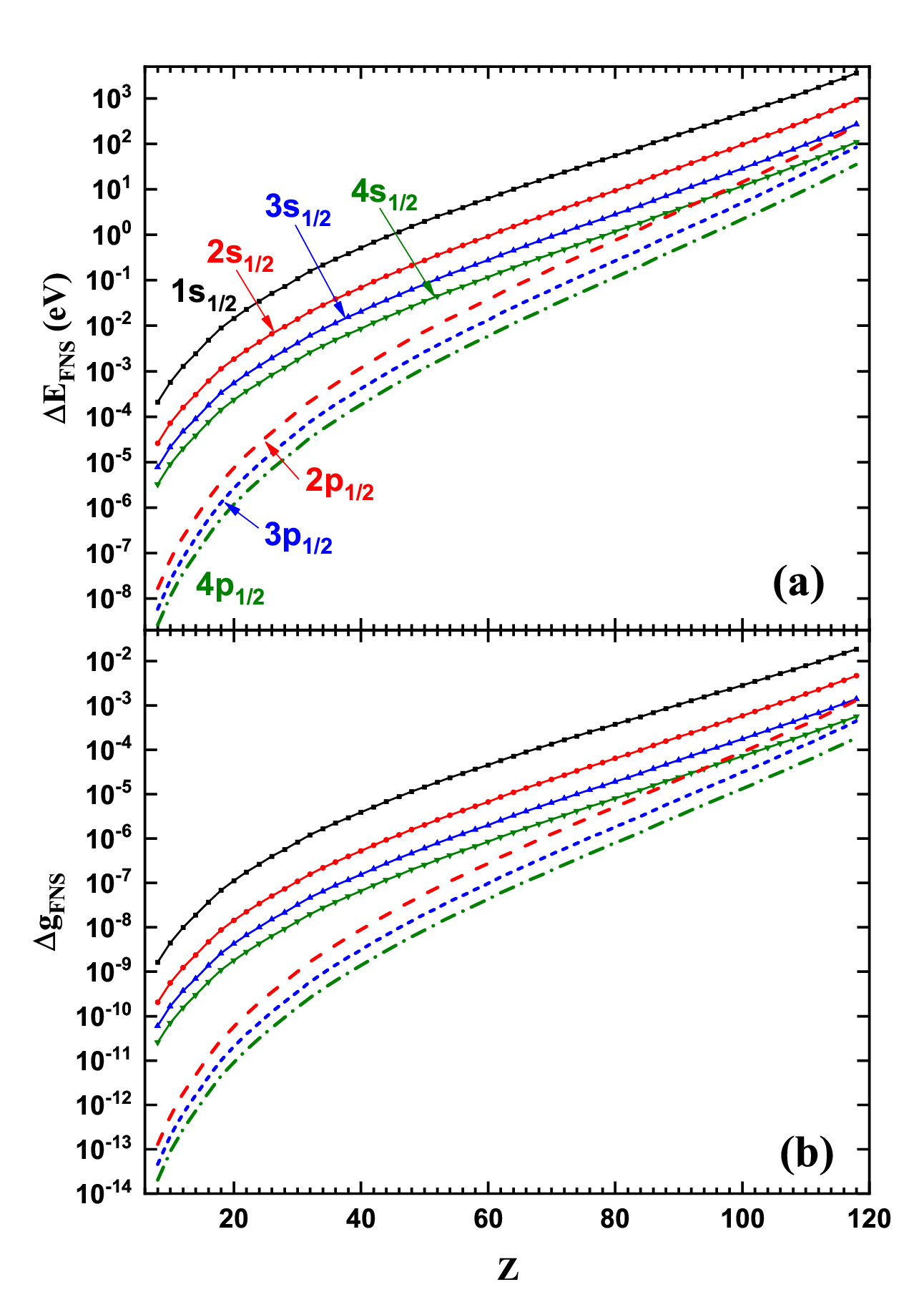}
\caption{The FNS corrections to the energy levels $\Delta E_{\rm FNS}$ (a) and bound-electron $g$ factors $\Delta g_{\rm FNS}$ (b) in the $ns_{1/2}$ and $np_{1/2}$ states with $n\le 4$ for hydrogen-like ions with $8 \le Z\le 118$.}
\label{fig-2}
\end{figure}

\begin{figure}[!ht]
\centering
\includegraphics[width=0.5\textwidth]{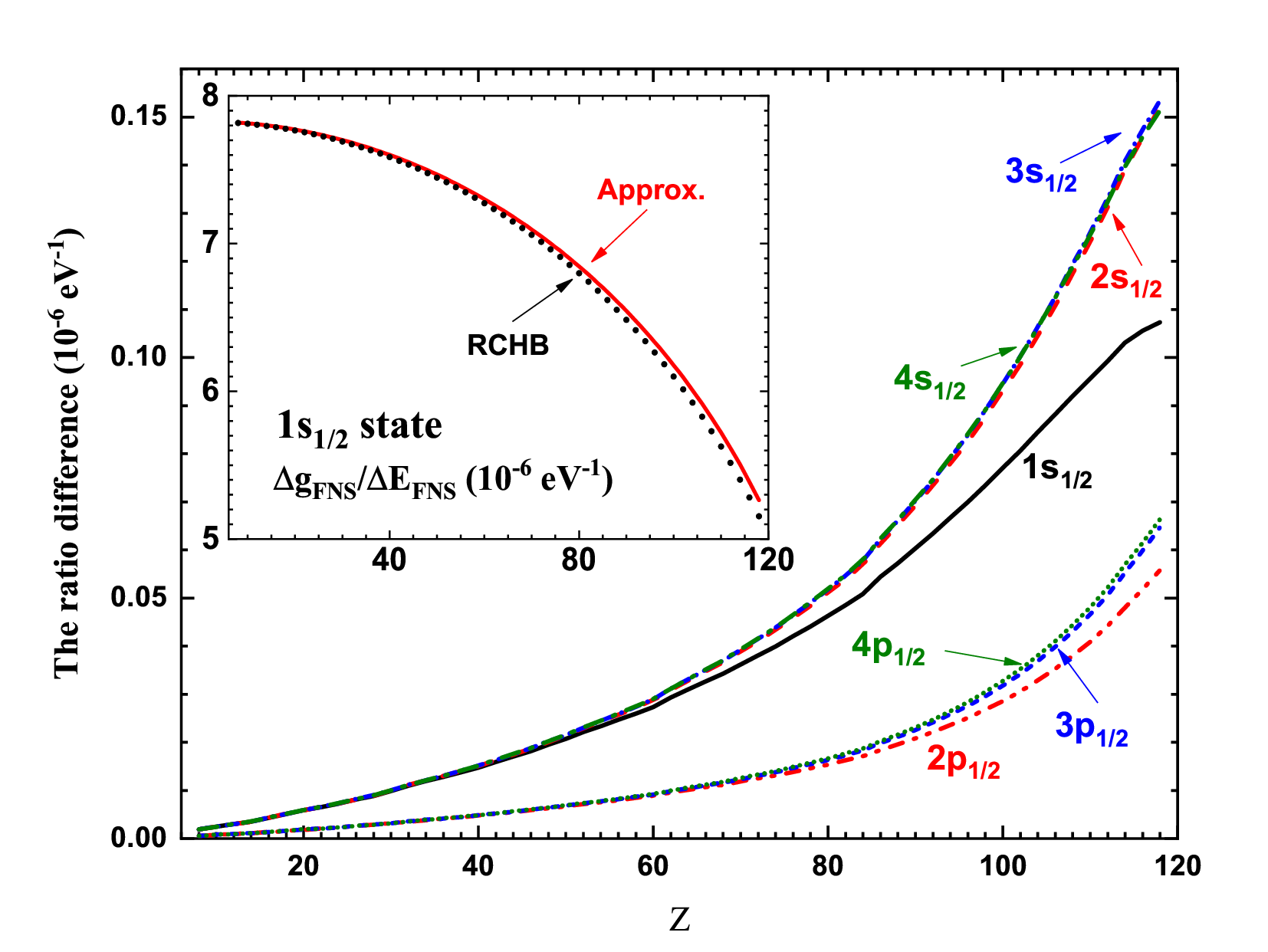}
\caption{The difference of the ratio $D=\frac{\Delta g_{\rm FNS}}{\Delta E_{\rm FNS}}$ between the approximate formula of Eq. (\ref{eq-ratio-g-E}) and the present RCHB numerical calculations for $ns_{1/2}$ and $np_{1/2}$ states with $n\le 4$. The inset shows the specific ratio for the ground state evaluated by Eq. (\ref{eq-ratio-g-E}) (\textit{red solid line}) and the corresponding RCHB numerical results (\textit{black square scatter}).}
\label{fig-3}
\end{figure}

To investigate the FNS effects in different hydrogen-like ions, we present in Figs. \ref{fig-2} (a) and (b), respectively, the FNS corrections $\Delta E_{\rm FNS}$ and $\Delta g_{\rm FNS}$ as functions of the nuclear charge number $Z$ for the $ns_{1/2}$ and $np_{1/2}$ bound states with $n\le 4$. The nuclear charge density distributions and corresponding electrostatic interaction potentials are calculated in the similar way as mentioned above. All numerical results are tabulated in the Supplementary Material for further reference. From Fig. \ref{fig-2} it can clearly be seen that the FNS corrections to both energy levels and $g$ factors increase rapidly as $Z$ increases. This is consistent with the fact that the energy levels of hydrogen-like ions are approximately proportional to $Z^2$, as well as with our findings in Table \ref{tab-1} that the contribution of FNS effects become increasingly important in heavy nuclei.

It can also be observed from Fig. \ref{fig-2} that for both $\Delta E_{\rm FNS}$ and $\Delta g_{\rm FNS}$, the magnitude of FNS corrections in $p$-wave states increases much faster than those in $s$-wave states. For example, $\Delta E_{\rm FNS}[np_{1/2}]$ is about three orders of magnitude smaller than $\Delta E_{\rm FNS}[ns_{1/2}]$ for light hydrogen-like ions below $Z = 20$, and it finally becomes one order of magnitude smaller as $Z$ approaches $118$. The faster increasing rate of the FNS corrections in higher orbital angular momentum states can be understood from the approximate formulas derived by Shabaev \cite{Shabaev_1993}. By employing the homogeneously charged sphere density distribution, the author has analytically derived (in atomic units)
\begin{align}
\label{eq-Shabaev1}
\Delta E_{ns_{1/2}}&=\frac{Z^2m_e^{2\gamma+1}}{10n}\left(\frac{2ZR_{s}}{n}\right)^{2\gamma}\left[1+(\alpha Z)^2f_{ns_{1/2}}(\alpha Z)\right],\\
\label{eq-Shabaev2}
\Delta E_{np_{1/2}}&=\frac{Z^4\alpha^2m_e^{2\gamma+1}}{40}\frac{n^2-1}{n^3}\left(\frac{2ZR_{s}}{n}\right)^{2\gamma}
\nonumber\\
&~~~\times\left[1+(\alpha Z)^2f_{np_{1/2}}(\alpha Z)\right],
\end{align}
where
\begin{equation}
\label{eq-Shabaev3}
f(\alpha Z)=b_0+b_1(\alpha Z)+b_2(\alpha Z)^2+b_3(\alpha Z)^3,
\end{equation}
and
\begin{equation}
\label{eq-Shabaev4}
\gamma=\sqrt{\kappa^2-(Z\alpha)^2}.
\end{equation}
The effective charge radius is given by $R_s=\sqrt{\frac{5}{3}}r_c$ and the coefficients $b_{0,1,2,3}$ for different bound states are available in Ref. \cite{Shabaev_1993}. The results obtained by using the above approximate formulas are indistinguishable from our numerical calculations in the figure scale.

By neglecting the factors $1+(\alpha Z)^2f(\alpha Z)$ in both Eqs. (\ref{eq-Shabaev1}) and (\ref{eq-Shabaev2}), we obtain the ratio between the FNS corrections for $p$- and $s$-wave state energies in the form
\begin{equation}
\label{eq-ratio-s-p}
\frac{\Delta E_{\rm FNS}[np_{1/2}]}{\Delta E_{\rm FNS}[ns_{1/2}]}\approx \frac{n^2-1}{4n^2}(Z\alpha)^2,
\end{equation}
which follows approximately a quadratic law with respect to $(Z\alpha)$. As a result, the FNS corrections in $p$-wave states increase much faster than those in the $s$-wave states.

\subsection{Ratio between $\Delta g_{\rm FNS}$ and $\Delta E_{\rm FNS}$}
\label{sec-3-c}

Considering the similarity between the FNS corrections $\Delta E_{\rm FNS}$ and $\Delta g_{\rm FNS}$ displayed in Fig. \ref{fig-2}, it is of great interest to investigate the quantitative relationship between them. Karshenboim \textit{et al.} \cite{PhysRevA.72.042101} have analytically derived that the FNS correction to the bound-electron $g$ factor approximately connects with the corresponding correction to the energy level via
\begin{equation}
\label{eq-ratio-g-E}
D=\frac{\Delta g_{\rm FNS}}{\Delta E_{\rm FNS}}=\frac{\kappa^2(2\gamma+1)\alpha^2}{j(j+1)}.
\end{equation}
The ratio $D$ only has a weak dependence on the nuclear charge $Z$ through the coefficient $\gamma$ shown in Eq. \ref{eq-Shabaev4}. Such a weak dependence can clearly be seen from the inset of Fig. \ref{fig-3} where the ratio for the ground state of hydrogen-like ions decreases smoothly from $7.8\times 10^{-6}$ to $5.2\times 10^{-6}$ $\text{eV}^{-1}$ as the nuclear charge $Z$ increases from $8$ to $118$. The present numerical calculations based on the RCHB nuclear charge densities only show small discrepancies with the approximate formula at large values of $Z$. Figure \ref{fig-3} depicts the difference of the ratio $D$ between the approximate formula and the present numerical calculations for the $ns_{1/2}$ and $np_{1/2}$ bound states with $n\le 4$. It is observed that although the discrepancies increase gradually along with increasing the nuclear charge, its magnitude does not exceed the $3\%$ proportion of the ratio for all nuclei considered here. Therefore, we may conclude that the approximate formula of Eq. (\ref{eq-ratio-g-E}) establishes a fairly accurate connection between the FNS correction to bound-electron $g$ factors and energy levels. The consequence of this connection leads to the similar behavior of $\Delta g_{\rm FNS}$ with respect to changing $Z$ as $\Delta E_{\rm FNS}$ shown in Fig. \ref{fig-2}.

\subsection{Comparison of direct calculations with approximate formulas}
\label{sec-3-d}

\begin{figure}[!ht]
\centering
\includegraphics[width=0.5\textwidth]{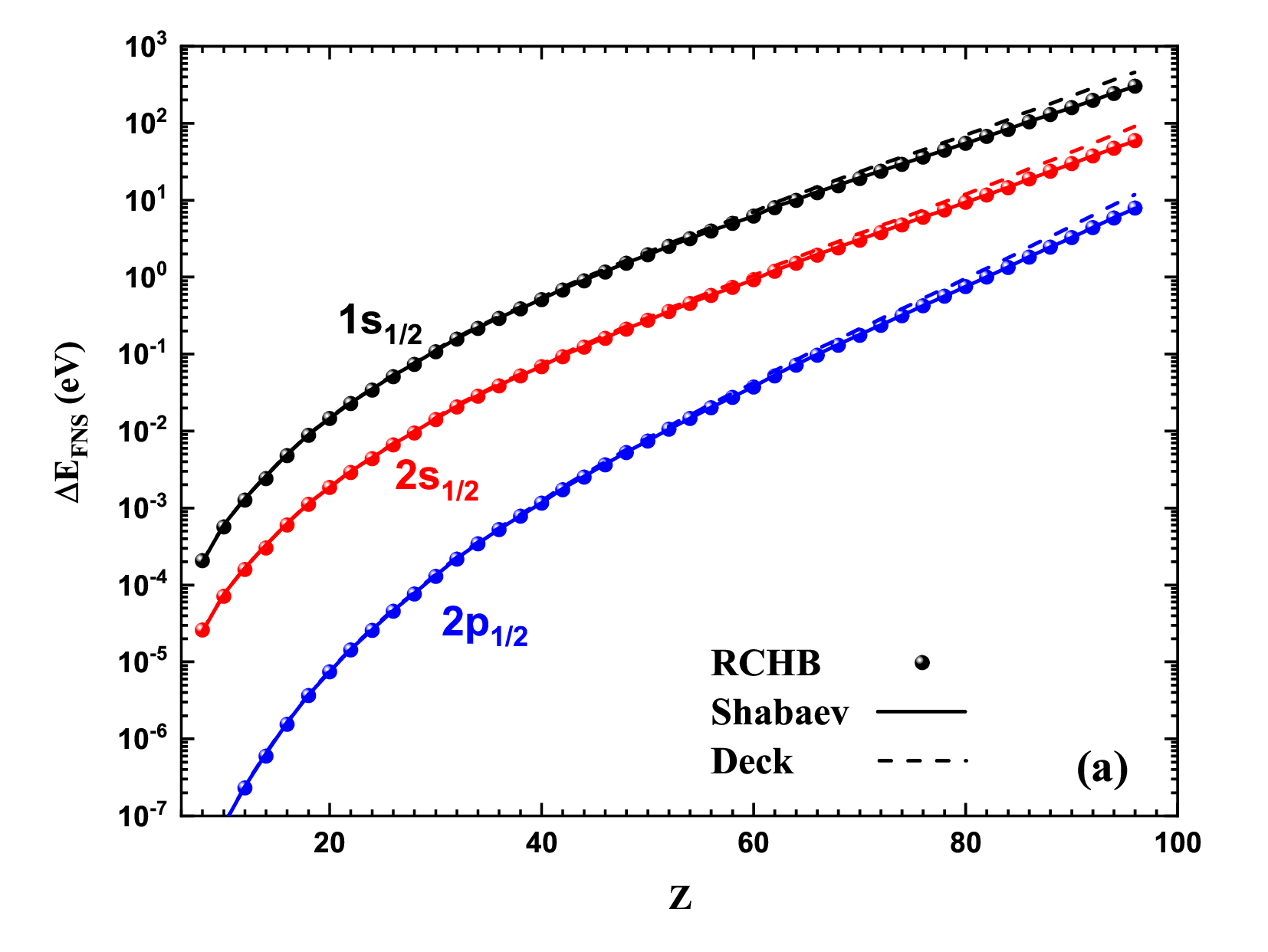}
\includegraphics[width=0.5\textwidth]{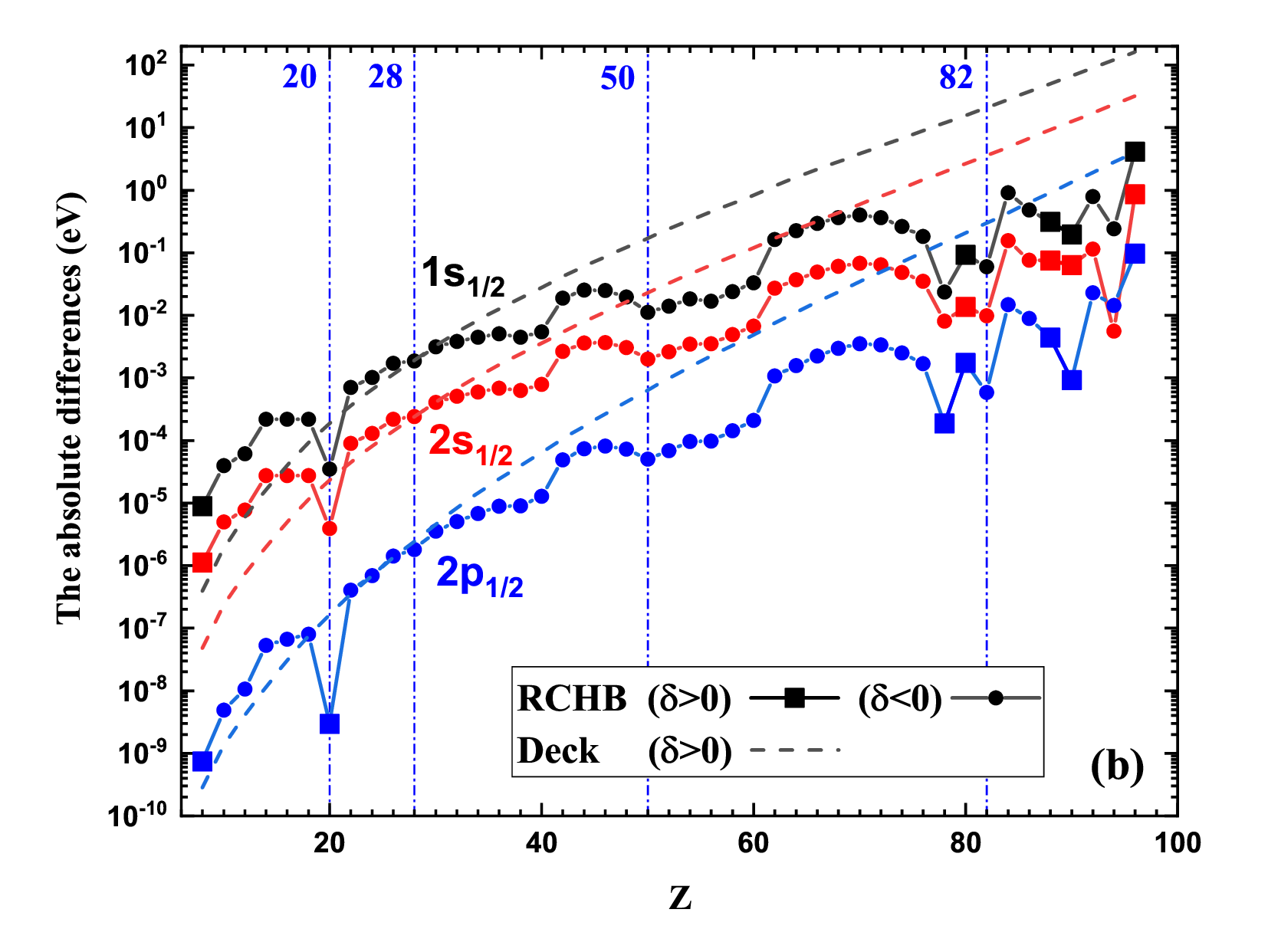}
\caption{Comparison of the present RCHB calculations and the two approximate formulas about $\Delta E_{\rm FNS}$ proposed by Shabaev \cite{Shabaev_1993} and Deck \textit{et al.} \cite{Deck_2005} for the $1s_{1/2}$, $2s_{1/2}$, and $2p_{1/2}$ states of hydrogen-like ions with $8\le Z\le 96$. (a) the direct comparison and (b) the absolute differences of the RCHB numerical calculations and Deck \textit{et al.}'s \cite{Deck_2005} estimation with respect to Shabaev's \cite{Shabaev_1993} approximation. The \textit{square} and \textit{circle} for RCHB results indicate, respectively, the positive and negative absolute difference of $\delta E=\Delta E_{\rm RCHB}-\Delta E_{\rm Shabaev}$. The dashed lines represent the positive absolute difference of $\delta E=\Delta E_{\rm Deck}-\Delta E_{\rm Shabaev}$. The four vertical dotted lines denote the nuclei with magic proton numbers of $20$, $28$, $50$, and $82$.}
\label{fig-4}
\end{figure}

In Fig. \ref{fig-4} (a), the FNS corrections to energy levels of $1s_{1/2}$, $2s_{1/2}$, and $2p_{1/2}$ states for hydrogen-like ions with $8\le Z\le 96$ based on the RCHB nuclear charge densities are compared with two approximate formulas proposed by Shabaev \cite{Shabaev_1993} and Deck \textit{et al.} \cite{Deck_2005}. The analytical formulas derived by Shabaev are shown in Eqs. (\ref{eq-Shabaev1}) and (\ref{eq-Shabaev2}), and those obtained by Deck \textit{et al.} are available as Eqs. (37) and (38) in Ref. \cite{Deck_2005}. Both of these two approximations are derived based on the homogeneously charged sphere nuclear model, except that the former author explicitly solved $\Delta E_{\rm FNS}$ as functions of the effective charge radius, while the latter authors employed the first-order perturbation theory in the treatment of the Coulomb potential in the interior of nucleus. In producing both approximate formulas in Fig. \ref{fig-4} (a), we employ the recent experimental rms nuclear charge radii summarized in Ref. \cite{ANGELI201369} where the largest nucleus is $Z=96$. It should be noted that in the isotopic chain of each element, only the isotope with the most natural abundance is chosen for comparison. We would like to leave the investigation of isotopic shift of FNS corrections in our future work. From Fig. \ref{fig-4} (a), it can be seen that the present RCHB numerical calculations are in good agreement with the results of two approximate formulas for light and medium-heavy nuclei, while for heavy nuclei our results are much closer to Shabaev's approximation \cite{Shabaev_1993}. The discrepancies between these two analytical predictions at heavy nuclei are probably attributed to the inaccuracy of the first-order perturbation theory \cite{Deck_2005}.

To further view the FNS effects introduced by different nuclear models, especially the difference between the present RCHB numerical calculations and Shabaev's analytical approximation, we present in Fig. \ref{fig-4} (b) the absolute difference $\delta E=\Delta E_{\rm RCHB}-\Delta E_{\rm Shabaev}$, as well as the difference between the two approximate formulas, i.e., $\delta E=\Delta E_{\rm Deck}-\Delta E_{\rm Shabaev}$. For convenience, we choose Shabaev's approximation as the referenced result. The systematic overestimation of the FNS corrections by Deck's perturbation method is constantly enhanced as the nuclear charge increases. The comparison between the RCHB calculations and Shabaev's estimation reveals apparent fluctuations of the FNS effects in different nuclei. An interesting phenomenon is observed for nuclei with magic proton numbers, e.g., $Z=20$, $28$, $50$, and $82$, where the absolute difference $\delta E$ acquires some local minima, indicating that the RCHB calculations are closer to Shabaev's approximation in these situations. Such a correspondence is consistent with the fact that the RCHB theory gives a better prediction of the charge radii for nuclei with magic proton numbers \cite{Xia2018} (A complete list of the charge radii for even-even nuclei calculated by the RCHB method is available in the Supplementary Material). Due to the good agreement with experiment for the RCHB calculations of important nuclear properties such as the binding and separation energies, rms radii of neutron and proton, and charge density distributions for both stable and unstable nuclei in the large nuclear landscape, we tentatively conjecture that the RCHB method should also be applicable in the investigation of FNS effects in systems with unstable nuclei, which still are challenging tasks for both theoretical development and experimental measurement.

\section{Conclusion}
\label{sec-4}

In the present work, the FNS effects to the energy levels and bound-electron $g$ factors have been investigated for hydrogen-like ions with $8\le Z\le 118$, where the nuclear charge density distributions are obtained by employing the RCHB method in the framework of covariant density functional theory. The nuclear charge density distributions and corresponding electronic interactions for $^{40}$Ca$^{19+}$ and $^{208}$Pb$^{81+}$ ions are calculated to demonstrate that the RCHB method reproduces very well the predictions from the Fourier-Bessel analysis of experimental data. For FNS corrections to energy levels and $g$ factors, a comparison among the RCHB, the three empirical nuclear charge models, the non-relativistic Skyrme-Hartree-Fock, and the referenced Fourier-Bessel results indicate that both the nuclear charge radius and the detailed shape of charge density distribution are responsible for accurate determination of the FNS effects. The present RCHB calculations are generally in good agreement with the referenced results, with the minor overestimation originating from the slightly larger estimation of the rms nuclear charge radius. We further demonstrated in $^{238}$U$^{91+}$ ion that the RCHB calculation of FNS correction to the ground state energy, combined with the contributions from other finite nuclear effects, are in good agreement with the experimental measurement of Lamb shift.

The systematic investigation of the FNS corrections to the energy levels and $g$ factors with respect to different nuclear charge reveals that not only the magnitude but also the relative contribution of the FNS corrections increase rapidly from light to heavy nuclei. Our numerical calculations for the $ns_{1/2}$ and $np_{1/2}$ states of hydrogen-like ions show good agreement with the approximate prediction of Shabaev, and the ratio between the FNS correction to $g$ factor and energy level reproduces very well the estimation by Karshenboim. The discrepancy between the RCHB calculations and the approximate predictions based on empirical nuclear charge models reveals the distinct contribution from microscopic nuclear structures.

There still exist many important open questions for the FNS effects in electronic structure calculations, e.g., the isotope shifts and the nuclear polarization and deformation effects. These contributions can also be taken into account through the RCHB method with further development. These research would be of great interest in our future studies.

\section*{Acknowledgments}

Financial support from the National Natural Science Foundation of China (Grant Nos. 11675063 and 12174147) is greatly acknowledged.

\appendix
\section{Empirical nuclear charge models}

The homogeneously charged sphere model reads
\begin{equation}
\rho_c(r)=\begin{cases}
\frac{3eZ}{4\pi R_0^3} \ ,& r\le R_0,\\
0 \ ,& r>R_0,
\end{cases}
\end{equation}
where $R_0=\sqrt{\frac{5}{3}}r_c$. The corresponding electrostatic potential is given by
\begin{equation}
V(r)=\begin{cases}
-\frac{eZ}{2R_0}\left(3-\frac{r^2}{R_0^2}\right) \ ,& r\le R_0,\\
-\frac{eZ}{r} \ ,& r>R_0.
\end{cases}
\end{equation}

The Gaussian charge distribution reads
\begin{equation}
\rho_c(r)=\rho_0^\mathrm G\exp(-\xi r^2),
\end{equation}
where $\xi=\frac{3}{2r_c^2}$ and $\rho_0^\mathrm G=eZ(\frac{\xi}{\pi})^{3/2}$. The potential is expressed as
\begin{equation}
V(r)=-\frac{eZ}{r}\mathrm{erf}(\sqrt{\xi}r),
\end{equation}
where $\mathrm{erf}(x)$ is the error function.

The two-parameter Fermi charge distribution reads
\begin{equation}
\rho_c(r)=\frac{\rho_0^{\mathrm F}}{1+\exp(4\ln3(r-C)/T)},
\end{equation}
where $C=\sqrt{R_0^2-\frac{7}{3}\left(\frac{\pi T}{4\ln 3}\right)^{2}}$ is the half charge radius, $T=2.3$ fm is the skin thickness parameter, and $\rho_0^\mathrm F$ is the normalized parameter determined from the number of total charge. The corresponding potential cannot be expressed in a simple analytic form but can be numerically obtained through Eq. \ref{eq-v}.

The parameters of the three nuclear charge models for $^{40}$Ca, $^{208}$Pb, and $^{238}$U nuclei are listed in Table \ref{tabapp}.

\begin{table}[!hb]
\caption{Parameters for the three nuclear charge models. \label{tabapp}}
\centering
\begin{tabular}{l|lllll}
\toprule
\multicolumn{1}{c}{Nucleus} \vline & \multicolumn{1}{c}{$r_c$} & \multicolumn{1}{c}{$R_0$} & \multicolumn{1}{c}{$\xi$} & \multicolumn{1}{c}{$\rho_0^\mathrm F$} & \multicolumn{1}{c}{$C$} \\
\hline
$^{40}$Ca & 3.4500 & 4.4539 & 0.1260 & 0.0900 & 3.6782 \\
$^{208}$Pb& 5.5032 & 7.1046 & 0.0495 & 0.0628 & 6.6458 \\
$^{238}$U & 5.8571 & 7.5615 & 0.0437 & 0.0575 & 7.1322 \\
\toprule
\end{tabular}
\end{table}


\bibliography{gpsfns}

\providecommand{\noopsort}[1]{}\providecommand{\singleletter}[1]{#1}%
\begin{thebibliography}{115}%
\makeatletter
\providecommand \@ifxundefined [1]{%
 \@ifx{#1\undefined}
}%
\providecommand \@ifnum [1]{%
 \ifnum #1\expandafter \@firstoftwo
 \else \expandafter \@secondoftwo
 \fi
}%
\providecommand \@ifx [1]{%
 \ifx #1\expandafter \@firstoftwo
 \else \expandafter \@secondoftwo
 \fi
}%
\providecommand \natexlab [1]{#1}%
\providecommand \enquote  [1]{``#1''}%
\providecommand \bibnamefont  [1]{#1}%
\providecommand \bibfnamefont [1]{#1}%
\providecommand \citenamefont [1]{#1}%
\providecommand \href@noop [0]{\@secondoftwo}%
\providecommand \href [0]{\begingroup \@sanitize@url \@href}%
\providecommand \@href[1]{\@@startlink{#1}\@@href}%
\providecommand \@@href[1]{\endgroup#1\@@endlink}%
\providecommand \@sanitize@url [0]{\catcode `\\12\catcode `\$12\catcode
  `\&12\catcode `\#12\catcode `\^12\catcode `\_12\catcode `\%12\relax}%
\providecommand \@@startlink[1]{}%
\providecommand \@@endlink[0]{}%
\providecommand \url  [0]{\begingroup\@sanitize@url \@url }%
\providecommand \@url [1]{\endgroup\@href {#1}{\urlprefix }}%
\providecommand \urlprefix  [0]{URL }%
\providecommand \Eprint [0]{\href }%
\providecommand \doibase [0]{https://doi.org/}%
\providecommand \selectlanguage [0]{\@gobble}%
\providecommand \bibinfo  [0]{\@secondoftwo}%
\providecommand \bibfield  [0]{\@secondoftwo}%
\providecommand \translation [1]{[#1]}%
\providecommand \BibitemOpen [0]{}%
\providecommand \bibitemStop [0]{}%
\providecommand \bibitemNoStop [0]{.\EOS\space}%
\providecommand \EOS [0]{\spacefactor3000\relax}%
\providecommand \BibitemShut  [1]{\csname bibitem#1\endcsname}%
\let\auto@bib@innerbib\@empty
\bibitem [{\citenamefont {Andrae}(2002)}]{andrae2002nuclear}%
  \BibitemOpen
  \bibfield  {author} {\bibinfo {author} {\bibfnamefont {D.}~\bibnamefont
  {Andrae}},\ }in\ \href@noop {} {\emph {\bibinfo {booktitle} {Theoretical and
  Computational Chemistry}}},\ Vol.~\bibinfo {volume} {11}\ (\bibinfo
  {publisher} {Elsevier},\ \bibinfo {year} {2002})\ p.\ \bibinfo {pages}
  {203}\BibitemShut {NoStop}%
\bibitem [{\citenamefont {Andrae}(2000)}]{ANDRAE2000413}%
  \BibitemOpen
  \bibfield  {author} {\bibinfo {author} {\bibfnamefont {D.}~\bibnamefont
  {Andrae}},\ }\href
  {https://doi.org/https://doi.org/10.1016/S0370-1573(00)00007-7} {\bibfield
  {journal} {\bibinfo  {journal} {Phys. Rep.}\ }\textbf {\bibinfo {volume}
  {336}},\ \bibinfo {pages} {413} (\bibinfo {year} {2000})}\BibitemShut
  {NoStop}%
\bibitem [{\citenamefont {Mohr}(2006)}]{Drake}%
  \BibitemOpen
  \bibfield  {author} {\bibinfo {author} {\bibfnamefont {P.~J.}\ \bibnamefont
  {Mohr}},\ }\href@noop {} {\emph {\bibinfo {title} {Springer Handbook of
  Atomic, Molecular, and Optical Physics}}},\ edited by\ \bibinfo {editor}
  {\bibfnamefont {G.~W.~F.}\ \bibnamefont {Drake}}\ (\bibinfo  {publisher}
  {Springer, New York},\ \bibinfo {year} {2006})\ Chap.\ \bibinfo {chapter}
  {28: Tests of Fundamental Physics}\BibitemShut {NoStop}%
\bibitem [{\citenamefont {Kozhedub}\ \emph {et~al.}(2008)\citenamefont
  {Kozhedub}, \citenamefont {Andreev}, \citenamefont {Shabaev}, \citenamefont
  {Tupitsyn}, \citenamefont {Brandau}, \citenamefont {Kozhuharov},
  \citenamefont {Plunien},\ and\ \citenamefont
  {St\"ohlker}}]{PhysRevA.77.032501}%
  \BibitemOpen
  \bibfield  {author} {\bibinfo {author} {\bibfnamefont {Y.~S.}\ \bibnamefont
  {Kozhedub}}, \bibinfo {author} {\bibfnamefont {O.~V.}\ \bibnamefont
  {Andreev}}, \bibinfo {author} {\bibfnamefont {V.~M.}\ \bibnamefont
  {Shabaev}}, \bibinfo {author} {\bibfnamefont {I.~I.}\ \bibnamefont
  {Tupitsyn}}, \bibinfo {author} {\bibfnamefont {C.}~\bibnamefont {Brandau}},
  \bibinfo {author} {\bibfnamefont {C.}~\bibnamefont {Kozhuharov}}, \bibinfo
  {author} {\bibfnamefont {G.}~\bibnamefont {Plunien}},\ and\ \bibinfo {author}
  {\bibfnamefont {T.}~\bibnamefont {St\"ohlker}},\ }\href
  {https://doi.org/10.1103/PhysRevA.77.032501} {\bibfield  {journal} {\bibinfo
  {journal} {Phys. Rev. A}\ }\textbf {\bibinfo {volume} {77}},\ \bibinfo
  {pages} {032501} (\bibinfo {year} {2008})}\BibitemShut {NoStop}%
\bibitem [{\citenamefont {Church}\ and\ \citenamefont
  {Weneser}(1960)}]{church1960nuclear}%
  \BibitemOpen
  \bibfield  {author} {\bibinfo {author} {\bibfnamefont {E.~L.}\ \bibnamefont
  {Church}}\ and\ \bibinfo {author} {\bibfnamefont {J.}~\bibnamefont
  {Weneser}},\ }\href@noop {} {\bibfield  {journal} {\bibinfo  {journal} {Annu.
  Rev. Nucl. Sci.}\ }\textbf {\bibinfo {volume} {10}},\ \bibinfo {pages} {193}
  (\bibinfo {year} {1960})}\BibitemShut {NoStop}%
\bibitem [{\citenamefont {Church}\ and\ \citenamefont
  {Weneser}(1956)}]{PhysRev.104.1382}%
  \BibitemOpen
  \bibfield  {author} {\bibinfo {author} {\bibfnamefont {E.~L.}\ \bibnamefont
  {Church}}\ and\ \bibinfo {author} {\bibfnamefont {J.}~\bibnamefont
  {Weneser}},\ }\href {https://doi.org/10.1103/PhysRev.104.1382} {\bibfield
  {journal} {\bibinfo  {journal} {Phys. Rev.}\ }\textbf {\bibinfo {volume}
  {104}},\ \bibinfo {pages} {1382} (\bibinfo {year} {1956})}\BibitemShut
  {NoStop}%
\bibitem [{\citenamefont {Behrens}\ and\ \citenamefont
  {Bühring}(1970)}]{BEHRENS1970481}%
  \BibitemOpen
  \bibfield  {author} {\bibinfo {author} {\bibfnamefont {H.}~\bibnamefont
  {Behrens}}\ and\ \bibinfo {author} {\bibfnamefont {W.}~\bibnamefont
  {Bühring}},\ }\href
  {https://doi.org/https://doi.org/10.1016/0375-9474(70)90413-6} {\bibfield
  {journal} {\bibinfo  {journal} {Nucl. Phys. A}\ }\textbf {\bibinfo {volume}
  {150}},\ \bibinfo {pages} {481} (\bibinfo {year} {1970})}\BibitemShut
  {NoStop}%
\bibitem [{\citenamefont {Rose}\ and\ \citenamefont
  {Holmes}(1951)}]{PhysRev.83.190.2}%
  \BibitemOpen
  \bibfield  {author} {\bibinfo {author} {\bibfnamefont {M.~E.}\ \bibnamefont
  {Rose}}\ and\ \bibinfo {author} {\bibfnamefont {D.~K.}\ \bibnamefont
  {Holmes}},\ }\href {https://doi.org/10.1103/PhysRev.83.190.2} {\bibfield
  {journal} {\bibinfo  {journal} {Phys. Rev.}\ }\textbf {\bibinfo {volume}
  {83}},\ \bibinfo {pages} {190} (\bibinfo {year} {1951})}\BibitemShut
  {NoStop}%
\bibitem [{\citenamefont {Hayen}\ \emph {et~al.}(2018)\citenamefont {Hayen},
  \citenamefont {Severijns}, \citenamefont {Bodek}, \citenamefont {Rozpedzik},\
  and\ \citenamefont {Mougeot}}]{RevModPhys.90.015008}%
  \BibitemOpen
  \bibfield  {author} {\bibinfo {author} {\bibfnamefont {L.}~\bibnamefont
  {Hayen}}, \bibinfo {author} {\bibfnamefont {N.}~\bibnamefont {Severijns}},
  \bibinfo {author} {\bibfnamefont {K.}~\bibnamefont {Bodek}}, \bibinfo
  {author} {\bibfnamefont {D.}~\bibnamefont {Rozpedzik}},\ and\ \bibinfo
  {author} {\bibfnamefont {X.}~\bibnamefont {Mougeot}},\ }\href
  {https://doi.org/10.1103/RevModPhys.90.015008} {\bibfield  {journal}
  {\bibinfo  {journal} {Rev. Mod. Phys.}\ }\textbf {\bibinfo {volume} {90}},\
  \bibinfo {pages} {015008} (\bibinfo {year} {2018})}\BibitemShut {NoStop}%
\bibitem [{\citenamefont {Sarriguren}(2020)}]{Sarriguren_2020}%
  \BibitemOpen
  \bibfield  {author} {\bibinfo {author} {\bibfnamefont {P.}~\bibnamefont
  {Sarriguren}},\ }\href {https://doi.org/10.1088/1361-6471/ab920d} {\bibfield
  {journal} {\bibinfo  {journal} {J. Phys. G: Nucl. Part. Phys.}\ }\textbf
  {\bibinfo {volume} {47}},\ \bibinfo {pages} {125107} (\bibinfo {year}
  {2020})}\BibitemShut {NoStop}%
\bibitem [{\citenamefont {Bambynek}\ \emph {et~al.}(1977)\citenamefont
  {Bambynek}, \citenamefont {Behrens}, \citenamefont {Chen}, \citenamefont
  {Crasemann}, \citenamefont {Fitzpatrick}, \citenamefont {Ledingham},
  \citenamefont {Genz}, \citenamefont {Mutterer},\ and\ \citenamefont
  {Intemann}}]{RevModPhys.49.77}%
  \BibitemOpen
  \bibfield  {author} {\bibinfo {author} {\bibfnamefont {W.}~\bibnamefont
  {Bambynek}}, \bibinfo {author} {\bibfnamefont {H.}~\bibnamefont {Behrens}},
  \bibinfo {author} {\bibfnamefont {M.~H.}\ \bibnamefont {Chen}}, \bibinfo
  {author} {\bibfnamefont {B.}~\bibnamefont {Crasemann}}, \bibinfo {author}
  {\bibfnamefont {M.~L.}\ \bibnamefont {Fitzpatrick}}, \bibinfo {author}
  {\bibfnamefont {K.~W.~D.}\ \bibnamefont {Ledingham}}, \bibinfo {author}
  {\bibfnamefont {H.}~\bibnamefont {Genz}}, \bibinfo {author} {\bibfnamefont
  {M.}~\bibnamefont {Mutterer}},\ and\ \bibinfo {author} {\bibfnamefont
  {R.~L.}\ \bibnamefont {Intemann}},\ }\href
  {https://doi.org/10.1103/RevModPhys.49.77} {\bibfield  {journal} {\bibinfo
  {journal} {Rev. Mod. Phys.}\ }\textbf {\bibinfo {volume} {49}},\ \bibinfo
  {pages} {77} (\bibinfo {year} {1977})}\BibitemShut {NoStop}%
\bibitem [{\citenamefont {Sturm}\ \emph {et~al.}(2013)\citenamefont {Sturm},
  \citenamefont {Wagner}, \citenamefont {Kretzschmar}, \citenamefont {Quint},
  \citenamefont {Werth},\ and\ \citenamefont {Blaum}}]{PhysRevA.87.030501}%
  \BibitemOpen
  \bibfield  {author} {\bibinfo {author} {\bibfnamefont {S.}~\bibnamefont
  {Sturm}}, \bibinfo {author} {\bibfnamefont {A.}~\bibnamefont {Wagner}},
  \bibinfo {author} {\bibfnamefont {M.}~\bibnamefont {Kretzschmar}}, \bibinfo
  {author} {\bibfnamefont {W.}~\bibnamefont {Quint}}, \bibinfo {author}
  {\bibfnamefont {G.}~\bibnamefont {Werth}},\ and\ \bibinfo {author}
  {\bibfnamefont {K.}~\bibnamefont {Blaum}},\ }\href
  {https://doi.org/10.1103/PhysRevA.87.030501} {\bibfield  {journal} {\bibinfo
  {journal} {Phys. Rev. A}\ }\textbf {\bibinfo {volume} {87}},\ \bibinfo
  {pages} {030501} (\bibinfo {year} {2013})}\BibitemShut {NoStop}%
\bibitem [{\citenamefont {Kraft-Bermuth}\ \emph {et~al.}(2017)\citenamefont
  {Kraft-Bermuth}, \citenamefont {Andrianov}, \citenamefont {Bleile},
  \citenamefont {Echler}, \citenamefont {Egelhof}, \citenamefont {Grabitz},
  \citenamefont {Ilieva}, \citenamefont {Kiselev}, \citenamefont {Kilbourne},
  \citenamefont {McCammon}, \citenamefont {Meier},\ and\ \citenamefont
  {Scholz}}]{Kraft_Bermuth_2017}%
  \BibitemOpen
  \bibfield  {author} {\bibinfo {author} {\bibfnamefont {S.}~\bibnamefont
  {Kraft-Bermuth}}, \bibinfo {author} {\bibfnamefont {V.}~\bibnamefont
  {Andrianov}}, \bibinfo {author} {\bibfnamefont {A.}~\bibnamefont {Bleile}},
  \bibinfo {author} {\bibfnamefont {A.}~\bibnamefont {Echler}}, \bibinfo
  {author} {\bibfnamefont {P.}~\bibnamefont {Egelhof}}, \bibinfo {author}
  {\bibfnamefont {P.}~\bibnamefont {Grabitz}}, \bibinfo {author} {\bibfnamefont
  {S.}~\bibnamefont {Ilieva}}, \bibinfo {author} {\bibfnamefont
  {O.}~\bibnamefont {Kiselev}}, \bibinfo {author} {\bibfnamefont
  {C.}~\bibnamefont {Kilbourne}}, \bibinfo {author} {\bibfnamefont
  {D.}~\bibnamefont {McCammon}}, \bibinfo {author} {\bibfnamefont {J.~P.}\
  \bibnamefont {Meier}},\ and\ \bibinfo {author} {\bibfnamefont
  {P.}~\bibnamefont {Scholz}},\ }\href
  {https://doi.org/10.1088/1361-6455/50/5/055603} {\bibfield  {journal}
  {\bibinfo  {journal} {J. Phys. B: At. Mol. Opt. Phys.}\ }\textbf {\bibinfo
  {volume} {50}},\ \bibinfo {pages} {055603} (\bibinfo {year}
  {2017})}\BibitemShut {NoStop}%
\bibitem [{\citenamefont {Beiersdorfer}\ \emph {et~al.}(1998)\citenamefont
  {Beiersdorfer}, \citenamefont {Osterheld}, \citenamefont {Scofield},
  \citenamefont {Crespo L\'opez-Urrutia},\ and\ \citenamefont
  {Widmann}}]{PhysRevLett.80.3022}%
  \BibitemOpen
  \bibfield  {author} {\bibinfo {author} {\bibfnamefont {P.}~\bibnamefont
  {Beiersdorfer}}, \bibinfo {author} {\bibfnamefont {A.~L.}\ \bibnamefont
  {Osterheld}}, \bibinfo {author} {\bibfnamefont {J.~H.}\ \bibnamefont
  {Scofield}}, \bibinfo {author} {\bibfnamefont {J.~R.}\ \bibnamefont {Crespo
  L\'opez-Urrutia}},\ and\ \bibinfo {author} {\bibfnamefont {K.}~\bibnamefont
  {Widmann}},\ }\href {https://doi.org/10.1103/PhysRevLett.80.3022} {\bibfield
  {journal} {\bibinfo  {journal} {Phys. Rev. Lett.}\ }\textbf {\bibinfo
  {volume} {80}},\ \bibinfo {pages} {3022} (\bibinfo {year}
  {1998})}\BibitemShut {NoStop}%
\bibitem [{\citenamefont {H\"affner}\ \emph {et~al.}(2000)\citenamefont
  {H\"affner}, \citenamefont {Beier}, \citenamefont {Hermanspahn},
  \citenamefont {Kluge}, \citenamefont {Quint}, \citenamefont {Stahl},
  \citenamefont {Verd\'u},\ and\ \citenamefont {Werth}}]{PhysRevLett.85.5308}%
  \BibitemOpen
  \bibfield  {author} {\bibinfo {author} {\bibfnamefont {H.}~\bibnamefont
  {H\"affner}}, \bibinfo {author} {\bibfnamefont {T.}~\bibnamefont {Beier}},
  \bibinfo {author} {\bibfnamefont {N.}~\bibnamefont {Hermanspahn}}, \bibinfo
  {author} {\bibfnamefont {H.-J.}\ \bibnamefont {Kluge}}, \bibinfo {author}
  {\bibfnamefont {W.}~\bibnamefont {Quint}}, \bibinfo {author} {\bibfnamefont
  {S.}~\bibnamefont {Stahl}}, \bibinfo {author} {\bibfnamefont
  {J.}~\bibnamefont {Verd\'u}},\ and\ \bibinfo {author} {\bibfnamefont
  {G.}~\bibnamefont {Werth}},\ }\href
  {https://doi.org/10.1103/PhysRevLett.85.5308} {\bibfield  {journal} {\bibinfo
   {journal} {Phys. Rev. Lett.}\ }\textbf {\bibinfo {volume} {85}},\ \bibinfo
  {pages} {5308} (\bibinfo {year} {2000})}\BibitemShut {NoStop}%
\bibitem [{\citenamefont {Verd\'u}\ \emph {et~al.}(2004)\citenamefont
  {Verd\'u}, \citenamefont {Djeki\ifmmode~\acute{c}\else \'{c}\fi{}},
  \citenamefont {Stahl}, \citenamefont {Valenzuela}, \citenamefont {Vogel},
  \citenamefont {Werth}, \citenamefont {Beier}, \citenamefont {Kluge},\ and\
  \citenamefont {Quint}}]{PhysRevLett.92.093002}%
  \BibitemOpen
  \bibfield  {author} {\bibinfo {author} {\bibfnamefont {J.}~\bibnamefont
  {Verd\'u}}, \bibinfo {author} {\bibfnamefont {S.}~\bibnamefont
  {Djeki\ifmmode~\acute{c}\else \'{c}\fi{}}}, \bibinfo {author} {\bibfnamefont
  {S.}~\bibnamefont {Stahl}}, \bibinfo {author} {\bibfnamefont
  {T.}~\bibnamefont {Valenzuela}}, \bibinfo {author} {\bibfnamefont
  {M.}~\bibnamefont {Vogel}}, \bibinfo {author} {\bibfnamefont
  {G.}~\bibnamefont {Werth}}, \bibinfo {author} {\bibfnamefont
  {T.}~\bibnamefont {Beier}}, \bibinfo {author} {\bibfnamefont {H.-J.}\
  \bibnamefont {Kluge}},\ and\ \bibinfo {author} {\bibfnamefont
  {W.}~\bibnamefont {Quint}},\ }\href
  {https://doi.org/10.1103/PhysRevLett.92.093002} {\bibfield  {journal}
  {\bibinfo  {journal} {Phys. Rev. Lett.}\ }\textbf {\bibinfo {volume} {92}},\
  \bibinfo {pages} {093002} (\bibinfo {year} {2004})}\BibitemShut {NoStop}%
\bibitem [{\citenamefont {Gumberidze}\ \emph {et~al.}(2005)\citenamefont
  {Gumberidze}, \citenamefont {St\"ohlker}, \citenamefont
  {Bana\ifmmode~\acute{s}\else \'{s}\fi{}}, \citenamefont {Beckert},
  \citenamefont {Beller}, \citenamefont {Beyer}, \citenamefont {Bosch},
  \citenamefont {Hagmann}, \citenamefont {Kozhuharov}, \citenamefont {Liesen},
  \citenamefont {Nolden}, \citenamefont {Ma}, \citenamefont {Mokler},
  \citenamefont {Steck}, \citenamefont {Sierpowski},\ and\ \citenamefont
  {Tashenov}}]{PhysRevLett.94.223001}%
  \BibitemOpen
  \bibfield  {author} {\bibinfo {author} {\bibfnamefont {A.}~\bibnamefont
  {Gumberidze}}, \bibinfo {author} {\bibfnamefont {T.}~\bibnamefont
  {St\"ohlker}}, \bibinfo {author} {\bibfnamefont {D.}~\bibnamefont
  {Bana\ifmmode~\acute{s}\else \'{s}\fi{}}}, \bibinfo {author} {\bibfnamefont
  {K.}~\bibnamefont {Beckert}}, \bibinfo {author} {\bibfnamefont
  {P.}~\bibnamefont {Beller}}, \bibinfo {author} {\bibfnamefont {H.~F.}\
  \bibnamefont {Beyer}}, \bibinfo {author} {\bibfnamefont {F.}~\bibnamefont
  {Bosch}}, \bibinfo {author} {\bibfnamefont {S.}~\bibnamefont {Hagmann}},
  \bibinfo {author} {\bibfnamefont {C.}~\bibnamefont {Kozhuharov}}, \bibinfo
  {author} {\bibfnamefont {D.}~\bibnamefont {Liesen}}, \bibinfo {author}
  {\bibfnamefont {F.}~\bibnamefont {Nolden}}, \bibinfo {author} {\bibfnamefont
  {X.}~\bibnamefont {Ma}}, \bibinfo {author} {\bibfnamefont {P.~H.}\
  \bibnamefont {Mokler}}, \bibinfo {author} {\bibfnamefont {M.}~\bibnamefont
  {Steck}}, \bibinfo {author} {\bibfnamefont {D.}~\bibnamefont {Sierpowski}},\
  and\ \bibinfo {author} {\bibfnamefont {S.}~\bibnamefont {Tashenov}},\ }\href
  {https://doi.org/10.1103/PhysRevLett.94.223001} {\bibfield  {journal}
  {\bibinfo  {journal} {Phys. Rev. Lett.}\ }\textbf {\bibinfo {volume} {94}},\
  \bibinfo {pages} {223001} (\bibinfo {year} {2005})}\BibitemShut {NoStop}%
\bibitem [{\citenamefont {Sturm}\ \emph {et~al.}(2011)\citenamefont {Sturm},
  \citenamefont {Wagner}, \citenamefont {Schabinger}, \citenamefont {Zatorski},
  \citenamefont {Harman}, \citenamefont {Quint}, \citenamefont {Werth},
  \citenamefont {Keitel},\ and\ \citenamefont
  {Blaum}}]{PhysRevLett.107.023002}%
  \BibitemOpen
  \bibfield  {author} {\bibinfo {author} {\bibfnamefont {S.}~\bibnamefont
  {Sturm}}, \bibinfo {author} {\bibfnamefont {A.}~\bibnamefont {Wagner}},
  \bibinfo {author} {\bibfnamefont {B.}~\bibnamefont {Schabinger}}, \bibinfo
  {author} {\bibfnamefont {J.}~\bibnamefont {Zatorski}}, \bibinfo {author}
  {\bibfnamefont {Z.}~\bibnamefont {Harman}}, \bibinfo {author} {\bibfnamefont
  {W.}~\bibnamefont {Quint}}, \bibinfo {author} {\bibfnamefont
  {G.}~\bibnamefont {Werth}}, \bibinfo {author} {\bibfnamefont {C.~H.}\
  \bibnamefont {Keitel}},\ and\ \bibinfo {author} {\bibfnamefont
  {K.}~\bibnamefont {Blaum}},\ }\href
  {https://doi.org/10.1103/PhysRevLett.107.023002} {\bibfield  {journal}
  {\bibinfo  {journal} {Phys. Rev. Lett.}\ }\textbf {\bibinfo {volume} {107}},\
  \bibinfo {pages} {023002} (\bibinfo {year} {2011})}\BibitemShut {NoStop}%
\bibitem [{\citenamefont {Arapoglou}\ \emph {et~al.}(2019)\citenamefont
  {Arapoglou}, \citenamefont {Egl}, \citenamefont {H\"ocker}, \citenamefont
  {Sailer}, \citenamefont {Tu}, \citenamefont {Weigel}, \citenamefont {Wolf},
  \citenamefont {Cakir}, \citenamefont {Yerokhin}, \citenamefont {Oreshkina},
  \citenamefont {Agababaev}, \citenamefont {Volotka}, \citenamefont {Zinenko},
  \citenamefont {Glazov}, \citenamefont {Harman}, \citenamefont {Keitel},
  \citenamefont {Sturm},\ and\ \citenamefont {Blaum}}]{PhysRevLett.122.253001}%
  \BibitemOpen
  \bibfield  {author} {\bibinfo {author} {\bibfnamefont {I.}~\bibnamefont
  {Arapoglou}}, \bibinfo {author} {\bibfnamefont {A.}~\bibnamefont {Egl}},
  \bibinfo {author} {\bibfnamefont {M.}~\bibnamefont {H\"ocker}}, \bibinfo
  {author} {\bibfnamefont {T.}~\bibnamefont {Sailer}}, \bibinfo {author}
  {\bibfnamefont {B.}~\bibnamefont {Tu}}, \bibinfo {author} {\bibfnamefont
  {A.}~\bibnamefont {Weigel}}, \bibinfo {author} {\bibfnamefont
  {R.}~\bibnamefont {Wolf}}, \bibinfo {author} {\bibfnamefont {H.}~\bibnamefont
  {Cakir}}, \bibinfo {author} {\bibfnamefont {V.~A.}\ \bibnamefont {Yerokhin}},
  \bibinfo {author} {\bibfnamefont {N.~S.}\ \bibnamefont {Oreshkina}}, \bibinfo
  {author} {\bibfnamefont {V.~A.}\ \bibnamefont {Agababaev}}, \bibinfo {author}
  {\bibfnamefont {A.~V.}\ \bibnamefont {Volotka}}, \bibinfo {author}
  {\bibfnamefont {D.~V.}\ \bibnamefont {Zinenko}}, \bibinfo {author}
  {\bibfnamefont {D.~A.}\ \bibnamefont {Glazov}}, \bibinfo {author}
  {\bibfnamefont {Z.}~\bibnamefont {Harman}}, \bibinfo {author} {\bibfnamefont
  {C.~H.}\ \bibnamefont {Keitel}}, \bibinfo {author} {\bibfnamefont
  {S.}~\bibnamefont {Sturm}},\ and\ \bibinfo {author} {\bibfnamefont
  {K.}~\bibnamefont {Blaum}},\ }\href
  {https://doi.org/10.1103/PhysRevLett.122.253001} {\bibfield  {journal}
  {\bibinfo  {journal} {Phys. Rev. Lett.}\ }\textbf {\bibinfo {volume} {122}},\
  \bibinfo {pages} {253001} (\bibinfo {year} {2019})}\BibitemShut {NoStop}%
\bibitem [{\citenamefont {Egl}\ \emph {et~al.}(2019)\citenamefont {Egl},
  \citenamefont {Arapoglou}, \citenamefont {H\"ocker}, \citenamefont {K\"onig},
  \citenamefont {Ratajczyk}, \citenamefont {Sailer}, \citenamefont {Tu},
  \citenamefont {Weigel}, \citenamefont {Blaum}, \citenamefont
  {N\"ortersh\"auser},\ and\ \citenamefont {Sturm}}]{PhysRevLett.123.123001}%
  \BibitemOpen
  \bibfield  {author} {\bibinfo {author} {\bibfnamefont {A.}~\bibnamefont
  {Egl}}, \bibinfo {author} {\bibfnamefont {I.}~\bibnamefont {Arapoglou}},
  \bibinfo {author} {\bibfnamefont {M.}~\bibnamefont {H\"ocker}}, \bibinfo
  {author} {\bibfnamefont {K.}~\bibnamefont {K\"onig}}, \bibinfo {author}
  {\bibfnamefont {T.}~\bibnamefont {Ratajczyk}}, \bibinfo {author}
  {\bibfnamefont {T.}~\bibnamefont {Sailer}}, \bibinfo {author} {\bibfnamefont
  {B.}~\bibnamefont {Tu}}, \bibinfo {author} {\bibfnamefont {A.}~\bibnamefont
  {Weigel}}, \bibinfo {author} {\bibfnamefont {K.}~\bibnamefont {Blaum}},
  \bibinfo {author} {\bibfnamefont {W.}~\bibnamefont {N\"ortersh\"auser}},\
  and\ \bibinfo {author} {\bibfnamefont {S.}~\bibnamefont {Sturm}},\ }\href
  {https://doi.org/10.1103/PhysRevLett.123.123001} {\bibfield  {journal}
  {\bibinfo  {journal} {Phys. Rev. Lett.}\ }\textbf {\bibinfo {volume} {123}},\
  \bibinfo {pages} {123001} (\bibinfo {year} {2019})}\BibitemShut {NoStop}%
\bibitem [{\citenamefont {Mohr}\ \emph {et~al.}(1998)\citenamefont {Mohr},
  \citenamefont {Plunien},\ and\ \citenamefont {Soff}}]{MOHR1998227}%
  \BibitemOpen
  \bibfield  {author} {\bibinfo {author} {\bibfnamefont {P.~J.}\ \bibnamefont
  {Mohr}}, \bibinfo {author} {\bibfnamefont {G.}~\bibnamefont {Plunien}},\ and\
  \bibinfo {author} {\bibfnamefont {G.}~\bibnamefont {Soff}},\ }\href
  {https://doi.org/https://doi.org/10.1016/S0370-1573(97)00046-X} {\bibfield
  {journal} {\bibinfo  {journal} {Phys. Rep.}\ }\textbf {\bibinfo {volume}
  {293}},\ \bibinfo {pages} {227} (\bibinfo {year} {1998})}\BibitemShut
  {NoStop}%
\bibitem [{\citenamefont {Yerokhin}\ and\ \citenamefont
  {Shabaev}(2015)}]{yerokhin2015lamb}%
  \BibitemOpen
  \bibfield  {author} {\bibinfo {author} {\bibfnamefont {V.~A.}\ \bibnamefont
  {Yerokhin}}\ and\ \bibinfo {author} {\bibfnamefont {V.~M.}\ \bibnamefont
  {Shabaev}},\ }\href@noop {} {\bibfield  {journal} {\bibinfo  {journal} {J.
  Phys. Chem. Ref. Data}\ }\textbf {\bibinfo {volume} {44}},\ \bibinfo {pages}
  {033103} (\bibinfo {year} {2015})}\BibitemShut {NoStop}%
\bibitem [{\citenamefont {Glazov}\ and\ \citenamefont
  {Shabaev}(2002)}]{GLAZOV2002408}%
  \BibitemOpen
  \bibfield  {author} {\bibinfo {author} {\bibfnamefont {D.~A.}\ \bibnamefont
  {Glazov}}\ and\ \bibinfo {author} {\bibfnamefont {V.~M.}\ \bibnamefont
  {Shabaev}},\ }\href
  {https://doi.org/https://doi.org/10.1016/S0375-9601(02)00021-X} {\bibfield
  {journal} {\bibinfo  {journal} {Phys. Lett. A}\ }\textbf {\bibinfo {volume}
  {297}},\ \bibinfo {pages} {408} (\bibinfo {year} {2002})}\BibitemShut
  {NoStop}%
\bibitem [{\citenamefont {Johnson}\ and\ \citenamefont
  {Soff}(1985)}]{johnson1985lamb}%
  \BibitemOpen
  \bibfield  {author} {\bibinfo {author} {\bibfnamefont {W.}~\bibnamefont
  {Johnson}}\ and\ \bibinfo {author} {\bibfnamefont {G.}~\bibnamefont {Soff}},\
  }\href {https://doi.org/10.1016/0092-640X(85)90010-5} {\bibfield  {journal}
  {\bibinfo  {journal} {At. Data. Nucl. Data Tables}\ }\textbf {\bibinfo
  {volume} {33}},\ \bibinfo {pages} {405} (\bibinfo {year} {1985})}\BibitemShut
  {NoStop}%
\bibitem [{\citenamefont {{I. Eides}}\ \emph {et~al.}(2001)\citenamefont {{I.
  Eides}}, \citenamefont {Grotch},\ and\ \citenamefont
  {Shelyuto}}]{IEIDES200163}%
  \BibitemOpen
  \bibfield  {author} {\bibinfo {author} {\bibfnamefont {M.}~\bibnamefont {{I.
  Eides}}}, \bibinfo {author} {\bibfnamefont {H.}~\bibnamefont {Grotch}},\ and\
  \bibinfo {author} {\bibfnamefont {V.~A.}\ \bibnamefont {Shelyuto}},\ }\href
  {https://doi.org/https://doi.org/10.1016/S0370-1573(00)00077-6} {\bibfield
  {journal} {\bibinfo  {journal} {Phys. Rep.}\ }\textbf {\bibinfo {volume}
  {342}},\ \bibinfo {pages} {63} (\bibinfo {year} {2001})}\BibitemShut
  {NoStop}%
\bibitem [{\citenamefont {Beier}(2000)}]{BEIER200079}%
  \BibitemOpen
  \bibfield  {author} {\bibinfo {author} {\bibfnamefont {T.}~\bibnamefont
  {Beier}},\ }\href
  {https://doi.org/https://doi.org/10.1016/S0370-1573(00)00071-5} {\bibfield
  {journal} {\bibinfo  {journal} {Phys. Rep.}\ }\textbf {\bibinfo {volume}
  {339}},\ \bibinfo {pages} {79} (\bibinfo {year} {2000})}\BibitemShut
  {NoStop}%
\bibitem [{\citenamefont {Shabaev}\ \emph {et~al.}(2018)\citenamefont
  {Shabaev}, \citenamefont {Bondarev}, \citenamefont {Glazov}, \citenamefont
  {Kaygorodov}, \citenamefont {Kozhedub}, \citenamefont {Maltsev},
  \citenamefont {Malyshev}, \citenamefont {Popov}, \citenamefont {Tupitsyn},\
  and\ \citenamefont {Zubova}}]{shabaev2018stringent}%
  \BibitemOpen
  \bibfield  {author} {\bibinfo {author} {\bibfnamefont {V.~M.}\ \bibnamefont
  {Shabaev}}, \bibinfo {author} {\bibfnamefont {A.~I.}\ \bibnamefont
  {Bondarev}}, \bibinfo {author} {\bibfnamefont {D.~A.}\ \bibnamefont
  {Glazov}}, \bibinfo {author} {\bibfnamefont {M.~Y.}\ \bibnamefont
  {Kaygorodov}}, \bibinfo {author} {\bibfnamefont {Y.~S.}\ \bibnamefont
  {Kozhedub}}, \bibinfo {author} {\bibfnamefont {I.~A.}\ \bibnamefont
  {Maltsev}}, \bibinfo {author} {\bibfnamefont {A.~V.}\ \bibnamefont
  {Malyshev}}, \bibinfo {author} {\bibfnamefont {R.~V.}\ \bibnamefont {Popov}},
  \bibinfo {author} {\bibfnamefont {I.~I.}\ \bibnamefont {Tupitsyn}},\ and\
  \bibinfo {author} {\bibfnamefont {N.~A.}\ \bibnamefont {Zubova}},\ }\href
  {https://doi.org/10.1007/s10751-018-1537-8} {\bibfield  {journal} {\bibinfo
  {journal} {Hyperfine Interact.}\ }\textbf {\bibinfo {volume} {239}},\
  \bibinfo {pages} {1} (\bibinfo {year} {2018})}\BibitemShut {NoStop}%
\bibitem [{\citenamefont {Paul}\ \emph {et~al.}(2021)\citenamefont {Paul},
  \citenamefont {Bian}, \citenamefont {Azuma}, \citenamefont {Okada},\ and\
  \citenamefont {Indelicato}}]{PhysRevLett.126.173001}%
  \BibitemOpen
  \bibfield  {author} {\bibinfo {author} {\bibfnamefont {N.}~\bibnamefont
  {Paul}}, \bibinfo {author} {\bibfnamefont {G.}~\bibnamefont {Bian}}, \bibinfo
  {author} {\bibfnamefont {T.}~\bibnamefont {Azuma}}, \bibinfo {author}
  {\bibfnamefont {S.}~\bibnamefont {Okada}},\ and\ \bibinfo {author}
  {\bibfnamefont {P.}~\bibnamefont {Indelicato}},\ }\href
  {https://doi.org/10.1103/PhysRevLett.126.173001} {\bibfield  {journal}
  {\bibinfo  {journal} {Phys. Rev. Lett.}\ }\textbf {\bibinfo {volume} {126}},\
  \bibinfo {pages} {173001} (\bibinfo {year} {2021})}\BibitemShut {NoStop}%
\bibitem [{\citenamefont {Volotka}\ and\ \citenamefont
  {Plunien}(2014{\natexlab{a}})}]{volotka2014nuclear}%
  \BibitemOpen
  \bibfield  {author} {\bibinfo {author} {\bibfnamefont {A.~V.}\ \bibnamefont
  {Volotka}}\ and\ \bibinfo {author} {\bibfnamefont {G.}~\bibnamefont
  {Plunien}},\ }\href {https://doi.org/10.1103/PhysRevLett.113.023002}
  {\bibfield  {journal} {\bibinfo  {journal} {Phys. Rev. Lett.}\ }\textbf
  {\bibinfo {volume} {113}},\ \bibinfo {pages} {023002} (\bibinfo {year}
  {2014}{\natexlab{a}})}\BibitemShut {NoStop}%
\bibitem [{\citenamefont {Indelicato}(2019)}]{Indelicato_2019}%
  \BibitemOpen
  \bibfield  {author} {\bibinfo {author} {\bibfnamefont {P.}~\bibnamefont
  {Indelicato}},\ }\href {https://doi.org/10.1088/1361-6455/ab42c9} {\bibfield
  {journal} {\bibinfo  {journal} {J. Phys. B: At. Mol. Opt. Phys.}\ }\textbf
  {\bibinfo {volume} {52}},\ \bibinfo {pages} {232001} (\bibinfo {year}
  {2019})}\BibitemShut {NoStop}%
\bibitem [{\citenamefont {Blaum}\ \emph {et~al.}(2020)\citenamefont {Blaum},
  \citenamefont {Eliseev},\ and\ \citenamefont {Sturm}}]{Blaum_2020}%
  \BibitemOpen
  \bibfield  {author} {\bibinfo {author} {\bibfnamefont {K.}~\bibnamefont
  {Blaum}}, \bibinfo {author} {\bibfnamefont {S.}~\bibnamefont {Eliseev}},\
  and\ \bibinfo {author} {\bibfnamefont {S.}~\bibnamefont {Sturm}},\ }\href
  {https://doi.org/10.1088/2058-9565/abbc75} {\bibfield  {journal} {\bibinfo
  {journal} {Quantum Sci. Technol.}\ }\textbf {\bibinfo {volume} {6}},\
  \bibinfo {pages} {014002} (\bibinfo {year} {2020})}\BibitemShut {NoStop}%
\bibitem [{\citenamefont {Sailer}\ \emph {et~al.}(2022)\citenamefont {Sailer},
  \citenamefont {Debierre}, \citenamefont {Harman}, \citenamefont {Heiße},
  \citenamefont {König}, \citenamefont {Morgner}, \citenamefont {Tu},
  \citenamefont {Volotka}, \citenamefont {Keitel}, \citenamefont {Blaum},\ and\
  \citenamefont {Sturm}}]{Sailer2022}%
  \BibitemOpen
  \bibfield  {author} {\bibinfo {author} {\bibfnamefont {T.}~\bibnamefont
  {Sailer}}, \bibinfo {author} {\bibfnamefont {V.}~\bibnamefont {Debierre}},
  \bibinfo {author} {\bibfnamefont {Z.}~\bibnamefont {Harman}}, \bibinfo
  {author} {\bibfnamefont {F.}~\bibnamefont {Heiße}}, \bibinfo {author}
  {\bibfnamefont {C.}~\bibnamefont {König}}, \bibinfo {author} {\bibfnamefont
  {J.}~\bibnamefont {Morgner}}, \bibinfo {author} {\bibfnamefont
  {B.}~\bibnamefont {Tu}}, \bibinfo {author} {\bibfnamefont {A.~V.}\
  \bibnamefont {Volotka}}, \bibinfo {author} {\bibfnamefont {C.~H.}\
  \bibnamefont {Keitel}}, \bibinfo {author} {\bibfnamefont {K.}~\bibnamefont
  {Blaum}},\ and\ \bibinfo {author} {\bibfnamefont {S.}~\bibnamefont {Sturm}},\
  }\href {https://doi.org/10.1038/s41586-022-04807-w} {\bibfield  {journal}
  {\bibinfo  {journal} {Nature}\ }\textbf {\bibinfo {volume} {606}},\ \bibinfo
  {pages} {479} (\bibinfo {year} {2022})}\BibitemShut {NoStop}%
\bibitem [{\citenamefont {Windberger}\ \emph {et~al.}(2015)\citenamefont
  {Windberger}, \citenamefont {Crespo L\'opez-Urrutia}, \citenamefont {Bekker},
  \citenamefont {Oreshkina}, \citenamefont {Berengut}, \citenamefont {Bock},
  \citenamefont {Borschevsky}, \citenamefont {Dzuba}, \citenamefont {Eliav},
  \citenamefont {Harman}, \citenamefont {Kaldor}, \citenamefont {Kaul},
  \citenamefont {Safronova}, \citenamefont {Flambaum}, \citenamefont {Keitel},
  \citenamefont {Schmidt}, \citenamefont {Ullrich},\ and\ \citenamefont
  {Versolato}}]{PhysRevLett.114.150801}%
  \BibitemOpen
  \bibfield  {author} {\bibinfo {author} {\bibfnamefont {A.}~\bibnamefont
  {Windberger}}, \bibinfo {author} {\bibfnamefont {J.~R.}\ \bibnamefont {Crespo
  L\'opez-Urrutia}}, \bibinfo {author} {\bibfnamefont {H.}~\bibnamefont
  {Bekker}}, \bibinfo {author} {\bibfnamefont {N.~S.}\ \bibnamefont
  {Oreshkina}}, \bibinfo {author} {\bibfnamefont {J.~C.}\ \bibnamefont
  {Berengut}}, \bibinfo {author} {\bibfnamefont {V.}~\bibnamefont {Bock}},
  \bibinfo {author} {\bibfnamefont {A.}~\bibnamefont {Borschevsky}}, \bibinfo
  {author} {\bibfnamefont {V.~A.}\ \bibnamefont {Dzuba}}, \bibinfo {author}
  {\bibfnamefont {E.}~\bibnamefont {Eliav}}, \bibinfo {author} {\bibfnamefont
  {Z.}~\bibnamefont {Harman}}, \bibinfo {author} {\bibfnamefont
  {U.}~\bibnamefont {Kaldor}}, \bibinfo {author} {\bibfnamefont
  {S.}~\bibnamefont {Kaul}}, \bibinfo {author} {\bibfnamefont {U.~I.}\
  \bibnamefont {Safronova}}, \bibinfo {author} {\bibfnamefont {V.~V.}\
  \bibnamefont {Flambaum}}, \bibinfo {author} {\bibfnamefont {C.~H.}\
  \bibnamefont {Keitel}}, \bibinfo {author} {\bibfnamefont {P.~O.}\
  \bibnamefont {Schmidt}}, \bibinfo {author} {\bibfnamefont {J.}~\bibnamefont
  {Ullrich}},\ and\ \bibinfo {author} {\bibfnamefont {O.~O.}\ \bibnamefont
  {Versolato}},\ }\href {https://doi.org/10.1103/PhysRevLett.114.150801}
  {\bibfield  {journal} {\bibinfo  {journal} {Phys. Rev. Lett.}\ }\textbf
  {\bibinfo {volume} {114}},\ \bibinfo {pages} {150801} (\bibinfo {year}
  {2015})}\BibitemShut {NoStop}%
\bibitem [{\citenamefont {Aoyama}\ \emph {et~al.}(2012)\citenamefont {Aoyama},
  \citenamefont {Hayakawa}, \citenamefont {Kinoshita},\ and\ \citenamefont
  {Nio}}]{PhysRevLett.109.111807}%
  \BibitemOpen
  \bibfield  {author} {\bibinfo {author} {\bibfnamefont {T.}~\bibnamefont
  {Aoyama}}, \bibinfo {author} {\bibfnamefont {M.}~\bibnamefont {Hayakawa}},
  \bibinfo {author} {\bibfnamefont {T.}~\bibnamefont {Kinoshita}},\ and\
  \bibinfo {author} {\bibfnamefont {M.}~\bibnamefont {Nio}},\ }\href
  {https://doi.org/10.1103/PhysRevLett.109.111807} {\bibfield  {journal}
  {\bibinfo  {journal} {Phys. Rev. Lett.}\ }\textbf {\bibinfo {volume} {109}},\
  \bibinfo {pages} {111807} (\bibinfo {year} {2012})}\BibitemShut {NoStop}%
\bibitem [{\citenamefont {Berengut}\ \emph {et~al.}(2010)\citenamefont
  {Berengut}, \citenamefont {Dzuba},\ and\ \citenamefont
  {Flambaum}}]{PhysRevLett.105.120801}%
  \BibitemOpen
  \bibfield  {author} {\bibinfo {author} {\bibfnamefont {J.~C.}\ \bibnamefont
  {Berengut}}, \bibinfo {author} {\bibfnamefont {V.~A.}\ \bibnamefont
  {Dzuba}},\ and\ \bibinfo {author} {\bibfnamefont {V.~V.}\ \bibnamefont
  {Flambaum}},\ }\href {https://doi.org/10.1103/PhysRevLett.105.120801}
  {\bibfield  {journal} {\bibinfo  {journal} {Phys. Rev. Lett.}\ }\textbf
  {\bibinfo {volume} {105}},\ \bibinfo {pages} {120801} (\bibinfo {year}
  {2010})}\BibitemShut {NoStop}%
\bibitem [{\citenamefont {Shabaev}\ \emph {et~al.}(2006)\citenamefont
  {Shabaev}, \citenamefont {Glazov}, \citenamefont {Oreshkina}, \citenamefont
  {Volotka}, \citenamefont {Plunien}, \citenamefont {Kluge},\ and\
  \citenamefont {Quint}}]{PhysRevLett.96.253002}%
  \BibitemOpen
  \bibfield  {author} {\bibinfo {author} {\bibfnamefont {V.~M.}\ \bibnamefont
  {Shabaev}}, \bibinfo {author} {\bibfnamefont {D.~A.}\ \bibnamefont {Glazov}},
  \bibinfo {author} {\bibfnamefont {N.~S.}\ \bibnamefont {Oreshkina}}, \bibinfo
  {author} {\bibfnamefont {A.~V.}\ \bibnamefont {Volotka}}, \bibinfo {author}
  {\bibfnamefont {G.}~\bibnamefont {Plunien}}, \bibinfo {author} {\bibfnamefont
  {H.-J.}\ \bibnamefont {Kluge}},\ and\ \bibinfo {author} {\bibfnamefont
  {W.}~\bibnamefont {Quint}},\ }\href
  {https://doi.org/10.1103/PhysRevLett.96.253002} {\bibfield  {journal}
  {\bibinfo  {journal} {Phys. Rev. Lett.}\ }\textbf {\bibinfo {volume} {96}},\
  \bibinfo {pages} {253002} (\bibinfo {year} {2006})}\BibitemShut {NoStop}%
\bibitem [{\citenamefont {Andreev}\ \emph {et~al.}(2005)\citenamefont
  {Andreev}, \citenamefont {Labzowsky}, \citenamefont {Plunien},\ and\
  \citenamefont {Soff}}]{PhysRevLett.94.243002}%
  \BibitemOpen
  \bibfield  {author} {\bibinfo {author} {\bibfnamefont {O.~Y.}\ \bibnamefont
  {Andreev}}, \bibinfo {author} {\bibfnamefont {L.~N.}\ \bibnamefont
  {Labzowsky}}, \bibinfo {author} {\bibfnamefont {G.}~\bibnamefont {Plunien}},\
  and\ \bibinfo {author} {\bibfnamefont {G.}~\bibnamefont {Soff}},\ }\href
  {https://doi.org/10.1103/PhysRevLett.94.243002} {\bibfield  {journal}
  {\bibinfo  {journal} {Phys. Rev. Lett.}\ }\textbf {\bibinfo {volume} {94}},\
  \bibinfo {pages} {243002} (\bibinfo {year} {2005})}\BibitemShut {NoStop}%
\bibitem [{\citenamefont {Beier}\ \emph {et~al.}(2001)\citenamefont {Beier},
  \citenamefont {H\"affner}, \citenamefont {Hermanspahn}, \citenamefont
  {Karshenboim}, \citenamefont {Kluge}, \citenamefont {Quint}, \citenamefont
  {Stahl}, \citenamefont {Verd\'u},\ and\ \citenamefont
  {Werth}}]{PhysRevLett.88.011603}%
  \BibitemOpen
  \bibfield  {author} {\bibinfo {author} {\bibfnamefont {T.}~\bibnamefont
  {Beier}}, \bibinfo {author} {\bibfnamefont {H.}~\bibnamefont {H\"affner}},
  \bibinfo {author} {\bibfnamefont {N.}~\bibnamefont {Hermanspahn}}, \bibinfo
  {author} {\bibfnamefont {S.~G.}\ \bibnamefont {Karshenboim}}, \bibinfo
  {author} {\bibfnamefont {H.-J.}\ \bibnamefont {Kluge}}, \bibinfo {author}
  {\bibfnamefont {W.}~\bibnamefont {Quint}}, \bibinfo {author} {\bibfnamefont
  {S.}~\bibnamefont {Stahl}}, \bibinfo {author} {\bibfnamefont
  {J.}~\bibnamefont {Verd\'u}},\ and\ \bibinfo {author} {\bibfnamefont
  {G.}~\bibnamefont {Werth}},\ }\href
  {https://doi.org/10.1103/PhysRevLett.88.011603} {\bibfield  {journal}
  {\bibinfo  {journal} {Phys. Rev. Lett.}\ }\textbf {\bibinfo {volume} {88}},\
  \bibinfo {pages} {011603} (\bibinfo {year} {2001})}\BibitemShut {NoStop}%
\bibitem [{\citenamefont {Oreshkina}\ \emph {et~al.}(2017)\citenamefont
  {Oreshkina}, \citenamefont {Cavaletto}, \citenamefont {Michel}, \citenamefont
  {Harman},\ and\ \citenamefont {Keitel}}]{PhysRevA.96.030501}%
  \BibitemOpen
  \bibfield  {author} {\bibinfo {author} {\bibfnamefont {N.~S.}\ \bibnamefont
  {Oreshkina}}, \bibinfo {author} {\bibfnamefont {S.~M.}\ \bibnamefont
  {Cavaletto}}, \bibinfo {author} {\bibfnamefont {N.}~\bibnamefont {Michel}},
  \bibinfo {author} {\bibfnamefont {Z.}~\bibnamefont {Harman}},\ and\ \bibinfo
  {author} {\bibfnamefont {C.~H.}\ \bibnamefont {Keitel}},\ }\href
  {https://doi.org/10.1103/PhysRevA.96.030501} {\bibfield  {journal} {\bibinfo
  {journal} {Phys. Rev. A}\ }\textbf {\bibinfo {volume} {96}},\ \bibinfo
  {pages} {030501} (\bibinfo {year} {2017})}\BibitemShut {NoStop}%
\bibitem [{\citenamefont {Sturm}\ \emph {et~al.}(2014)\citenamefont {Sturm},
  \citenamefont {K{\"o}hler}, \citenamefont {Zatorski}, \citenamefont {Wagner},
  \citenamefont {Harman}, \citenamefont {Werth}, \citenamefont {Quint},
  \citenamefont {Keitel},\ and\ \citenamefont {Blaum}}]{sturm2014high}%
  \BibitemOpen
  \bibfield  {author} {\bibinfo {author} {\bibfnamefont {S.}~\bibnamefont
  {Sturm}}, \bibinfo {author} {\bibfnamefont {F.}~\bibnamefont {K{\"o}hler}},
  \bibinfo {author} {\bibfnamefont {J.}~\bibnamefont {Zatorski}}, \bibinfo
  {author} {\bibfnamefont {A.}~\bibnamefont {Wagner}}, \bibinfo {author}
  {\bibfnamefont {Z.}~\bibnamefont {Harman}}, \bibinfo {author} {\bibfnamefont
  {G.}~\bibnamefont {Werth}}, \bibinfo {author} {\bibfnamefont
  {W.}~\bibnamefont {Quint}}, \bibinfo {author} {\bibfnamefont {C.~H.}\
  \bibnamefont {Keitel}},\ and\ \bibinfo {author} {\bibfnamefont
  {K.}~\bibnamefont {Blaum}},\ }\href {https://doi.org/10.1038/nature13026}
  {\bibfield  {journal} {\bibinfo  {journal} {Nature}\ }\textbf {\bibinfo
  {volume} {506}},\ \bibinfo {pages} {467} (\bibinfo {year}
  {2014})}\BibitemShut {NoStop}%
\bibitem [{\citenamefont {Debierre}\ \emph {et~al.}(2020)\citenamefont
  {Debierre}, \citenamefont {Keitel},\ and\ \citenamefont
  {Harman}}]{DEBIERRE2020135527}%
  \BibitemOpen
  \bibfield  {author} {\bibinfo {author} {\bibfnamefont {V.}~\bibnamefont
  {Debierre}}, \bibinfo {author} {\bibfnamefont {C.}~\bibnamefont {Keitel}},\
  and\ \bibinfo {author} {\bibfnamefont {Z.}~\bibnamefont {Harman}},\ }\href
  {https://doi.org/https://doi.org/10.1016/j.physletb.2020.135527} {\bibfield
  {journal} {\bibinfo  {journal} {Phys. Lett. B}\ }\textbf {\bibinfo {volume}
  {807}},\ \bibinfo {pages} {135527} (\bibinfo {year} {2020})}\BibitemShut
  {NoStop}%
\bibitem [{\citenamefont {Safronova}\ \emph {et~al.}(2018)\citenamefont
  {Safronova}, \citenamefont {Budker}, \citenamefont {DeMille}, \citenamefont
  {Kimball}, \citenamefont {Derevianko},\ and\ \citenamefont
  {Clark}}]{RevModPhys.90.025008}%
  \BibitemOpen
  \bibfield  {author} {\bibinfo {author} {\bibfnamefont {M.~S.}\ \bibnamefont
  {Safronova}}, \bibinfo {author} {\bibfnamefont {D.}~\bibnamefont {Budker}},
  \bibinfo {author} {\bibfnamefont {D.}~\bibnamefont {DeMille}}, \bibinfo
  {author} {\bibfnamefont {D.~F.~J.}\ \bibnamefont {Kimball}}, \bibinfo
  {author} {\bibfnamefont {A.}~\bibnamefont {Derevianko}},\ and\ \bibinfo
  {author} {\bibfnamefont {C.~W.}\ \bibnamefont {Clark}},\ }\href
  {https://doi.org/10.1103/RevModPhys.90.025008} {\bibfield  {journal}
  {\bibinfo  {journal} {Rev. Mod. Phys.}\ }\textbf {\bibinfo {volume} {90}},\
  \bibinfo {pages} {025008} (\bibinfo {year} {2018})}\BibitemShut {NoStop}%
\bibitem [{\citenamefont {Shabaev}(1993)}]{Shabaev_1993}%
  \BibitemOpen
  \bibfield  {author} {\bibinfo {author} {\bibfnamefont {V.~M.}\ \bibnamefont
  {Shabaev}},\ }\href {https://doi.org/10.1088/0953-4075/26/6/011} {\bibfield
  {journal} {\bibinfo  {journal} {J. Phys. B: At. Mol. Opt. Phys.}\ }\textbf
  {\bibinfo {volume} {26}},\ \bibinfo {pages} {1103} (\bibinfo {year}
  {1993})}\BibitemShut {NoStop}%
\bibitem [{\citenamefont {Deck}\ \emph {et~al.}(2005)\citenamefont {Deck},
  \citenamefont {Amar},\ and\ \citenamefont {Fralick}}]{Deck_2005}%
  \BibitemOpen
  \bibfield  {author} {\bibinfo {author} {\bibfnamefont {R.~T.}\ \bibnamefont
  {Deck}}, \bibinfo {author} {\bibfnamefont {J.~G.}\ \bibnamefont {Amar}},\
  and\ \bibinfo {author} {\bibfnamefont {G.}~\bibnamefont {Fralick}},\ }\href
  {https://doi.org/10.1088/0953-4075/38/13/010} {\bibfield  {journal} {\bibinfo
   {journal} {J. Phys. B: At. Mol. Opt. Phys.}\ }\textbf {\bibinfo {volume}
  {38}},\ \bibinfo {pages} {2173} (\bibinfo {year} {2005})}\BibitemShut
  {NoStop}%
\bibitem [{\citenamefont {Karshenboim}\ \emph {et~al.}(2005)\citenamefont
  {Karshenboim}, \citenamefont {Lee},\ and\ \citenamefont
  {Milstein}}]{PhysRevA.72.042101}%
  \BibitemOpen
  \bibfield  {author} {\bibinfo {author} {\bibfnamefont {S.~G.}\ \bibnamefont
  {Karshenboim}}, \bibinfo {author} {\bibfnamefont {R.~N.}\ \bibnamefont
  {Lee}},\ and\ \bibinfo {author} {\bibfnamefont {A.~I.}\ \bibnamefont
  {Milstein}},\ }\href {https://doi.org/10.1103/PhysRevA.72.042101} {\bibfield
  {journal} {\bibinfo  {journal} {Phys. Rev. A}\ }\textbf {\bibinfo {volume}
  {72}},\ \bibinfo {pages} {042101} (\bibinfo {year} {2005})}\BibitemShut
  {NoStop}%
\bibitem [{\citenamefont {Karshenboim}\ \emph {et~al.}(2019)\citenamefont
  {Karshenboim}, \citenamefont {Korzinin}, \citenamefont {Shelyuto},\ and\
  \citenamefont {Ivanov}}]{PhysRevA.99.032508}%
  \BibitemOpen
  \bibfield  {author} {\bibinfo {author} {\bibfnamefont {S.~G.}\ \bibnamefont
  {Karshenboim}}, \bibinfo {author} {\bibfnamefont {E.~Y.}\ \bibnamefont
  {Korzinin}}, \bibinfo {author} {\bibfnamefont {V.~A.}\ \bibnamefont
  {Shelyuto}},\ and\ \bibinfo {author} {\bibfnamefont {V.~G.}\ \bibnamefont
  {Ivanov}},\ }\href {https://doi.org/10.1103/PhysRevA.99.032508} {\bibfield
  {journal} {\bibinfo  {journal} {Phys. Rev. A}\ }\textbf {\bibinfo {volume}
  {99}},\ \bibinfo {pages} {032508} (\bibinfo {year} {2019})}\BibitemShut
  {NoStop}%
\bibitem [{\citenamefont {Babak Nadiri~Niri}(2018)}]{babak2018}%
  \BibitemOpen
  \bibfield  {author} {\bibinfo {author} {\bibfnamefont {A.~A.}\ \bibnamefont
  {Babak Nadiri~Niri}},\ }\href
  {https://doi.org/https://doi.org/10.11648/j.ns.20180301.11} {\bibfield
  {journal} {\bibinfo  {journal} {Nucl. Sci.}\ }\textbf {\bibinfo {volume}
  {3}},\ \bibinfo {pages} {1} (\bibinfo {year} {2018})}\BibitemShut {NoStop}%
\bibitem [{\citenamefont {Borisoglebsky}\ and\ \citenamefont
  {Trofimenko}(1979)}]{BORISOGLEBSKY1979175}%
  \BibitemOpen
  \bibfield  {author} {\bibinfo {author} {\bibfnamefont {L.}~\bibnamefont
  {Borisoglebsky}}\ and\ \bibinfo {author} {\bibfnamefont {E.}~\bibnamefont
  {Trofimenko}},\ }\href
  {https://doi.org/https://doi.org/10.1016/0370-2693(79)90516-1} {\bibfield
  {journal} {\bibinfo  {journal} {Phys. Lett. B}\ }\textbf {\bibinfo {volume}
  {81}},\ \bibinfo {pages} {175} (\bibinfo {year} {1979})}\BibitemShut
  {NoStop}%
\bibitem [{\citenamefont {Valuev}\ \emph {et~al.}(2020)\citenamefont {Valuev},
  \citenamefont {Harman}, \citenamefont {Keitel},\ and\ \citenamefont
  {Oreshkina}}]{PRA101}%
  \BibitemOpen
  \bibfield  {author} {\bibinfo {author} {\bibfnamefont {I.~A.}\ \bibnamefont
  {Valuev}}, \bibinfo {author} {\bibfnamefont {Z.}~\bibnamefont {Harman}},
  \bibinfo {author} {\bibfnamefont {C.~H.}\ \bibnamefont {Keitel}},\ and\
  \bibinfo {author} {\bibfnamefont {N.~S.}\ \bibnamefont {Oreshkina}},\ }\href
  {https://doi.org/10.1103/PhysRevA.101.062502} {\bibfield  {journal} {\bibinfo
   {journal} {Phys. Rev. A}\ }\textbf {\bibinfo {volume} {101}},\ \bibinfo
  {pages} {062502} (\bibinfo {year} {2020})}\BibitemShut {NoStop}%
\bibitem [{\citenamefont {Visscher}\ and\ \citenamefont
  {Dyall}(1997)}]{VISSCHER1997207}%
  \BibitemOpen
  \bibfield  {author} {\bibinfo {author} {\bibfnamefont {L.}~\bibnamefont
  {Visscher}}\ and\ \bibinfo {author} {\bibfnamefont {K.}~\bibnamefont
  {Dyall}},\ }\href {https://doi.org/https://doi.org/10.1006/adnd.1997.0751}
  {\bibfield  {journal} {\bibinfo  {journal} {At. Data. Nucl. Data Tables}\
  }\textbf {\bibinfo {volume} {67}},\ \bibinfo {pages} {207} (\bibinfo {year}
  {1997})}\BibitemShut {NoStop}%
\bibitem [{\citenamefont {Lee}\ \emph {et~al.}(2007)\citenamefont {Lee},
  \citenamefont {Milstein}, \citenamefont {Terekhov},\ and\ \citenamefont
  {Karshenboim}}]{lee2007g}%
  \BibitemOpen
  \bibfield  {author} {\bibinfo {author} {\bibfnamefont {R.~N.}\ \bibnamefont
  {Lee}}, \bibinfo {author} {\bibfnamefont {A.~I.}\ \bibnamefont {Milstein}},
  \bibinfo {author} {\bibfnamefont {I.~S.}\ \bibnamefont {Terekhov}},\ and\
  \bibinfo {author} {\bibfnamefont {S.~G.}\ \bibnamefont {Karshenboim}},\
  }\href {https://doi.org/10.1139/p07-024} {\bibfield  {journal} {\bibinfo
  {journal} {Can. J. Phys.}\ }\textbf {\bibinfo {volume} {85}},\ \bibinfo
  {pages} {541} (\bibinfo {year} {2007})}\BibitemShut {NoStop}%
\bibitem [{\citenamefont {Karplus}\ \emph {et~al.}(1952)\citenamefont
  {Karplus}, \citenamefont {Klein},\ and\ \citenamefont
  {Schwinger}}]{PhysRev.86.288}%
  \BibitemOpen
  \bibfield  {author} {\bibinfo {author} {\bibfnamefont {R.}~\bibnamefont
  {Karplus}}, \bibinfo {author} {\bibfnamefont {A.}~\bibnamefont {Klein}},\
  and\ \bibinfo {author} {\bibfnamefont {J.}~\bibnamefont {Schwinger}},\ }\href
  {https://doi.org/10.1103/PhysRev.86.288} {\bibfield  {journal} {\bibinfo
  {journal} {Phys. Rev.}\ }\textbf {\bibinfo {volume} {86}},\ \bibinfo {pages}
  {288} (\bibinfo {year} {1952})}\BibitemShut {NoStop}%
\bibitem [{\citenamefont {Erickson}\ and\ \citenamefont
  {Yennie}(1965)}]{ERICKSON1965271}%
  \BibitemOpen
  \bibfield  {author} {\bibinfo {author} {\bibfnamefont {G.~W.}\ \bibnamefont
  {Erickson}}\ and\ \bibinfo {author} {\bibfnamefont {D.~R.}\ \bibnamefont
  {Yennie}},\ }\href
  {https://doi.org/https://doi.org/10.1016/0003-4916(65)90081-3} {\bibfield
  {journal} {\bibinfo  {journal} {Ann. Phys. (NY)}\ }\textbf {\bibinfo {volume}
  {35}},\ \bibinfo {pages} {271} (\bibinfo {year} {1965})}\BibitemShut
  {NoStop}%
\bibitem [{\citenamefont {Friar}(1979)}]{FRIAR1979151}%
  \BibitemOpen
  \bibfield  {author} {\bibinfo {author} {\bibfnamefont {J.~L.}\ \bibnamefont
  {Friar}},\ }\href
  {https://doi.org/https://doi.org/10.1016/0003-4916(79)90300-2} {\bibfield
  {journal} {\bibinfo  {journal} {Ann. Phys. (NY)}\ }\textbf {\bibinfo {volume}
  {122}},\ \bibinfo {pages} {151} (\bibinfo {year} {1979})}\BibitemShut
  {NoStop}%
\bibitem [{\citenamefont {Ring}(1996)}]{RING1996193}%
  \BibitemOpen
  \bibfield  {author} {\bibinfo {author} {\bibfnamefont {P.}~\bibnamefont
  {Ring}},\ }\href
  {https://doi.org/https://doi.org/10.1016/0146-6410(96)00054-3} {\bibfield
  {journal} {\bibinfo  {journal} {Prog. Part. Nucl. Phys.}\ }\textbf {\bibinfo
  {volume} {37}},\ \bibinfo {pages} {193} (\bibinfo {year} {1996})}\BibitemShut
  {NoStop}%
\bibitem [{\citenamefont {Meng}\ \emph {et~al.}(2006)\citenamefont {Meng},
  \citenamefont {Toki}, \citenamefont {Zhou}, \citenamefont {Zhang},
  \citenamefont {Long},\ and\ \citenamefont {Geng}}]{MENG2006470}%
  \BibitemOpen
  \bibfield  {author} {\bibinfo {author} {\bibfnamefont {J.}~\bibnamefont
  {Meng}}, \bibinfo {author} {\bibfnamefont {H.}~\bibnamefont {Toki}}, \bibinfo
  {author} {\bibfnamefont {S.~G.}\ \bibnamefont {Zhou}}, \bibinfo {author}
  {\bibfnamefont {S.~Q.}\ \bibnamefont {Zhang}}, \bibinfo {author}
  {\bibfnamefont {W.~H.}\ \bibnamefont {Long}},\ and\ \bibinfo {author}
  {\bibfnamefont {L.~S.}\ \bibnamefont {Geng}},\ }\href
  {https://doi.org/https://doi.org/10.1016/j.ppnp.2005.06.001} {\bibfield
  {journal} {\bibinfo  {journal} {Prog. Part. Nucl. Phys.}\ }\textbf {\bibinfo
  {volume} {57}},\ \bibinfo {pages} {470} (\bibinfo {year} {2006})}\BibitemShut
  {NoStop}%
\bibitem [{\citenamefont {Meng}(2015)}]{Meng2015}%
  \BibitemOpen
  \bibfield  {author} {\bibinfo {author} {\bibfnamefont {J.}~\bibnamefont
  {Meng}},\ }\href {https://doi.org/10.1142/9872} {\emph {\bibinfo {title}
  {Relativistic Density Functional for Nuclear Structure}}}\ (\bibinfo
  {publisher} {World Scientific, Singapore},\ \bibinfo {year}
  {2015})\BibitemShut {NoStop}%
\bibitem [{\citenamefont {Vretenar}\ \emph {et~al.}(2005)\citenamefont
  {Vretenar}, \citenamefont {Afanasjev}, \citenamefont {Lalazissis},\ and\
  \citenamefont {Ring}}]{VRETENAR2005101}%
  \BibitemOpen
  \bibfield  {author} {\bibinfo {author} {\bibfnamefont {D.}~\bibnamefont
  {Vretenar}}, \bibinfo {author} {\bibfnamefont {A.}~\bibnamefont {Afanasjev}},
  \bibinfo {author} {\bibfnamefont {G.}~\bibnamefont {Lalazissis}},\ and\
  \bibinfo {author} {\bibfnamefont {P.}~\bibnamefont {Ring}},\ }\href
  {https://doi.org/https://doi.org/10.1016/j.physrep.2004.10.001} {\bibfield
  {journal} {\bibinfo  {journal} {Phys. Rep.}\ }\textbf {\bibinfo {volume}
  {409}},\ \bibinfo {pages} {101} (\bibinfo {year} {2005})}\BibitemShut
  {NoStop}%
\bibitem [{\citenamefont {Nikšić}\ \emph {et~al.}(2011)\citenamefont
  {Nikšić}, \citenamefont {Vretenar},\ and\ \citenamefont
  {Ring}}]{NIKSIC2011519}%
  \BibitemOpen
  \bibfield  {author} {\bibinfo {author} {\bibfnamefont {T.}~\bibnamefont
  {Nikšić}}, \bibinfo {author} {\bibfnamefont {D.}~\bibnamefont {Vretenar}},\
  and\ \bibinfo {author} {\bibfnamefont {P.}~\bibnamefont {Ring}},\ }\href
  {https://doi.org/https://doi.org/10.1016/j.ppnp.2011.01.055} {\bibfield
  {journal} {\bibinfo  {journal} {Prog. Part. Nucl. Phys.}\ }\textbf {\bibinfo
  {volume} {66}},\ \bibinfo {pages} {519} (\bibinfo {year} {2011})}\BibitemShut
  {NoStop}%
\bibitem [{\citenamefont {Meng}\ \emph {et~al.}(2013)\citenamefont {Meng},
  \citenamefont {Peng}, \citenamefont {Zhang},\ and\ \citenamefont
  {Zhao}}]{meng2013progress}%
  \BibitemOpen
  \bibfield  {author} {\bibinfo {author} {\bibfnamefont {J.}~\bibnamefont
  {Meng}}, \bibinfo {author} {\bibfnamefont {J.}~\bibnamefont {Peng}}, \bibinfo
  {author} {\bibfnamefont {S.-Q.}\ \bibnamefont {Zhang}},\ and\ \bibinfo
  {author} {\bibfnamefont {P.-W.}\ \bibnamefont {Zhao}},\ }\href
  {https://doi.org/10.1007/s11467-013-0287-y} {\bibfield  {journal} {\bibinfo
  {journal} {Front. Phys.}\ }\textbf {\bibinfo {volume} {8}},\ \bibinfo {pages}
  {55} (\bibinfo {year} {2013})}\BibitemShut {NoStop}%
\bibitem [{\citenamefont {Sharma}\ \emph {et~al.}(1993)\citenamefont {Sharma},
  \citenamefont {Lalazissis},\ and\ \citenamefont {Ring}}]{SHARMA19939}%
  \BibitemOpen
  \bibfield  {author} {\bibinfo {author} {\bibfnamefont {M.~M.}\ \bibnamefont
  {Sharma}}, \bibinfo {author} {\bibfnamefont {G.~A.}\ \bibnamefont
  {Lalazissis}},\ and\ \bibinfo {author} {\bibfnamefont {P.}~\bibnamefont
  {Ring}},\ }\href
  {https://doi.org/https://doi.org/10.1016/0370-2693(93)91561-Z} {\bibfield
  {journal} {\bibinfo  {journal} {Phys. Lett. B}\ }\textbf {\bibinfo {volume}
  {317}},\ \bibinfo {pages} {9} (\bibinfo {year} {1993})}\BibitemShut {NoStop}%
\bibitem [{\citenamefont {Zhou}\ \emph {et~al.}(2003)\citenamefont {Zhou},
  \citenamefont {Meng},\ and\ \citenamefont {Ring}}]{PhysRevLett.91.262501}%
  \BibitemOpen
  \bibfield  {author} {\bibinfo {author} {\bibfnamefont {S.-G.}\ \bibnamefont
  {Zhou}}, \bibinfo {author} {\bibfnamefont {J.}~\bibnamefont {Meng}},\ and\
  \bibinfo {author} {\bibfnamefont {P.}~\bibnamefont {Ring}},\ }\href
  {https://doi.org/10.1103/PhysRevLett.91.262501} {\bibfield  {journal}
  {\bibinfo  {journal} {Phys. Rev. Lett.}\ }\textbf {\bibinfo {volume} {91}},\
  \bibinfo {pages} {262501} (\bibinfo {year} {2003})}\BibitemShut {NoStop}%
\bibitem [{\citenamefont {Liang}\ \emph {et~al.}(2010)\citenamefont {Liang},
  \citenamefont {Long}, \citenamefont {Meng},\ and\ \citenamefont
  {Van~Giai}}]{liang2010spin}%
  \BibitemOpen
  \bibfield  {author} {\bibinfo {author} {\bibfnamefont {H.}~\bibnamefont
  {Liang}}, \bibinfo {author} {\bibfnamefont {W.~H.}\ \bibnamefont {Long}},
  \bibinfo {author} {\bibfnamefont {J.}~\bibnamefont {Meng}},\ and\ \bibinfo
  {author} {\bibfnamefont {N.}~\bibnamefont {Van~Giai}},\ }\href
  {https://doi.org/10.1140/epja/i2010-10938-6} {\bibfield  {journal} {\bibinfo
  {journal} {Eur. Phys. J. A}\ }\textbf {\bibinfo {volume} {44}},\ \bibinfo
  {pages} {119} (\bibinfo {year} {2010})}\BibitemShut {NoStop}%
\bibitem [{\citenamefont {Liang}\ \emph {et~al.}(2015)\citenamefont {Liang},
  \citenamefont {Meng},\ and\ \citenamefont {Zhou}}]{LIANG20151}%
  \BibitemOpen
  \bibfield  {author} {\bibinfo {author} {\bibfnamefont {H.}~\bibnamefont
  {Liang}}, \bibinfo {author} {\bibfnamefont {J.}~\bibnamefont {Meng}},\ and\
  \bibinfo {author} {\bibfnamefont {S.-G.}\ \bibnamefont {Zhou}},\ }\href
  {https://doi.org/https://doi.org/10.1016/j.physrep.2014.12.005} {\bibfield
  {journal} {\bibinfo  {journal} {Phys. Rep.}\ }\textbf {\bibinfo {volume}
  {570}},\ \bibinfo {pages} {1} (\bibinfo {year} {2015})}\BibitemShut {NoStop}%
\bibitem [{\citenamefont {Li}\ \emph {et~al.}(2013)\citenamefont {Li},
  \citenamefont {Wei}, \citenamefont {Hu}, \citenamefont {Ring},\ and\
  \citenamefont {Meng}}]{lijian1}%
  \BibitemOpen
  \bibfield  {author} {\bibinfo {author} {\bibfnamefont {J.}~\bibnamefont
  {Li}}, \bibinfo {author} {\bibfnamefont {J.~X.}\ \bibnamefont {Wei}},
  \bibinfo {author} {\bibfnamefont {J.~N.}\ \bibnamefont {Hu}}, \bibinfo
  {author} {\bibfnamefont {P.}~\bibnamefont {Ring}},\ and\ \bibinfo {author}
  {\bibfnamefont {J.}~\bibnamefont {Meng}},\ }\href@noop {} {\bibfield
  {journal} {\bibinfo  {journal} {Phys. Rev. C}\ }\textbf {\bibinfo {volume}
  {88}},\ \bibinfo {pages} {064307} (\bibinfo {year} {2013})}\BibitemShut
  {NoStop}%
\bibitem [{\citenamefont {Li}\ and\ \citenamefont {Meng}(2018)}]{lijian2}%
  \BibitemOpen
  \bibfield  {author} {\bibinfo {author} {\bibfnamefont {J.}~\bibnamefont
  {Li}}\ and\ \bibinfo {author} {\bibfnamefont {J.}~\bibnamefont {Meng}},\
  }\href@noop {} {\bibfield  {journal} {\bibinfo  {journal} {Front. Phys.}\
  }\textbf {\bibinfo {volume} {13}},\ \bibinfo {pages} {132109} (\bibinfo
  {year} {2018})}\BibitemShut {NoStop}%
\bibitem [{\citenamefont {Meng}\ and\ \citenamefont
  {Ring}(1996)}]{PhysRevLett.77.3963}%
  \BibitemOpen
  \bibfield  {author} {\bibinfo {author} {\bibfnamefont {J.}~\bibnamefont
  {Meng}}\ and\ \bibinfo {author} {\bibfnamefont {P.}~\bibnamefont {Ring}},\
  }\href {https://doi.org/10.1103/PhysRevLett.77.3963} {\bibfield  {journal}
  {\bibinfo  {journal} {Phys. Rev. Lett.}\ }\textbf {\bibinfo {volume} {77}},\
  \bibinfo {pages} {3963} (\bibinfo {year} {1996})}\BibitemShut {NoStop}%
\bibitem [{\citenamefont {Meng}(1998)}]{Meng1998NPA}%
  \BibitemOpen
  \bibfield  {author} {\bibinfo {author} {\bibfnamefont {J.}~\bibnamefont
  {Meng}},\ }\href
  {https://doi.org/https://doi.org/10.1016/S0375-9474(98)00178-X} {\bibfield
  {journal} {\bibinfo  {journal} {Nucl. Phys. A}\ }\textbf {\bibinfo {volume}
  {635}},\ \bibinfo {pages} {3} (\bibinfo {year} {1998})}\BibitemShut {NoStop}%
\bibitem [{\citenamefont {Meng}\ and\ \citenamefont
  {Ring}(1998)}]{PhysRevLett.80.460}%
  \BibitemOpen
  \bibfield  {author} {\bibinfo {author} {\bibfnamefont {J.}~\bibnamefont
  {Meng}}\ and\ \bibinfo {author} {\bibfnamefont {P.}~\bibnamefont {Ring}},\
  }\href {https://doi.org/10.1103/PhysRevLett.80.460} {\bibfield  {journal}
  {\bibinfo  {journal} {Phys. Rev. Lett.}\ }\textbf {\bibinfo {volume} {80}},\
  \bibinfo {pages} {460} (\bibinfo {year} {1998})}\BibitemShut {NoStop}%
\bibitem [{\citenamefont {Meng}\ \emph
  {et~al.}(2002{\natexlab{a}})\citenamefont {Meng}, \citenamefont {Toki},
  \citenamefont {Zeng}, \citenamefont {Zhang},\ and\ \citenamefont
  {Zhou}}]{PhysRevC.65.041302}%
  \BibitemOpen
  \bibfield  {author} {\bibinfo {author} {\bibfnamefont {J.}~\bibnamefont
  {Meng}}, \bibinfo {author} {\bibfnamefont {H.}~\bibnamefont {Toki}}, \bibinfo
  {author} {\bibfnamefont {J.~Y.}\ \bibnamefont {Zeng}}, \bibinfo {author}
  {\bibfnamefont {S.~Q.}\ \bibnamefont {Zhang}},\ and\ \bibinfo {author}
  {\bibfnamefont {S.-G.}\ \bibnamefont {Zhou}},\ }\href
  {https://doi.org/10.1103/PhysRevC.65.041302} {\bibfield  {journal} {\bibinfo
  {journal} {Phys. Rev. C}\ }\textbf {\bibinfo {volume} {65}},\ \bibinfo
  {pages} {041302} (\bibinfo {year} {2002}{\natexlab{a}})}\BibitemShut
  {NoStop}%
\bibitem [{\citenamefont {Zhang}\ \emph {et~al.}(2002)\citenamefont {Zhang},
  \citenamefont {Meng}, \citenamefont {Zhou},\ and\ \citenamefont
  {Zeng}}]{Shuang_Quan_2002}%
  \BibitemOpen
  \bibfield  {author} {\bibinfo {author} {\bibfnamefont {S.-Q.}\ \bibnamefont
  {Zhang}}, \bibinfo {author} {\bibfnamefont {J.}~\bibnamefont {Meng}},
  \bibinfo {author} {\bibfnamefont {S.-G.}\ \bibnamefont {Zhou}},\ and\
  \bibinfo {author} {\bibfnamefont {J.-Y.}\ \bibnamefont {Zeng}},\ }\href
  {https://doi.org/10.1088/0256-307x/19/3/308} {\bibfield  {journal} {\bibinfo
  {journal} {Chin. Phys. Lett.}\ }\textbf {\bibinfo {volume} {19}},\ \bibinfo
  {pages} {312} (\bibinfo {year} {2002})}\BibitemShut {NoStop}%
\bibitem [{\citenamefont {Meng}\ and\ \citenamefont {Zhou}(2015)}]{Meng_2015}%
  \BibitemOpen
  \bibfield  {author} {\bibinfo {author} {\bibfnamefont {J.}~\bibnamefont
  {Meng}}\ and\ \bibinfo {author} {\bibfnamefont {S.~G.}\ \bibnamefont
  {Zhou}},\ }\href {https://doi.org/10.1088/0954-3899/42/9/093101} {\bibfield
  {journal} {\bibinfo  {journal} {J. Phys. G: Nucl. Part. Phys.}\ }\textbf
  {\bibinfo {volume} {42}},\ \bibinfo {pages} {093101} (\bibinfo {year}
  {2015})}\BibitemShut {NoStop}%
\bibitem [{\citenamefont {Meng}\ \emph {et~al.}(1998)\citenamefont {Meng},
  \citenamefont {Tanihata},\ and\ \citenamefont {Yamaji}}]{MENG19981}%
  \BibitemOpen
  \bibfield  {author} {\bibinfo {author} {\bibfnamefont {J.}~\bibnamefont
  {Meng}}, \bibinfo {author} {\bibfnamefont {I.}~\bibnamefont {Tanihata}},\
  and\ \bibinfo {author} {\bibfnamefont {S.}~\bibnamefont {Yamaji}},\ }\href
  {https://doi.org/https://doi.org/10.1016/S0370-2693(97)01386-5} {\bibfield
  {journal} {\bibinfo  {journal} {Phys. Lett. B}\ }\textbf {\bibinfo {volume}
  {419}},\ \bibinfo {pages} {1} (\bibinfo {year} {1998})}\BibitemShut {NoStop}%
\bibitem [{\citenamefont {Meng}\ \emph
  {et~al.}(2002{\natexlab{b}})\citenamefont {Meng}, \citenamefont {Zhou},\ and\
  \citenamefont {Tanihata}}]{MENG2002209}%
  \BibitemOpen
  \bibfield  {author} {\bibinfo {author} {\bibfnamefont {J.}~\bibnamefont
  {Meng}}, \bibinfo {author} {\bibfnamefont {S.-G.}\ \bibnamefont {Zhou}},\
  and\ \bibinfo {author} {\bibfnamefont {I.}~\bibnamefont {Tanihata}},\ }\href
  {https://doi.org/https://doi.org/10.1016/S0370-2693(02)01574-5} {\bibfield
  {journal} {\bibinfo  {journal} {Phys. Lett. B}\ }\textbf {\bibinfo {volume}
  {532}},\ \bibinfo {pages} {209} (\bibinfo {year}
  {2002}{\natexlab{b}})}\BibitemShut {NoStop}%
\bibitem [{\citenamefont {Zhao}\ \emph {et~al.}(2010)\citenamefont {Zhao},
  \citenamefont {Li}, \citenamefont {Yao},\ and\ \citenamefont
  {Meng}}]{PhysRevC.82.054319}%
  \BibitemOpen
  \bibfield  {author} {\bibinfo {author} {\bibfnamefont {P.~W.}\ \bibnamefont
  {Zhao}}, \bibinfo {author} {\bibfnamefont {Z.~P.}\ \bibnamefont {Li}},
  \bibinfo {author} {\bibfnamefont {J.~M.}\ \bibnamefont {Yao}},\ and\ \bibinfo
  {author} {\bibfnamefont {J.}~\bibnamefont {Meng}},\ }\href
  {https://doi.org/10.1103/PhysRevC.82.054319} {\bibfield  {journal} {\bibinfo
  {journal} {Phys. Rev. C}\ }\textbf {\bibinfo {volume} {82}},\ \bibinfo
  {pages} {054319} (\bibinfo {year} {2010})}\BibitemShut {NoStop}%
\bibitem [{\citenamefont {Ring}\ and\ \citenamefont {Schuck}(1980)}]{Ring1980}%
  \BibitemOpen
  \bibfield  {author} {\bibinfo {author} {\bibfnamefont {P.}~\bibnamefont
  {Ring}}\ and\ \bibinfo {author} {\bibfnamefont {P.}~\bibnamefont {Schuck}},\
  }\href@noop {} {\emph {\bibinfo {title} {The Nuclear Many-Body Problem}}}\
  (\bibinfo  {publisher} {Springer-Varlag},\ \bibinfo {year}
  {1980})\BibitemShut {NoStop}%
\bibitem [{\citenamefont {Xia}\ \emph {et~al.}(2018)\citenamefont {Xia},
  \citenamefont {Lim}, \citenamefont {Zhao}, \citenamefont {Liang},
  \citenamefont {Qu}, \citenamefont {Chen}, \citenamefont {Liu}, \citenamefont
  {Zhang}, \citenamefont {Zhang}, \citenamefont {Kim},\ and\ \citenamefont
  {Meng}}]{Xia2018}%
  \BibitemOpen
  \bibfield  {author} {\bibinfo {author} {\bibfnamefont {X.~W.}\ \bibnamefont
  {Xia}}, \bibinfo {author} {\bibfnamefont {Y.}~\bibnamefont {Lim}}, \bibinfo
  {author} {\bibfnamefont {P.~W.}\ \bibnamefont {Zhao}}, \bibinfo {author}
  {\bibfnamefont {H.~Z.}\ \bibnamefont {Liang}}, \bibinfo {author}
  {\bibfnamefont {X.~Y.}\ \bibnamefont {Qu}}, \bibinfo {author} {\bibfnamefont
  {Y.}~\bibnamefont {Chen}}, \bibinfo {author} {\bibfnamefont {H.}~\bibnamefont
  {Liu}}, \bibinfo {author} {\bibfnamefont {L.~F.}\ \bibnamefont {Zhang}},
  \bibinfo {author} {\bibfnamefont {S.~Q.}\ \bibnamefont {Zhang}}, \bibinfo
  {author} {\bibfnamefont {Y.}~\bibnamefont {Kim}},\ and\ \bibinfo {author}
  {\bibfnamefont {J.}~\bibnamefont {Meng}},\ }\href
  {https://doi.org/10.1016/j.adt.2017.09.001} {\bibfield  {journal} {\bibinfo
  {journal} {Atom Data Nucl Data}\ }\textbf {\bibinfo {volume} {121-122}},\
  \bibinfo {pages} {1} (\bibinfo {year} {2018})}\BibitemShut {NoStop}%
\bibitem [{\citenamefont {Friar}\ and\ \citenamefont
  {Negele}(1975)}]{Friar1975}%
  \BibitemOpen
  \bibfield  {author} {\bibinfo {author} {\bibfnamefont {J.~L.}\ \bibnamefont
  {Friar}}\ and\ \bibinfo {author} {\bibfnamefont {J.~W.}\ \bibnamefont
  {Negele}},\ }\href {https://doi.org/10.1007/978-1-4757-4398-2_3} {\bibfield
  {journal} {\bibinfo  {journal} {Adv. Nucl. Phys.}\ }\textbf {\bibinfo
  {volume} {8}},\ \bibinfo {pages} {219} (\bibinfo {year} {1975})}\BibitemShut
  {NoStop}%
\bibitem [{\citenamefont {Kurasawa}\ and\ \citenamefont
  {Suzuki}(2000)}]{PhysRevC.62.054303}%
  \BibitemOpen
  \bibfield  {author} {\bibinfo {author} {\bibfnamefont {H.}~\bibnamefont
  {Kurasawa}}\ and\ \bibinfo {author} {\bibfnamefont {T.}~\bibnamefont
  {Suzuki}},\ }\href {https://doi.org/10.1103/PhysRevC.62.054303} {\bibfield
  {journal} {\bibinfo  {journal} {Phys. Rev. C}\ }\textbf {\bibinfo {volume}
  {62}},\ \bibinfo {pages} {054303} (\bibinfo {year} {2000})}\BibitemShut
  {NoStop}%
\bibitem [{\citenamefont {Reinhard}\ and\ \citenamefont
  {Nazarewicz}(2021)}]{PhysRevC.103.054310}%
  \BibitemOpen
  \bibfield  {author} {\bibinfo {author} {\bibfnamefont {P.-G.}\ \bibnamefont
  {Reinhard}}\ and\ \bibinfo {author} {\bibfnamefont {W.}~\bibnamefont
  {Nazarewicz}},\ }\href {https://doi.org/10.1103/PhysRevC.103.054310}
  {\bibfield  {journal} {\bibinfo  {journal} {Phys. Rev. C}\ }\textbf {\bibinfo
  {volume} {103}},\ \bibinfo {pages} {054310} (\bibinfo {year}
  {2021})}\BibitemShut {NoStop}%
\bibitem [{\citenamefont {Kurasawa}\ and\ \citenamefont
  {Suzuki}(2019)}]{kurasawa2019n}%
  \BibitemOpen
  \bibfield  {author} {\bibinfo {author} {\bibfnamefont {H.}~\bibnamefont
  {Kurasawa}}\ and\ \bibinfo {author} {\bibfnamefont {T.}~\bibnamefont
  {Suzuki}},\ }\href {https://doi.org/10.1093/ptep/ptz121} {\bibfield
  {journal} {\bibinfo  {journal} {Prog. Theor. Exp. Phys.}\ }\textbf {\bibinfo
  {volume} {2019}},\ \bibinfo {pages} {113D01} (\bibinfo {year}
  {2019})}\BibitemShut {NoStop}%
\bibitem [{\citenamefont {Tiesinga}\ \emph {et~al.}(2021)\citenamefont
  {Tiesinga}, \citenamefont {Mohr}, \citenamefont {Newell},\ and\ \citenamefont
  {Taylor}}]{RevModPhys.93.025010}%
  \BibitemOpen
  \bibfield  {author} {\bibinfo {author} {\bibfnamefont {E.}~\bibnamefont
  {Tiesinga}}, \bibinfo {author} {\bibfnamefont {P.~J.}\ \bibnamefont {Mohr}},
  \bibinfo {author} {\bibfnamefont {D.~B.}\ \bibnamefont {Newell}},\ and\
  \bibinfo {author} {\bibfnamefont {B.~N.}\ \bibnamefont {Taylor}},\ }\href
  {https://doi.org/10.1103/RevModPhys.93.025010} {\bibfield  {journal}
  {\bibinfo  {journal} {Rev. Mod. Phys.}\ }\textbf {\bibinfo {volume} {93}},\
  \bibinfo {pages} {025010} (\bibinfo {year} {2021})}\BibitemShut {NoStop}%
\bibitem [{\citenamefont {Atac}\ \emph {et~al.}(2021)\citenamefont {Atac},
  \citenamefont {Constantinou}, \citenamefont {Meziani}, \citenamefont
  {Paolone},\ and\ \citenamefont {Sparveris}}]{atac2021measurement}%
  \BibitemOpen
  \bibfield  {author} {\bibinfo {author} {\bibfnamefont {H.}~\bibnamefont
  {Atac}}, \bibinfo {author} {\bibfnamefont {M.}~\bibnamefont {Constantinou}},
  \bibinfo {author} {\bibfnamefont {Z.-E.}\ \bibnamefont {Meziani}}, \bibinfo
  {author} {\bibfnamefont {M.}~\bibnamefont {Paolone}},\ and\ \bibinfo {author}
  {\bibfnamefont {N.}~\bibnamefont {Sparveris}},\ }\href
  {https://doi.org/10.1038/s41467-021-22028-z} {\bibfield  {journal} {\bibinfo
  {journal} {Nat. Commun.}\ }\textbf {\bibinfo {volume} {12}},\ \bibinfo
  {pages} {1} (\bibinfo {year} {2021})}\BibitemShut {NoStop}%
\bibitem [{\citenamefont {Engel}(2002)}]{engel2002relativistic}%
  \BibitemOpen
  \bibfield  {author} {\bibinfo {author} {\bibfnamefont {E.}~\bibnamefont
  {Engel}},\ }\href@noop {} {\emph {\bibinfo {title} {Relativistic Electronic
  Structure Theory, Part 1. Fundamentals}}}\ (\bibinfo  {publisher} {Elsevier,
  Amsterdam},\ \bibinfo {year} {2002})\ pp.\ \bibinfo {pages}
  {524--624}\BibitemShut {NoStop}%
\bibitem [{\citenamefont {Grant}(2007)}]{grant2007relativistic}%
  \BibitemOpen
  \bibfield  {author} {\bibinfo {author} {\bibfnamefont {I.~P.}\ \bibnamefont
  {Grant}},\ }\href {https://doi.org/10.1007/978-0-387-35069-1} {\emph
  {\bibinfo {title} {Relativistic quantum theory of atoms and molecules: theory
  and computation}}},\ Vol.~\bibinfo {volume} {40}\ (\bibinfo  {publisher}
  {Springer, New York},\ \bibinfo {year} {2007})\BibitemShut {NoStop}%
\bibitem [{\citenamefont {Rosman}\ and\ \citenamefont
  {Taylor}(1999)}]{rosman1999table}%
  \BibitemOpen
  \bibfield  {author} {\bibinfo {author} {\bibfnamefont {K.~J.~R.}\
  \bibnamefont {Rosman}}\ and\ \bibinfo {author} {\bibfnamefont {P.~D.~P.}\
  \bibnamefont {Taylor}},\ }\href@noop {} {\bibfield  {journal} {\bibinfo
  {journal} {Pure Appl. Chem.}\ }\textbf {\bibinfo {volume} {71}},\ \bibinfo
  {pages} {1593} (\bibinfo {year} {1999})}\BibitemShut {NoStop}%
\bibitem [{\citenamefont {Jiao}\ \emph
  {et~al.}(2021{\natexlab{a}})\citenamefont {Jiao}, \citenamefont {He},
  \citenamefont {Liu}, \citenamefont {Zhang},\ and\ \citenamefont
  {Ho}}]{PhysRevA.104.022801}%
  \BibitemOpen
  \bibfield  {author} {\bibinfo {author} {\bibfnamefont {L.~G.}\ \bibnamefont
  {Jiao}}, \bibinfo {author} {\bibfnamefont {Y.~Y.}\ \bibnamefont {He}},
  \bibinfo {author} {\bibfnamefont {A.}~\bibnamefont {Liu}}, \bibinfo {author}
  {\bibfnamefont {Y.~Z.}\ \bibnamefont {Zhang}},\ and\ \bibinfo {author}
  {\bibfnamefont {Y.~K.}\ \bibnamefont {Ho}},\ }\href
  {https://doi.org/10.1103/PhysRevA.104.022801} {\bibfield  {journal} {\bibinfo
   {journal} {Phys. Rev. A}\ }\textbf {\bibinfo {volume} {104}},\ \bibinfo
  {pages} {022801} (\bibinfo {year} {2021}{\natexlab{a}})}\BibitemShut
  {NoStop}%
\bibitem [{\citenamefont {Chu}\ and\ \citenamefont {Telnov}(2004)}]{SIChu}%
  \BibitemOpen
  \bibfield  {author} {\bibinfo {author} {\bibfnamefont {S.-I.}\ \bibnamefont
  {Chu}}\ and\ \bibinfo {author} {\bibfnamefont {D.~A.}\ \bibnamefont
  {Telnov}},\ }\href
  {https://doi.org/https://doi.org/10.1016/j.physrep.2003.10.001} {\bibfield
  {journal} {\bibinfo  {journal} {Phys. Rep.}\ }\textbf {\bibinfo {volume}
  {390}},\ \bibinfo {pages} {1} (\bibinfo {year} {2004})}\BibitemShut {NoStop}%
\bibitem [{\citenamefont {Xie}\ \emph {et~al.}(2021)\citenamefont {Xie},
  \citenamefont {Jiao}, \citenamefont {Liu},\ and\ \citenamefont
  {Ho}}]{Xie_IJQC}%
  \BibitemOpen
  \bibfield  {author} {\bibinfo {author} {\bibfnamefont {H.~H.}\ \bibnamefont
  {Xie}}, \bibinfo {author} {\bibfnamefont {L.~G.}\ \bibnamefont {Jiao}},
  \bibinfo {author} {\bibfnamefont {A.}~\bibnamefont {Liu}},\ and\ \bibinfo
  {author} {\bibfnamefont {Y.~K.}\ \bibnamefont {Ho}},\ }\href
  {https://doi.org/https://doi.org/10.1002/qua.26653} {\bibfield  {journal}
  {\bibinfo  {journal} {Int. J. Quantum Chem.}\ }\textbf {\bibinfo {volume}
  {121}},\ \bibinfo {pages} {e26653} (\bibinfo {year} {2021})}\BibitemShut
  {NoStop}%
\bibitem [{\citenamefont {Jiao}\ \emph
  {et~al.}(2021{\natexlab{b}})\citenamefont {Jiao}, \citenamefont {Xie},
  \citenamefont {Liu}, \citenamefont {Montgomery},\ and\ \citenamefont
  {Ho}}]{Jiao_2021}%
  \BibitemOpen
  \bibfield  {author} {\bibinfo {author} {\bibfnamefont {L.~G.}\ \bibnamefont
  {Jiao}}, \bibinfo {author} {\bibfnamefont {H.~H.}\ \bibnamefont {Xie}},
  \bibinfo {author} {\bibfnamefont {A.}~\bibnamefont {Liu}}, \bibinfo {author}
  {\bibfnamefont {H.~E.}\ \bibnamefont {Montgomery}},\ and\ \bibinfo {author}
  {\bibfnamefont {Y.~K.}\ \bibnamefont {Ho}},\ }\href
  {https://doi.org/10.1088/1361-6455/ac259c} {\bibfield  {journal} {\bibinfo
  {journal} {J. Phys. B: At. Mol. Opt. Phys.}\ }\textbf {\bibinfo {volume}
  {54}},\ \bibinfo {pages} {175002} (\bibinfo {year}
  {2021}{\natexlab{b}})}\BibitemShut {NoStop}%
\bibitem [{\citenamefont {{De Vries}}\ \emph {et~al.}(1987)\citenamefont {{De
  Vries}}, \citenamefont {{De Jager}},\ and\ \citenamefont {{De
  Vries}}}]{DEVRIES1987495}%
  \BibitemOpen
  \bibfield  {author} {\bibinfo {author} {\bibfnamefont {H.}~\bibnamefont {{De
  Vries}}}, \bibinfo {author} {\bibfnamefont {C.}~\bibnamefont {{De Jager}}},\
  and\ \bibinfo {author} {\bibfnamefont {C.}~\bibnamefont {{De Vries}}},\
  }\href {https://doi.org/https://doi.org/10.1016/0092-640X(87)90013-1}
  {\bibfield  {journal} {\bibinfo  {journal} {At. Data. Nucl. Data Tables}\
  }\textbf {\bibinfo {volume} {36}},\ \bibinfo {pages} {495} (\bibinfo {year}
  {1987})}\BibitemShut {NoStop}%
\bibitem [{\citenamefont {Angeli}\ and\ \citenamefont
  {Marinova}(2013)}]{ANGELI201369}%
  \BibitemOpen
  \bibfield  {author} {\bibinfo {author} {\bibfnamefont {I.}~\bibnamefont
  {Angeli}}\ and\ \bibinfo {author} {\bibfnamefont {K.}~\bibnamefont
  {Marinova}},\ }\href
  {https://doi.org/https://doi.org/10.1016/j.adt.2011.12.006} {\bibfield
  {journal} {\bibinfo  {journal} {At. Data. Nucl. Data Tables}\ }\textbf
  {\bibinfo {volume} {99}},\ \bibinfo {pages} {69} (\bibinfo {year}
  {2013})}\BibitemShut {NoStop}%
\bibitem [{\citenamefont {Yerokhin}\ \emph {et~al.}(2002)\citenamefont
  {Yerokhin}, \citenamefont {Indelicato},\ and\ \citenamefont
  {Shabaev}}]{PhysRevLett.89.143001}%
  \BibitemOpen
  \bibfield  {author} {\bibinfo {author} {\bibfnamefont {V.~A.}\ \bibnamefont
  {Yerokhin}}, \bibinfo {author} {\bibfnamefont {P.}~\bibnamefont
  {Indelicato}},\ and\ \bibinfo {author} {\bibfnamefont {V.~M.}\ \bibnamefont
  {Shabaev}},\ }\href {https://doi.org/10.1103/PhysRevLett.89.143001}
  {\bibfield  {journal} {\bibinfo  {journal} {Phys. Rev. Lett.}\ }\textbf
  {\bibinfo {volume} {89}},\ \bibinfo {pages} {143001} (\bibinfo {year}
  {2002})}\BibitemShut {NoStop}%
\bibitem [{\citenamefont {Yerokhin}\ \emph {et~al.}(2003)\citenamefont
  {Yerokhin}, \citenamefont {Indelicato},\ and\ \citenamefont
  {Shabaev}}]{PhysRevLett.91.073001}%
  \BibitemOpen
  \bibfield  {author} {\bibinfo {author} {\bibfnamefont {V.~A.}\ \bibnamefont
  {Yerokhin}}, \bibinfo {author} {\bibfnamefont {P.}~\bibnamefont
  {Indelicato}},\ and\ \bibinfo {author} {\bibfnamefont {V.~M.}\ \bibnamefont
  {Shabaev}},\ }\href {https://doi.org/10.1103/PhysRevLett.91.073001}
  {\bibfield  {journal} {\bibinfo  {journal} {Phys. Rev. Lett.}\ }\textbf
  {\bibinfo {volume} {91}},\ \bibinfo {pages} {073001} (\bibinfo {year}
  {2003})}\BibitemShut {NoStop}%
\bibitem [{\citenamefont {Yerokhin}\ and\ \citenamefont
  {Shabaev}(2001)}]{PhysRevA.64.062507}%
  \BibitemOpen
  \bibfield  {author} {\bibinfo {author} {\bibfnamefont {V.~A.}\ \bibnamefont
  {Yerokhin}}\ and\ \bibinfo {author} {\bibfnamefont {V.~M.}\ \bibnamefont
  {Shabaev}},\ }\href {https://doi.org/10.1103/PhysRevA.64.062507} {\bibfield
  {journal} {\bibinfo  {journal} {Phys. Rev. A}\ }\textbf {\bibinfo {volume}
  {64}},\ \bibinfo {pages} {062507} (\bibinfo {year} {2001})}\BibitemShut
  {NoStop}%
\bibitem [{\citenamefont {Mohr}(1974)}]{MOHR197426}%
  \BibitemOpen
  \bibfield  {author} {\bibinfo {author} {\bibfnamefont {P.~J.}\ \bibnamefont
  {Mohr}},\ }\href
  {https://doi.org/https://doi.org/10.1016/0003-4916(74)90398-4} {\bibfield
  {journal} {\bibinfo  {journal} {Ann. Phys. (NY)}\ }\textbf {\bibinfo {volume}
  {88}},\ \bibinfo {pages} {26} (\bibinfo {year} {1974})}\BibitemShut {NoStop}%
\bibitem [{\citenamefont {Desiderio}\ and\ \citenamefont
  {Johnson}(1971)}]{PhysRevA.3.1267}%
  \BibitemOpen
  \bibfield  {author} {\bibinfo {author} {\bibfnamefont {A.~M.}\ \bibnamefont
  {Desiderio}}\ and\ \bibinfo {author} {\bibfnamefont {W.~R.}\ \bibnamefont
  {Johnson}},\ }\href {https://doi.org/10.1103/PhysRevA.3.1267} {\bibfield
  {journal} {\bibinfo  {journal} {Phys. Rev. A}\ }\textbf {\bibinfo {volume}
  {3}},\ \bibinfo {pages} {1267} (\bibinfo {year} {1971})}\BibitemShut
  {NoStop}%
\bibitem [{\citenamefont {Mohr}(1982)}]{PhysRevA.26.2338}%
  \BibitemOpen
  \bibfield  {author} {\bibinfo {author} {\bibfnamefont {P.~J.}\ \bibnamefont
  {Mohr}},\ }\href {https://doi.org/10.1103/PhysRevA.26.2338} {\bibfield
  {journal} {\bibinfo  {journal} {Phys. Rev. A}\ }\textbf {\bibinfo {volume}
  {26}},\ \bibinfo {pages} {2338} (\bibinfo {year} {1982})}\BibitemShut
  {NoStop}%
\bibitem [{\citenamefont {Shabaev}(1985)}]{shabaev1985mass}%
  \BibitemOpen
  \bibfield  {author} {\bibinfo {author} {\bibfnamefont {V.~M.}\ \bibnamefont
  {Shabaev}},\ }\href {https://doi.org/10.1007/BF01017505} {\bibfield
  {journal} {\bibinfo  {journal} {Theor. Math. Phys.}\ }\textbf {\bibinfo
  {volume} {63}},\ \bibinfo {pages} {588} (\bibinfo {year} {1985})}\BibitemShut
  {NoStop}%
\bibitem [{\citenamefont {Shabaev}\ \emph
  {et~al.}(1998{\natexlab{a}})\citenamefont {Shabaev}, \citenamefont
  {Artemyev}, \citenamefont {Beier},\ and\ \citenamefont
  {Soff}}]{Shabaev_1998}%
  \BibitemOpen
  \bibfield  {author} {\bibinfo {author} {\bibfnamefont {V.~M.}\ \bibnamefont
  {Shabaev}}, \bibinfo {author} {\bibfnamefont {A.~N.}\ \bibnamefont
  {Artemyev}}, \bibinfo {author} {\bibfnamefont {T.}~\bibnamefont {Beier}},\
  and\ \bibinfo {author} {\bibfnamefont {G.}~\bibnamefont {Soff}},\ }\href
  {https://doi.org/10.1088/0953-4075/31/8/002} {\bibfield  {journal} {\bibinfo
  {journal} {J. Phys. B: At. Mol. Opt. Phys.}\ }\textbf {\bibinfo {volume}
  {31}},\ \bibinfo {pages} {L337} (\bibinfo {year}
  {1998}{\natexlab{a}})}\BibitemShut {NoStop}%
\bibitem [{\citenamefont {Artemyev}\ \emph {et~al.}(1995)\citenamefont
  {Artemyev}, \citenamefont {Shabaev},\ and\ \citenamefont
  {Yerokhin}}]{PhysRevA.52.1884}%
  \BibitemOpen
  \bibfield  {author} {\bibinfo {author} {\bibfnamefont {A.~N.}\ \bibnamefont
  {Artemyev}}, \bibinfo {author} {\bibfnamefont {V.~M.}\ \bibnamefont
  {Shabaev}},\ and\ \bibinfo {author} {\bibfnamefont {V.~A.}\ \bibnamefont
  {Yerokhin}},\ }\href {https://doi.org/10.1103/PhysRevA.52.1884} {\bibfield
  {journal} {\bibinfo  {journal} {Phys. Rev. A}\ }\textbf {\bibinfo {volume}
  {52}},\ \bibinfo {pages} {1884} (\bibinfo {year} {1995})}\BibitemShut
  {NoStop}%
\bibitem [{\citenamefont {Shabaev}\ \emph
  {et~al.}(1998{\natexlab{b}})\citenamefont {Shabaev}, \citenamefont
  {Artemyev}, \citenamefont {Beier}, \citenamefont {Plunien}, \citenamefont
  {Yerokhin},\ and\ \citenamefont {Soff}}]{PhysRevA.57.4235}%
  \BibitemOpen
  \bibfield  {author} {\bibinfo {author} {\bibfnamefont {V.~M.}\ \bibnamefont
  {Shabaev}}, \bibinfo {author} {\bibfnamefont {A.~N.}\ \bibnamefont
  {Artemyev}}, \bibinfo {author} {\bibfnamefont {T.}~\bibnamefont {Beier}},
  \bibinfo {author} {\bibfnamefont {G.}~\bibnamefont {Plunien}}, \bibinfo
  {author} {\bibfnamefont {V.~A.}\ \bibnamefont {Yerokhin}},\ and\ \bibinfo
  {author} {\bibfnamefont {G.}~\bibnamefont {Soff}},\ }\href
  {https://doi.org/10.1103/PhysRevA.57.4235} {\bibfield  {journal} {\bibinfo
  {journal} {Phys. Rev. A}\ }\textbf {\bibinfo {volume} {57}},\ \bibinfo
  {pages} {4235} (\bibinfo {year} {1998}{\natexlab{b}})}\BibitemShut {NoStop}%
\bibitem [{\citenamefont {Zatorski}\ \emph {et~al.}(2012)\citenamefont
  {Zatorski}, \citenamefont {Oreshkina}, \citenamefont {Keitel},\ and\
  \citenamefont {Harman}}]{PhysRevLett.108.063005}%
  \BibitemOpen
  \bibfield  {author} {\bibinfo {author} {\bibfnamefont {J.}~\bibnamefont
  {Zatorski}}, \bibinfo {author} {\bibfnamefont {N.~S.}\ \bibnamefont
  {Oreshkina}}, \bibinfo {author} {\bibfnamefont {C.~H.}\ \bibnamefont
  {Keitel}},\ and\ \bibinfo {author} {\bibfnamefont {Z.}~\bibnamefont
  {Harman}},\ }\href {https://doi.org/10.1103/PhysRevLett.108.063005}
  {\bibfield  {journal} {\bibinfo  {journal} {Phys. Rev. Lett.}\ }\textbf
  {\bibinfo {volume} {108}},\ \bibinfo {pages} {063005} (\bibinfo {year}
  {2012})}\BibitemShut {NoStop}%
\bibitem [{\citenamefont {Michel}\ \emph {et~al.}(2019)\citenamefont {Michel},
  \citenamefont {Zatorski}, \citenamefont {Oreshkina},\ and\ \citenamefont
  {Keitel}}]{PhysRevA.99.012505}%
  \BibitemOpen
  \bibfield  {author} {\bibinfo {author} {\bibfnamefont {N.}~\bibnamefont
  {Michel}}, \bibinfo {author} {\bibfnamefont {J.}~\bibnamefont {Zatorski}},
  \bibinfo {author} {\bibfnamefont {N.~S.}\ \bibnamefont {Oreshkina}},\ and\
  \bibinfo {author} {\bibfnamefont {C.~H.}\ \bibnamefont {Keitel}},\ }\href
  {https://doi.org/10.1103/PhysRevA.99.012505} {\bibfield  {journal} {\bibinfo
  {journal} {Phys. Rev. A}\ }\textbf {\bibinfo {volume} {99}},\ \bibinfo
  {pages} {012505} (\bibinfo {year} {2019})}\BibitemShut {NoStop}%
\bibitem [{\citenamefont {Nefiodov}\ \emph {et~al.}(1996)\citenamefont
  {Nefiodov}, \citenamefont {Labzowsky}, \citenamefont {Plunien},\ and\
  \citenamefont {Soff}}]{NEFIODOV1996227}%
  \BibitemOpen
  \bibfield  {author} {\bibinfo {author} {\bibfnamefont {A.~V.}\ \bibnamefont
  {Nefiodov}}, \bibinfo {author} {\bibfnamefont {L.~N.}\ \bibnamefont
  {Labzowsky}}, \bibinfo {author} {\bibfnamefont {G.}~\bibnamefont {Plunien}},\
  and\ \bibinfo {author} {\bibfnamefont {G.}~\bibnamefont {Soff}},\ }\href
  {https://doi.org/https://doi.org/10.1016/0375-9601(96)00650-0} {\bibfield
  {journal} {\bibinfo  {journal} {Phys. Lett. A}\ }\textbf {\bibinfo {volume}
  {222}},\ \bibinfo {pages} {227} (\bibinfo {year} {1996})}\BibitemShut
  {NoStop}%
\bibitem [{\citenamefont {Plunien}\ \emph {et~al.}(1991)\citenamefont
  {Plunien}, \citenamefont {M\"uller}, \citenamefont {Greiner},\ and\
  \citenamefont {Soff}}]{PhysRevA.43.5853}%
  \BibitemOpen
  \bibfield  {author} {\bibinfo {author} {\bibfnamefont {G.}~\bibnamefont
  {Plunien}}, \bibinfo {author} {\bibfnamefont {B.}~\bibnamefont {M\"uller}},
  \bibinfo {author} {\bibfnamefont {W.}~\bibnamefont {Greiner}},\ and\ \bibinfo
  {author} {\bibfnamefont {G.}~\bibnamefont {Soff}},\ }\href
  {https://doi.org/10.1103/PhysRevA.43.5853} {\bibfield  {journal} {\bibinfo
  {journal} {Phys. Rev. A}\ }\textbf {\bibinfo {volume} {43}},\ \bibinfo
  {pages} {5853} (\bibinfo {year} {1991})}\BibitemShut {NoStop}%
\bibitem [{\citenamefont {Pachucki}\ and\ \citenamefont
  {Moro}(2007)}]{PhysRevA.75.032521}%
  \BibitemOpen
  \bibfield  {author} {\bibinfo {author} {\bibfnamefont {K.}~\bibnamefont
  {Pachucki}}\ and\ \bibinfo {author} {\bibfnamefont {A.~M.}\ \bibnamefont
  {Moro}},\ }\href {https://doi.org/10.1103/PhysRevA.75.032521} {\bibfield
  {journal} {\bibinfo  {journal} {Phys. Rev. A}\ }\textbf {\bibinfo {volume}
  {75}},\ \bibinfo {pages} {032521} (\bibinfo {year} {2007})}\BibitemShut
  {NoStop}%
\bibitem [{\citenamefont {Flambaum}\ \emph {et~al.}(2021)\citenamefont
  {Flambaum}, \citenamefont {Samsonov}, \citenamefont {Tan},\ and\
  \citenamefont {Viatkina}}]{PhysRevA.103.032811}%
  \BibitemOpen
  \bibfield  {author} {\bibinfo {author} {\bibfnamefont {V.~V.}\ \bibnamefont
  {Flambaum}}, \bibinfo {author} {\bibfnamefont {I.~B.}\ \bibnamefont
  {Samsonov}}, \bibinfo {author} {\bibfnamefont {H.~B.~T.}\ \bibnamefont
  {Tan}},\ and\ \bibinfo {author} {\bibfnamefont {A.~V.}\ \bibnamefont
  {Viatkina}},\ }\href {https://doi.org/10.1103/PhysRevA.103.032811} {\bibfield
   {journal} {\bibinfo  {journal} {Phys. Rev. A}\ }\textbf {\bibinfo {volume}
  {103}},\ \bibinfo {pages} {032811} (\bibinfo {year} {2021})}\BibitemShut
  {NoStop}%
\bibitem [{\citenamefont {Volotka}\ and\ \citenamefont
  {Plunien}(2014{\natexlab{b}})}]{PhysRevLett.113.023002}%
  \BibitemOpen
  \bibfield  {author} {\bibinfo {author} {\bibfnamefont {A.~V.}\ \bibnamefont
  {Volotka}}\ and\ \bibinfo {author} {\bibfnamefont {G.}~\bibnamefont
  {Plunien}},\ }\href {https://doi.org/10.1103/PhysRevLett.113.023002}
  {\bibfield  {journal} {\bibinfo  {journal} {Phys. Rev. Lett.}\ }\textbf
  {\bibinfo {volume} {113}},\ \bibinfo {pages} {023002} (\bibinfo {year}
  {2014}{\natexlab{b}})}\BibitemShut {NoStop}%
\bibitem [{\citenamefont {Plunien}\ and\ \citenamefont
  {Soff}(1996)}]{PhysRevA.53.4614.2}%
  \BibitemOpen
  \bibfield  {author} {\bibinfo {author} {\bibfnamefont {G.}~\bibnamefont
  {Plunien}}\ and\ \bibinfo {author} {\bibfnamefont {G.}~\bibnamefont {Soff}},\
  }\href {https://doi.org/10.1103/PhysRevA.53.4614.2} {\bibfield  {journal}
  {\bibinfo  {journal} {Phys. Rev. A}\ }\textbf {\bibinfo {volume} {53}},\
  \bibinfo {pages} {4614} (\bibinfo {year} {1996})}\BibitemShut {NoStop}%
\bibitem [{\citenamefont {Beier}\ \emph {et~al.}(1998)\citenamefont {Beier},
  \citenamefont {Mohr}, \citenamefont {Persson},\ and\ \citenamefont
  {Soff}}]{PhysRevA.58.954}%
  \BibitemOpen
  \bibfield  {author} {\bibinfo {author} {\bibfnamefont {T.}~\bibnamefont
  {Beier}}, \bibinfo {author} {\bibfnamefont {P.~J.}\ \bibnamefont {Mohr}},
  \bibinfo {author} {\bibfnamefont {H.}~\bibnamefont {Persson}},\ and\ \bibinfo
  {author} {\bibfnamefont {G.}~\bibnamefont {Soff}},\ }\href
  {https://doi.org/10.1103/PhysRevA.58.954} {\bibfield  {journal} {\bibinfo
  {journal} {Phys. Rev. A}\ }\textbf {\bibinfo {volume} {58}},\ \bibinfo
  {pages} {954} (\bibinfo {year} {1998})}\BibitemShut {NoStop}%
\bibitem [{\citenamefont {Bondarev}\ \emph {et~al.}(2010)\citenamefont
  {Bondarev}, \citenamefont {Kozhedub},\ and\ \citenamefont
  {Oreshkina}}]{bondarev2010finite}%
  \BibitemOpen
  \bibfield  {author} {\bibinfo {author} {\bibfnamefont {A.~I.}\ \bibnamefont
  {Bondarev}}, \bibinfo {author} {\bibfnamefont {Y.~S.}\ \bibnamefont
  {Kozhedub}},\ and\ \bibinfo {author} {\bibfnamefont {N.~S.}\ \bibnamefont
  {Oreshkina}},\ }\href {https://doi.org/10.1134/S0030400X10120027} {\bibfield
  {journal} {\bibinfo  {journal} {Opt. Spectrosc.}\ }\textbf {\bibinfo {volume}
  {109}},\ \bibinfo {pages} {823} (\bibinfo {year} {2010})}\BibitemShut
  {NoStop}%
\bibitem [{\citenamefont {Eides}\ and\ \citenamefont
  {Grotch}(1997)}]{PhysRevA.56.R2507}%
  \BibitemOpen
  \bibfield  {author} {\bibinfo {author} {\bibfnamefont {M.~I.}\ \bibnamefont
  {Eides}}\ and\ \bibinfo {author} {\bibfnamefont {H.}~\bibnamefont {Grotch}},\
  }\href {https://doi.org/10.1103/PhysRevA.56.R2507} {\bibfield  {journal}
  {\bibinfo  {journal} {Phys. Rev. A}\ }\textbf {\bibinfo {volume} {56}},\
  \bibinfo {pages} {R2507} (\bibinfo {year} {1997})}\BibitemShut {NoStop}%
\bibitem [{\citenamefont {Milstein}\ \emph {et~al.}(2004)\citenamefont
  {Milstein}, \citenamefont {Sushkov},\ and\ \citenamefont
  {Terekhov}}]{PhysRevA.69.022114}%
  \BibitemOpen
  \bibfield  {author} {\bibinfo {author} {\bibfnamefont {A.~I.}\ \bibnamefont
  {Milstein}}, \bibinfo {author} {\bibfnamefont {O.~P.}\ \bibnamefont
  {Sushkov}},\ and\ \bibinfo {author} {\bibfnamefont {I.~S.}\ \bibnamefont
  {Terekhov}},\ }\href {https://doi.org/10.1103/PhysRevA.69.022114} {\bibfield
  {journal} {\bibinfo  {journal} {Phys. Rev. A}\ }\textbf {\bibinfo {volume}
  {69}},\ \bibinfo {pages} {022114} (\bibinfo {year} {2004})}\BibitemShut
  {NoStop}%
\bibitem [{\citenamefont {Yerokhin}(2011)}]{PhysRevA.83.012507}%
  \BibitemOpen
  \bibfield  {author} {\bibinfo {author} {\bibfnamefont {V.~A.}\ \bibnamefont
  {Yerokhin}},\ }\href {https://doi.org/10.1103/PhysRevA.83.012507} {\bibfield
  {journal} {\bibinfo  {journal} {Phys. Rev. A}\ }\textbf {\bibinfo {volume}
  {83}},\ \bibinfo {pages} {012507} (\bibinfo {year} {2011})}\BibitemShut
  {NoStop}%
\end{thebibliography}%

\end{document}


\title{Supplementary Materials: \\Finite-nuclear-size effect in hydrogen-like ions with relativistic nuclear structure}

\author{Hui Hui Xie}
\author{Jian Li}\email{jianli@jlu.edu.cn}
\author{Li Guang Jiao}\email{lgjiao@jlu.edu.cn}
\affiliation{College of Physics, Jilin University, Changchun 130012}

\author{Yew Kam Ho}
\affiliation{Institute of Atomic and Molecular Sciences, Academia Sinica, Taipei 10617}

\date{\today}

\maketitle

Table \ref{tabsm} presents the FNS corrections to atomic energy levels $\Delta E_{\rm FNS}$ and bound-electron $g$ factors $\Delta g_{\rm FNS}$ for the $1s_{1/2}$, $2s_{1/2}$, and $2p_{1/2}$ states of hydrogen-like ions with even-even nuclei in the rang of $8\le Z\le 108$.

\newpage

\setlength{\tabcolsep}{5pt}
\begin{longtable*}{lll|lll|llll}
\caption{FNS corrections to atomic energy levels $\Delta E_{\rm FNS}$ (in units of eV) and bound-electron $g$ factors $\Delta g_{\rm FNS}$ for the $1s_{1/2}$, $2s_{1/2}$, and $2p_{1/2}$ states for hydrogen-like ions. The root-mean-square radius $r_c$ are given in units of fm. Numbers in parentheses represent the power of ten.}
\label{tabsm}\\
\toprule
&&&\multicolumn{3}{c}{{$\Delta E_{\rm FNS}(\rm eV)$}}\vline & \multicolumn{3}{c}{$\Delta g_{\rm FNS}$} \\
\hline
$Z$ & $A$ & \multicolumn{1}{c}{$r_c$} \vline &\multicolumn{1}{c}{$1s_{1/2}$} &\multicolumn{1}{c}{$2s_{1/2}$}  &\multicolumn{1}{c}{$ 2p_{1/2}$} \vline &\multicolumn{1}{c}{$1s_{1/2}$} &\multicolumn{1}{c}{$2s_{1/2}$}         &\multicolumn{1}{c}{$ 2p_{1/2}$} \\
\hline
		8  &16  &2.7597 &2.0771(-4) &2.6041(-5) &1.6680(-8) &1.6236(-9) &2.0356(-10) &1.3041(-13) \\
		10 &20  &2.9062 &5.7039(-4) &7.1634(-5) &7.1777(-8) &4.4556(-9) &5.5956(-10) &5.6080(-13) \\
		12 &24  &2.9859 &1.2699(-3) &1.5981(-4) &2.3092(-7) &9.9112(-9) &1.2473(-9) &1.8027(-12) \\
		14 &28  &2.9894 &2.4053(-3) &3.0344(-4) &5.9776(-7) &1.8754(-8) &2.3659(-9) &4.6621(-12) \\
		16 &32  &3.1901 &4.7740(-3) &6.0396(-4) &1.5570(-6) &3.7179(-8) &4.7035(-9) &1.2130(-11) \\
		18 &40  &3.3869 &8.8270(-3) &1.1203(-3) &3.6633(-6) &6.8652(-8) &8.7131(-9) &2.8504(-11) \\
		20 &40  &3.4750 &1.4543(-2) &1.8523(-3) &7.4960(-6) &1.1294(-7) &1.4385(-8) &5.8244(-11) \\
		22 &48  &3.5384 &2.2726(-2) &2.9061(-3) &1.4268(-5) &1.7621(-7) &2.2533(-8) &1.1069(-10) \\
		24 &52  &3.5930 &3.4246(-2) &4.3983(-3) &2.5773(-5) &2.6506(-7) &3.4042(-8) &1.9961(-10) \\
		26 &56  &3.6769 &5.1068(-2) &6.5899(-3) &4.5464(-5) &3.9450(-7) &5.0906(-8) &3.5145(-10) \\
		28 &58  &3.7284 &7.3201(-2) &9.4942(-3) &7.6225(-5) &5.6429(-7) &7.3188(-8) &5.8806(-10) \\
		30 &64  &3.8726 &1.0796(-1) &1.4079(-2) &1.3025(-4) &8.3032(-7) &1.0829(-7) &1.0026(-9) \\
		32 &74  &4.0252 &1.5693(-1) &2.0586(-2) &2.1756(-4) &1.2040(-6) &1.5795(-7) &1.6708(-9) \\
		34 &80  &4.0976 &2.1609(-1) &2.8525(-2) &3.4176(-4) &1.6536(-6) &2.1829(-7) &2.6182(-9) \\
		36 &84  &4.1522 &2.9143(-1) &3.8728(-2) &5.2255(-4) &2.2241(-6) &2.9555(-7) &3.9925(-9) \\
		38 &88  &4.1999 &3.8766(-1) &5.1881(-2) &7.8370(-4) &2.9499(-6) &3.9477(-7) &5.9708(-9) \\
		40 &90  &4.2466 &5.1068(-1) &6.8854(-2) &1.1583(-3) &3.8739(-6) &5.2229(-7) &8.7984(-9) \\
		42 &98  &4.3471 &6.8323(-1) &9.2845(-2) &1.7314(-3) &5.1657(-6) &7.0194(-7) &1.3109(-8) \\
		44 &102 &4.4160 &8.9453(-1) &1.2256(-1) &2.5228(-3) &6.7398(-6) &9.2340(-7) &1.9037(-8) \\
		46 &106 &4.4819 &1.1618 	&1.6057(-1) &3.6341(-3) &8.7216(-6) &1.2053(-6) &2.7325(-8) \\
		48 &114 &4.5784 &1.5189  	&2.1182(-1) &5.2533(-3) &1.1358(-5) &1.5838(-6) &3.9352(-8) \\
		50 &120 &4.6389 &1.9457 	&2.7393(-1) &7.4206(-3) &1.4490(-5) &2.0398(-6) &5.5367(-8) \\
		52 &130 &4.7293 &2.5102  	&3.5689(-1) &1.0531(-2) &1.8614(-5) &2.6462(-6) &7.8247(-8) \\
		54 &132 &4.7723 &3.1649 	&4.5463(-1) &1.4572(-2) &2.3364(-5) &3.3557(-6) &1.0781(-7) \\
		56 &138 &4.8289 &3.9960 	&5.8020(-1) &2.0155(-2) &2.9361(-5) &4.2624(-6) &1.4842(-7) \\
		58 &140 &4.8661 &4.9948 	&7.3332(-1) &2.7546(-2) &3.6519(-5) &5.3607(-6) &2.0189(-7) \\
		60 &142 &4.9003 &6.2170 	&9.2336(-1) &3.7432(-2) &4.5221(-5) &6.7149(-6) &2.7296(-7) \\
		62 &152 &5.0273 &7.9819 	&1.1997 	&5.2400(-2) &5.7740(-5) &8.6768(-6) &3.8009(-7) \\
		64 &158 &5.0964 &10.008 	&1.5231 	&7.1545(-2) &7.1987(-5) &1.0952(-5) &5.1610(-7) \\
		66 &164 &5.1569 &12.485 	&1.9245 	&9.7077(-2) &8.9266(-5) &1.3756(-5) &6.9624(-7) \\
		68 &166 &5.1863 &15.387 	&2.4036 	&1.3001(-1) &1.0933(-4) &1.7073(-5) &9.2679(-7) \\
		70 &174 &5.2529 &19.172 	&3.0363 	&1.7589(-1) &1.3535(-4) &2.1427(-5) &1.2460(-6) \\
		72 &180 &5.3052 &23.748 	&3.8147 	&2.3640(-1) &1.6651(-4) &2.6735(-5) &1.6635(-6) \\
		74 &184 &5.3435 &29.258 	&4.7690 	&3.1582(-1) &2.0368(-4) &3.3184(-5) &2.2071(-6) \\
		76 &192 &5.4029 &36.259 	&6.0001 	&4.2425(-1) &2.5054(-4) &4.1435(-5) &2.9434(-6) \\
		78 &196 &5.4356 &44.545 	&7.4869 	&5.6472(-1) &3.0540(-4) &5.1297(-5) &3.8883(-6) \\
		80 &202 &5.4773 &54.859 	&9.3694 	&7.5336(-1) &3.7304(-4) &6.3666(-5) &5.1461(-6) \\
		82 &208 &5.5055 &67.278 	&11.682 	&1.0007 	&4.5358(-4) &7.8695(-5) &6.7787(-6) \\
		84 &210 &5.5384 &82.665 	&14.600 	&1.3316  	&5.5233(-4) &9.7460(-5) &8.9422(-6) \\
		86 &222 &5.6820 &104.74 	&18.825 	&1.8280 	&6.9318(-4) &1.2445(-4) &1.2163(-5) \\
		88 &226 &5.7367 &129.49 	&23.695 	&2.4483 	&8.4850(-4) &1.5507(-4) &1.6135(-5) \\
		90 &232 &5.7954 &160.33 	&29.884 	&3.2850 	&1.0396(-3) &1.9351(-4) &2.1431(-5) \\
		92 &238 &5.8462 &198.24 	&37.659 	&4.4033 	&1.2714(-3) &2.4115(-4) &2.8422(-5) \\
		94 &244 &5.8947 &245.19 	&47.494 	&5.9070 	&1.5545(-3) &3.0057(-4) &3.7704(-5) \\
		96 &248 &5.9286 &302.56 	&59.791 	&7.9104  	&1.8951(-3) &3.7374(-4) &4.9900(-5) \\
		98 &252 &5.9643 &374.10 	&75.461 	&10.622  	&2.3133(-3) &4.6555(-4) &6.6179(-5) \\
		100&258 &6.0111 &464.53 	&95.694 	&14.338  	&2.8337(-3) &5.8223(-4) &8.8159(-5) \\
		102&260 &6.0341 &574.89 	&121.01 	&19.307  	&3.4572(-3) &7.2556(-4) &1.1707(-4) \\
		104&264 &6.0676 &714.78 	&153.82 	&26.149  	&4.2336(-3) &9.0799(-4) &1.5623(-4) \\
		108&270 &6.1174 &1110.0 	&250.06 	&48.383  	&6.3597(-3) &1.4264(-3) &2.7982(-4) \\
		110&272 &6.1356 &1388.2 	&320.22 	&66.207 	&7.8096(-3) &1.7924(-3) &3.7609(-4) \\
		112&278 &6.1698 &1747.7 	&413.02 	&91.385  	&9.6421(-3) &2.2654(-3) &5.0920(-4) \\
		114&290 &6.2401 &2223.2 	&538.43 	&127.73   	&1.2010(-2) &2.8893(-3) &6.9705(-4) \\
		116&290 &6.2512 &2812.3 	&698.46 	&177.95   	&1.4855(-2) &3.6613(-3) &9.4950(-4) \\
		118&294 &6.2719 &3583.1 	&912.91 	&250.37   	&1.8472(-2) &4.6656(-3) &1.3036(-3) \\
\toprule
\end{longtable*}
